\documentclass[a4paper]{aa}
\usepackage{graphicx}
\usepackage{natbib}
\usepackage{multirow}
\usepackage{longtable,lscape}

\usepackage{txfonts}  

\def\atlas9{{\sc ATLAS9}}

\begin{document}

\title{Spectral optical monitoring of a double-peaked emission line AGN Arp 102B:
I. Variability of spectral lines and continuum}

\author{A. I. Shapovalova\inst{1} \and L.\v C. Popovi\'c\inst{2,3}
\and A. N. Burenkov\inst{1} \and V.H. Chavushyan\inst{4} \and D. Ili\'c
\inst{3,5} \and W. Kollatschny\inst{6} \and A. Kova\v
cevi\'c\inst{3,5} \and N. G. Bochkarev\inst{7}
\and  J. R. Vald\'es\inst{4} \and  J. Torrealba\inst{4}
\and V. Pati\~no-\'Alvarez\inst{4} \and J. Le\'on-Tavares\inst{8, 9} \and E. Benitez\inst{10}
\and L. Carrasco\inst{4} \and D. Dultzin\inst{10} \and A.
Mercado\inst{11} \and V. E. Zhdanova\inst{1}}

\titlerunning{Spectral monitoring of Arp 102B}
\authorrunning{A.I. Shapovalova et al.}
\offprints{A.I. Shapovalova, \\ \email{ashap@sao.ru}\\ }

\institute{Special Astrophysical Observatory of the Russian AS,
Nizhnij Arkhyz, Karachaevo-Cherkesia 369167, Russia \and
Astronomical Observatory, Volgina 7, 11160 Belgrade 74, Serbia \and
Isaac Newton Institute of Chile, Yugoslavia Branch \and Instituto
Nacional de Astrof\'{\i}sica, \'{O}ptica y Electr\'onica, Apartado
Postal 51, CP 72000, Puebla, Pue. M\'exico \and
 Department of Astronomy, Faculty of Mathematics, University
of Belgrade, Studentski trg 16, 11000 Belgrade, Serbia \and Institut
f\"ur Astrophysik, Friedrich-Hund-Platz 1, G\"ottingen, Germany \and
Sternberg Astronomical Institute, Moscow, Russia
\and
Finnish Centre for Astronomy with ESO (FINCA), University of Turku,
V\"ais\"al\"antie 20, FI-21500 Piikki\"o, Finland
\and
Aalto University Mets\"ahovi Radio Observatory, Mets\"ahovintie 114, FIN-02540 Kylm\"al\"a, Finland
\and
Instituto de Astronom\'ia, Universidad Nacional Aut\'onoma de M\'exico, Apartado Postal 70-264,
M\'exico, D.F. 04510, M\'exico
\and
Universidad Polit\'ecnica de Baja California, Av. de la Industrial 291, 21010 Mexicali, B.C., M\'exico
}

\date{Received  / Accepted }

\abstract
{Here we present results of the long-term (1987-2010) optical spectral
monitoring of the broad line radio galaxy Arp 102B, a prototype of
active galactic nuclei with the double-peaked broad emission lines, usually
assumed to be emitted from an accretion disk.}
{To explore the structure of the broad line region (BLR), we analyze the
light curves of the broad H$\alpha$ and H$\beta$ lines and the
continuum flux. We aim to estimate the dimensions of the broad-line emitting regions and
the mass of the central black hole.}
{We use the CCF to find lags between the lines and continuum variations. We investigate in
more details the correlation between line and continuum fluxes, moreover we explore
periodical variations of the red-to-blue line flux ratio using Lomb-Scargle periodograms.}
{The line and continuum light curves show several flare-like events. The fluxes in lines
and in the continuum are not showing a big change (around 20\%) during the monitoring period.
We found a small correlation between the line and continuum flux variation, that may indicate that
variation in lines has weak connection with the variation of the central photoionization source.
In spite of a low line-continuum correlation, using several methods, we estimated a time lag
for H$\beta$ around 20 days. The correlation between the H$\beta$ and H$\alpha$ flux variation
is significantly  higher than between lines and continuum.
During the monitoring period, the H$\beta$ and H$\alpha$ lines show double-peaked profiles and
we found an indication for a periodical oscillation in the red-to-blue flux ratio of the H$\alpha$
line. The estimated mass of the central black hole is {  $\sim 1.1 \times 10^8M\odot$ that is in an 
agreement with the mass estimated from the $M-\sigma*$ relation.}}
{}


\maketitle

\section{Introduction}

The broad-line Active Galactic Nuclei (AGNs) showing two peaks in
the broad-line component are assumed to have the broad-line emission
originating in an accretion disk \citep[][]{eh94,e97,er09}. They
also often show a variability in the broad-emission lines and
continuum \citep[see][]{ge07,fl08,sh10,pop11,di12}. The region where broad lines
are formed (hereinafter the BLR -- Broad Line Region) is  close to
the central super-massive black hole and may hold basic information
about the formation and fueling of AGNs. Additionally, the shapes of
broad lines and their variability can give information about the BLR
geometry \citep[see][]{sul00,er09,ga09}.

A long-term spectral monitoring in the optical band of the nucleus
of some AGNs has revealed a time lag in the response of the
broad-emission line fluxes relative to the change in the continuum
flux, that depends on the size, geometry and physical conditions of
the BLR. Therefore, the search for correlations between the nuclear
continuum changes and broad line flux variations  may serve as a
tool for mapping the geometrical and dynamical structure of the BLR
\citep[see e.g.][and reference therein]{pet93}.

Arp 102B is a subluminous, radio-loud LINER 1.8 galaxy at z=0.024
which displays the presence of double-peaked Balmer emission lines
\citep{s83}. This AGN was the first one, where assumption of an
accretion disk BLR geometry is applied by \cite{c89} and
\cite{ch89}, where  a geometrically thin, optically thick accretion
disk model has been used to fit the double-peaked Balmer lines. The
accretion disk emission in the broad line region of Arp 102B has
been widely accepted \citep[see][]{ch89, g04}. Additionally,
\cite{s00} monitored the broad H$\alpha$ line from 1992 to 1996, and
found the variations in the profile that correspond to gas rotating
in a disk with inhomogeneities in the surface brightness. On the
other side \cite{h96} found that high ionized lines, such as
Ly$\alpha$ and C IV $\lambda$1550 do not show the disk-like profile
(two peaks). They found that broad Mg II $\lambda$2798 is present
with nearly the same profile as the Balmer lines (peaks separated by
$\sim$ 12,000 km $s^{-1}$), and a typical Mg II/H$\beta$ ratio of 1,
but they found a little, if any C III] $\lambda$ 1909 or C IV
$\lambda$1550 emission corresponding to the displaced Balmer-line
peaks. Most important, they found that there is no double-peaked
component detected in Ly$\alpha$.

Also, \cite{n97} studied the profile variability of the
double-peaked H$\alpha$ line in Arp 102B over 13 years, and found a
sinusoidal variation of the red-to-blue flux ratio from 1990 to
1994. {  A similar period was found by \cite{ge07}}.
 The authors  modeled this variation as a transient orbiting
hot spot in the accretion disk. Similar variation  can also be a
consequence of gravitational lensing from a massive body close to
the primary black hole \citep[][]{pop01}. Additionally, there are
some contradictions in the disk model
\citep[see][]{mp90,su90,a96,c97,g04}. Moreover, recent \cite{fat11}
discovered a two-armed mini-spiral structure within the inner
kiloparsec resolved in the H$\alpha$ line, that indicates dramatic
processes happening on larger scales in the galaxy.

Studies of the variations in the continuum as well as in the broad
emission line profiles and their correlations can give us
information about the BLR physics \citep[see e.g.][]{sh09}. Here we
present our investigation of a long-term optical spectral variations
of Arp 102B. In this paper, we present the results of the  spectral
(H$\alpha$ and H$\beta$) monitoring of Arp 102B during the period
between 1987 and 2010, discussing the broad line and continuum flux
variability. In Paper II (Shapovalova et al. 2013, in preparation)
we will give more details about the broad line profile variability
and discuss the structure and geometry of the BLR.  The paper is
organized as follows: in \S 2  we report on our observations and
describe the data reduction; in \S 3 we describe the performed data
analysis, and in \S 4 we discuss our results; finally in \S 5 we
outline our conclusions.

\section{Observations and data reduction}

\subsection{Spectral observations}

Spectra of Arp 102B  (during 142 nights) were taken with the 6 m and
1 m telescopes of the SAO RAS (Russia, 1998--2010), the INAOE's 2.1
m telescope of the ''Guillermo Haro Observatory'' (GHO) at Cananea,
Sonora, M\'exico (1998--2007), the 2.1 m telescope of the
Observatorio Astron\'omico Nacional at San Pedro Martir (OAN-SPM),
Baja California, M\'exico (2005--2007), and the 3.5 m and 2.2 m
telescopes of Calar Alto observatory, Spain (1987--1994).
Information on the source of spectroscopic observations are listed
in Table~\ref{tab1}.

The SAO's and Mexican spectra were obtained with long--slit
spectrographs equipped with CCDs. The typical observed wavelength
range was 4000--7500 \AA , the spectral resolution was R=8--15  \AA
, and the S/N ratio was 20--50 in the continuum near H$\alpha$ and
H$\beta$. Additionally, we collected the observations taken with the
Calar Alto 3.5 m and 2.2 m telescopes at 10 epochs between June 1987
and September 1994. For these observations, Boller \& Chivens
spectrographs were attached to the telescopes in most cases (for two
epochs the TWIN spectrograph was used for the 3.5m telescope), and
different CCD detectors (RCA, GEC, Tektronix) were used. The
individual spectra covers different wavelength ranges from
3630\,\AA\ to 9100~\AA\ , and the spectral resolution was 10--15
\AA. The observations were taken with exposure times from 20 to 138
minutes. The typical slit width was
2\arcsec\hspace*{-1ex}.\hspace*{0.3ex}0 projected on the sky. HeAr
spectra were taken after each object exposure to enable a wavelength
calibration.  Further information about observing conditions
with Calar Alto telescopes can be found in \cite{ko00}. From 2004
to 2007, the spectral observations with two Mexican 2.1 m telescopes
were carried out with two observational setups. We used the
following configurations: 1) with a grating of 150 l/mm (spectral
resolution of R=15 \AA, a resolution similar to the observations of
1998--2003); 2) with a grating of 300 l/mm (moderate spectral
resolution of R=8\AA). As a rule, spectra with R$\sim$15 \AA\ have a
very good quality (S/N$>$50), but spectra with R$\sim$8 \AA\ are
with more noise (S/N$\sim$20-40).

From 2004 to 2010 spectral observations with 1 m Zeiss telescope of
the SAO (19 nights) were carried with the CCD 2k$\times$2k
EEV CCD42-40, allowing us to observe the entire wavelength range
(4000--8000) \AA\, with a spectral resolution of R=8--10 \AA.
But some of these spectra had bad S/N ratio in the blue part and have not been
used in the analysis presented here. Although the red part is with the
relatively high S/N ratio, in any case, the spectra taken with the 1 m
Zeiss telescope (code Z2K in Table \ref{tab1}) have to be treated with caution.

Spectrophotometric standard stars were observed every night. The log
of the spectroscopic observations is given in Table~\ref{tab2}
(available electronically only). The spectrophotometric data
reduction was carried out either with the software developed at the
SAO RAS or with the IRAF package for the spectra obtained in
M\'exico and at Calar Alto. The image reduction process included
bias and flat-field corrections, cosmic ray removal, 2D wavelength
linearization, sky spectrum subtraction, addition of the spectra for
every night, and relative flux calibration based on observations of
standard star.

In the analysis, about 10\% of spectra were discarded for several different
reasons (e.g. large noise, badly corrected spectral sensitivity, poor spectral
resolution ($>$15\AA), etc.), thus our final data set consisted of 118 blue and
90 red spectra, which were used in the further analysis.

\begin{figure}
\centering
\includegraphics[width=9cm]{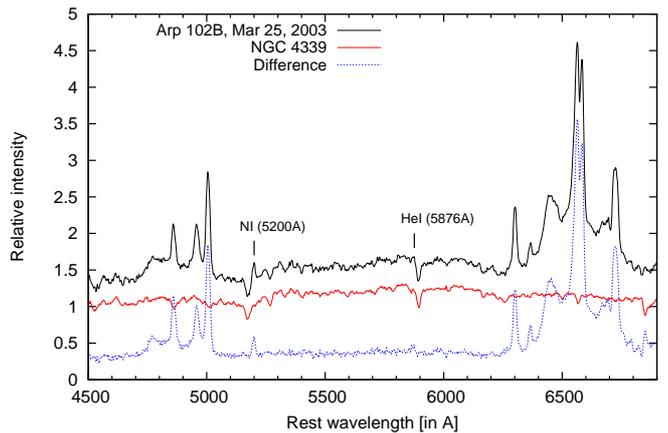}
\caption{  The spectra of Arp 102B (top) and NGC 4339 (75\% flux of Arp 102B, middle) 
taken on Mar 25, 2003, and their difference (bottom) giving the AGN-continuum of Arp 102B.
} \label{host}
\end{figure}

\begin{figure}
\centering
\includegraphics[width=9cm]{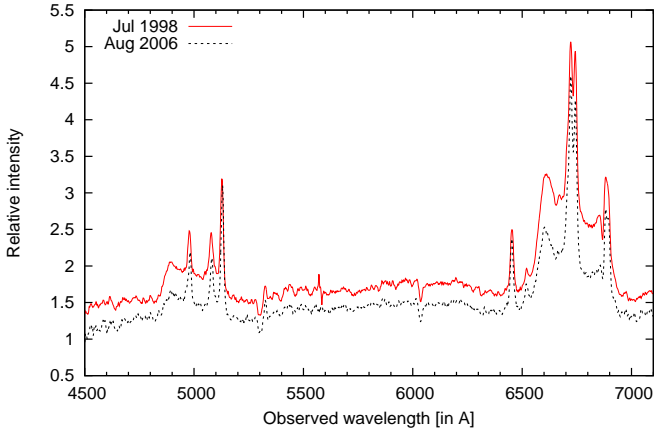}
\caption{Two examples of the total optical spectrum of Arp 102B,
when the object was in a high activity state in July 1998 (solid line), 
and when it was in a low activity state in Aug 2006 (dashed line).} \label{arp}
\end{figure}

\begin{figure*}
\centering
\includegraphics[width=15cm]{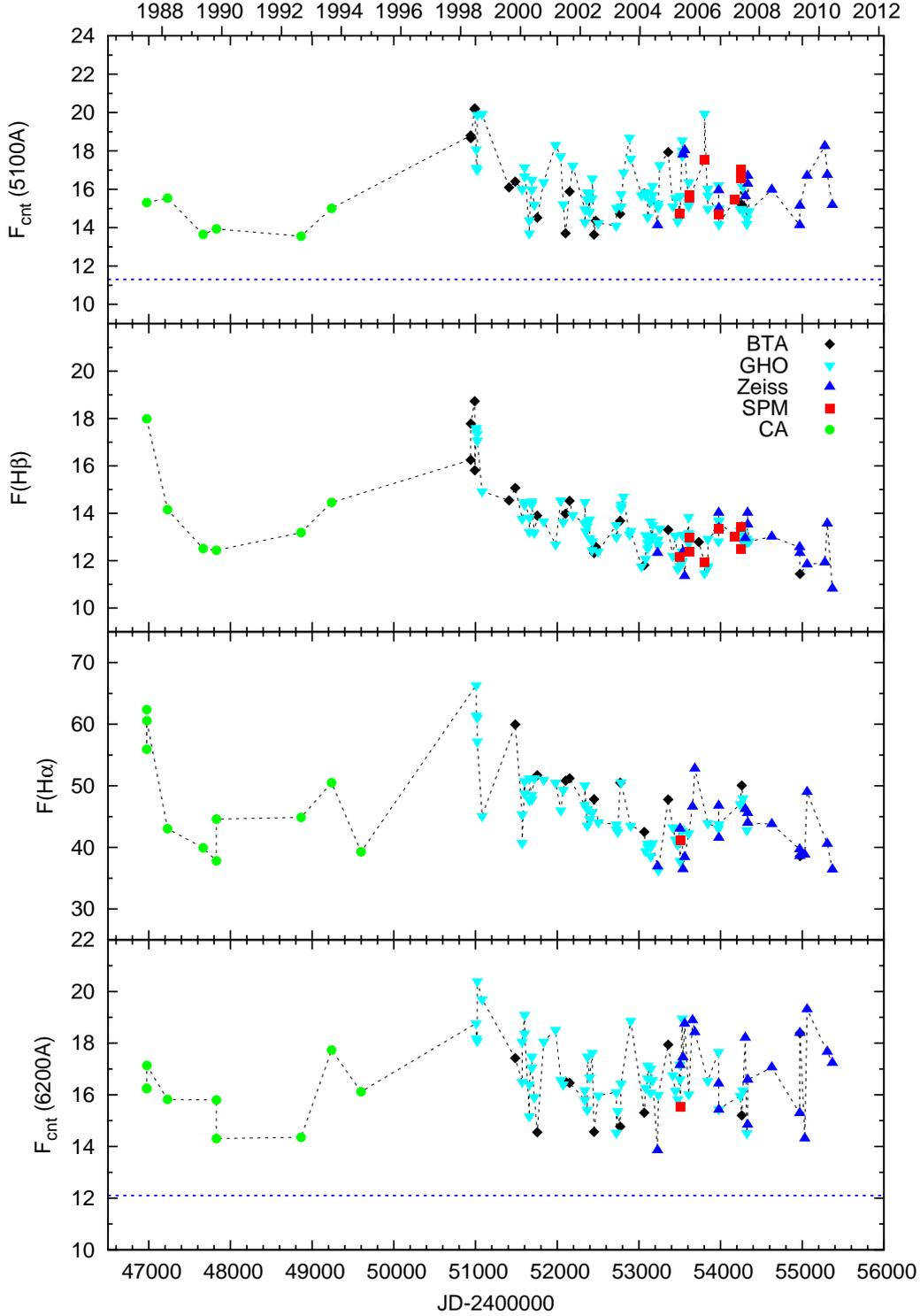}
\caption{Light curves (from top to bottom) for the blue continuum
flux, H$\beta$ and H$\alpha$ line flux, and red continuum flux.
Observations with different telescopes are denoted with different
symbols given in the middle plot. The continuum flux is in units
10$^{-16}$ erg cm$^{−2}$s$^{−1}$A$^{−1}$, and the line flux in
10$^{-14}$ erg cm$^{−2}$s$^{−1}$. {  The dashed line in the blue
and red continuum light curves mark the contribution of the 
starlight-continuum of the host galaxy.}} \label{lc}
\end{figure*}

\begin{table*}
\begin{center}
\caption[]{Sources of spectroscopic observations. }\label{tab1}
\begin{tabular}{lcccc}
\hline \hline
Observatory   &  Code &    Tel.aperture + equipment  &  Aperture  & Focus \\
 $\,\,\,\,\,\,\,\,\,\,\,\,1$            &   2  &     3                       &  4       &  5  \\
\hline
SAO(Russia)                  & L(N)    & 6 m + Long slit  &  2.0$\times$6.0   & Nasmith    \\
SAO(Russia)                  & L(U)    & 6 m + UAGS       &  2.0$\times$6.0   & Prime      \\
SAO(Russia)                  & L(Sc)   & 6 m + Scorpio    &  1.0$\times$6.07  & Prime      \\
Gullermo Haro (M\'exico)     & GHO     & 2.1 m + B\&C     &  2.5$\times$6.0   & Cassegrain   \\
San Pedro Martir (M\'exico)  & SPM     & 2.1 m + B\&C     &  2.5$\times$6.0   & Cassegrain   \\
SAO(Russia)                  & Z2K     & 1 m + GAD        &  4.0$\times$9.45  & Cassegrain   \\
Calar Alto(Spain)            & CA1     & 3.5 m + B\&C / TWIN  &  (1.5--2.1)$\times$3.5&   Cassegrain \\
Calar Alto(Spain)            & CA2     & 2.2 m + B\&C     &  2.0$\times$3.5&   Cassegrain \\
\hline
\end{tabular}
\tablefoot{-- Col.(1): Observatory. Col.(2): Code assigned to each combination of
telescope + equipment used throughout this paper. Col.(3): Telescope aperture and spectrograph.
Col.(4): Projected spectrograph entrance apertures (slit width$\times$slit length in
arcsec). Col.(5): Focus of the telescope.\\
}
\end{center}
\end{table*}

\onllongtab{2}{
\begin{longtable}{clccccc}
\caption[]{\label{tab2} The log of spectroscopic observations.}\\
\hline \hline
 N& UT-date& JD      & CODE\tablefootmark{*} & Aperture & Sp.range&  Seeing\\
    &       & 2400000+ &  & [arcsec] & [${\rm \AA}$]  & [arcsec]       \\
\hline1&2&3&4&5&6&7\\
\hline
\endfirsthead
\caption{Continued.}\\
\hline
 N& UT-date& JD      & CODE\tablefootmark{*} & Aperture & Sp.range&  Seeing\\
    &       & 2400000+ &  & [arcsec] & [${\rm \AA}$]  & [arcsec]       \\
\hline1&2&3&4&5&6&7\\
\hline
\endhead
\hline
\endfoot
\hline
\endlastfoot
  1 &  1987Jun28 & 46975.00 & CA1 & 2.0x3.5  & 3690-7080  & 1.5 - 2.5    \\
  2 &  1987Jun29 & 46976.00 & CA1 & 1.5x3.5  & 6360-7245  & 1.5 - 2.5    \\
  3 &  1987Jun30 & 46977.00 & CA1 & 1.5x3.5  & 6360-7245  & 1.5 - 2.5    \\
  4 &  1988Mar08 & 47229.00 & CA1 & 1.6x3.5  & 3720-7105  & 1.5 - 2.5    \\
  5 &  1989May18 & 47665.00 & CA1 & 2.1x3.5  &   H$\beta$, H$\alpha$   & 1.5 - 2.5    \\
  6 &  1989Oct27 & 47827.00 & CA1 & 2.0x3.5  & 4094-7440  & 1.5 - 2.5    \\
  7 &  1989Oct28 & 47828.00 & CA1 & 2.0x3.5  & 4094-7440  & 1.5 - 2.5    \\
  8 &  1992Aug30 & 48865.00 & CA2 & 2.0x3.5  & 3850-9100  & 1.5 - 2.5    \\
  9 &  1993Sep07 & 49238.00 & CA1 & 2.1x3.5  & 3630-8310  & 1.5 - 2.5    \\
 10 &  1994Sep02 & 49598.00 & CA2 & 2.0x3.5  & 4540-8520  & 1.5 - 2.5    \\
 11 &  1998May06 & 50940.34 & L(N) & 2.0x6.0  &3700-6200 &  3.0    \\
 12 &  1998May08 & 50942.33 & L(N) & 2.0x6.0  &3700-6200 &  2.0    \\
 13 &  1998Jun25 & 50990.29 & L(N) & 2.0x6.0  &3600-6100 &  3.0    \\
 14 &  1998Jun26 & 50991.20 & L(N) & 2.0x6.0  &3600-6100 &  3.0    \\
 15 &  1998Jul13 & 51008.30 & GHO  & 2.5x6.0  &3970-7224 &  2.1    \\
 16 &  1998Jul16 & 51011.23 & GHO  & 2.5x6.0  &4209-7479 &  2.8    \\
 17 &  1998Jul25 & 51020.26 & GHO  & 2.5x6.0  &3964-7235 &  2.5    \\
 18 &  1998Jul26 & 51021.27 & GHO  & 2.5x6.0  &3927-7222 &  2.1    \\
 19 &  1998Sep23 & 51079.50 & GHO  & 2.5x6.0  &4240-7538 &  2.5    \\
 20 &  1999Aug19 & 51410.30 & L(U) & 2.0x6.0  &4500-5736 &  2.0    \\
 21 &  1999Nov03 & 51486.14 & L(U) & 2.0x6.0  &3600-8350 &  1.3    \\
 22 &  2000Jan26 & 51569.97 & GHO  & 2.5x6.0  &4100-7400 &  2.7    \\
 23 &  2000Jan27 & 51570.96 & GHO  & 2.5x6.0  &4100-7400 &  1.8    \\
 24 &  2000Feb25 & 51599.92 & GHO  & 2.5x6.0  &4550-7850 &  3.0    \\
 25 &  2000Feb26 & 51600.88 & GHO  & 2.5x6.0  &4300-7600 &  3.0    \\
 26 &  2000Apr24 & 51658.87 & GHO  & 2.5x6.0  &4200-7500 &  3.0    \\
 27 &  2000Apr25 & 51659.83 & GHO  & 2.5x6.0  &4200-7500 &  3.0    \\
 28 &  2000May24 & 51689.47 & GHO  & 2.5x6.0  &4200-7500 &  2.5    \\
 29 &  2000May25 & 51689.54 & GHO  & 2.5x6.0  &4200-7500 &  2.5    \\
 30 &  2000Jun24 & 51719.74 & GHO  & 2.5x6.0  &4691-7991 &  3.0    \\
 31 &  2000Jul30 & 51756.26 & L(U) & 2.0x6.0  &3670-6310 &  1.6    \\
 32 &  2000Oct17 & 51834.58 & GHO  & 2.5x6.0  &4000-7300 &  2.7    \\
 33 &  2001Mar12 & 51980.61 & GHO  & 2.5x6.0  &4200-7500 &  3.0    \\
 34 &  2001May11 & 52041.88 & GHO  & 2.5x6.0  &3600-6900 &  3.2    \\
 35 &  2001May13 & 52043.94 & GHO  & 2.5x6.0  &4100-7400 &  3.0    \\
 36 &  2001Jun13 & 52073.80 & GHO  & 2.5x6.0  &4100-7400 &  4.1    \\
 37 &  2001Jul10 & 52101.46 & L(U) & 2.0x6.0  &5700-8124 &  2.0    \\
 38 &  2001Jul11 & 52102.48 & L(U) & 2.0x6.0  &3600-6024 &  1.2    \\
 39 &  2001Aug29 & 52151.30 & L(U) & 2.0x6.0  &3680-6104 &  2.5    \\
 40 &  2001Oct08 & 52190.62 & GHO  & 2.5x6.0  &3600-6900 &  2.7    \\
 41 &  2002Mar04 & 52337.91 & GHO  & 2.5x6.0  &3800-7100 &  2.0    \\
 42 &  2002Mar05 & 52338.91 & GHO  & 2.5x6.0  &5700-7460 &  2.0    \\
 43 &  2002Mar06 & 52339.97 & GHO  & 2.5x6.0  &4300-5900 &  2.0    \\
 44 &  2002Mar16 & 52349.94 & GHO  & 2.5x6.0  &4300-5900 &  2.0    \\
 45 &  2002Apr02 & 52366.85 & GHO  & 2.5x6.0  &4300-5900 &  1.8    \\
 46 &  2002Apr03 & 52367.83 & GHO  & 2.5x6.0  &5700-7460 &  1.5    \\
 47 &  2002Apr05 & 52369.88 & GHO  & 2.5x6.0  &3800-7100 &  1.5    \\
 48 &  2002May02 & 52396.82 & GHO  & 2.5x6.0  &4200-5960 &  2.3    \\
 49 &  2002May03 & 52397.76 & GHO  & 2.5x6.0  &5700-7460 &  2.0    \\
 50 &  2002May04 & 52398.77 & GHO  & 2.5x6.0  &4300-5900 &  2.0    \\
 51 &  2002Jun01 & 52426.82 & GHO  & 2.5x6.0  &4200-5850 &  2.7    \\
 52 &  2002Jun02 & 52427.78 & GHO  & 2.5x6.0  &5700-7460 &  3.2    \\
 53 &  2002Jun04 & 52429.77 & GHO  & 2.5x6.0  &4200-5960 &  3.2    \\
 54 &  2002Jun24 & 52450.37 & L(U) & 2.0x6.0  &3430-5854 &  5.0    \\
 55 &  2002Jul15 & 52471.42 & L(U) & 2.0x6.0  &3470-5894 &  2.0    \\
 56 &  2002Aug15 & 52501.65 & GHO  & 2.5x6.0  &4200-5700 &  2.3    \\
 57 &  2002Aug17 & 52503.68 & GHO  & 2.5x6.0  &5700-7460 &  2.0    \\
 58 &  2003Mar24 & 52722.93 & GHO  & 2.5x6.0  &4200-5850 &  2.7    \\
 59 &  2003Mar25 & 52723.92 & GHO  & 2.5x6.0  &3800-7100 &  2.7    \\
 60 &  2003Mar26 & 52724.96 & GHO  & 2.5x6.0  &5600-7462 &  2.3    \\
 61 &  2003Apr11 & 52740.86 & GHO  & 2.5x6.0  &5600-7462 &  3.6    \\
 62 &  2003May09 & 52769.26 & L(U) & 2.0x6.0  &3650-6070 &  1.5    \\
 63 &  2003May11 & 52771.41 & L(U) & 2.0x6.0  &5730-8151 &  1.5    \\
 64 &  2003May22 & 52781.79 & GHO  & 2.5x6.0  &3700-7430 &  3.6    \\
 65 &  2003May23 & 52782.87 & GHO  & 2.5x6.0  &4200-6085 &  2.3    \\
 66 &  2003Jun21 & 52811.85 & GHO  & 2.5x6.0  &4200-5960 &  2.3    \\
 67 &  2003Sep03 & 52885.65 & GHO  & 2.5x6.0  &4300-5900 &  4.1    \\
 68 &  2003Sep17 & 52899.69 & GHO  & 2.5x6.0  &3800-7100 &  2.3    \\
 69 &  2004Jan27 & 53032.00 & GHO  & 2.5x6.0  &4300-6060 &  2.5    \\
 70 &  2004Mar02 & 53066.59 & L(U) & 2.0x6.0  &3700-6124 &  2.0    \\
 71 &  2004Mar16 & 53080.94 & GHO  & 2.5x6.0  &4300-6060 &  5.0    \\
 72 &  2004Mar18 & 53082.94 & GHO  & 2.5x6.0  &3800-7100 &  5.0    \\
 73 &  2004Apr11 & 53106.94 & GHO  & 2.5x6.0  &3800-7100 &  3.5    \\
 74 &  2004Apr12 & 53107.84 & GHO  & 2.5x6.0  &4300-6060 &  2.5    \\
 75 &  2004Apr13 & 53108.92 & GHO  & 2.5x6.0  &5700-7460 &  2.5    \\
 76 &  2004May18 & 53143.83 & GHO  & 2.5x6.0  &3800-7100 &  2.3    \\
 77 &  2004May19 & 53144.82 & GHO  & 2.5x6.0  &4300-6060 &  2.3    \\
 78 &  2004May20 & 53145.78 & GHO  & 2.5x6.0  &5700-7460 &  2.7    \\
 79 &  2004Jun10 & 53166.80 & GHO  & 2.5x6.0  &3800-7100 &  1.8    \\
 80 &  2004Jun11 & 53167.77 & GHO  & 2.5x6.0  &4300-6060 &  1.8    \\
 81 &  2004Aug10 & 53228.32 & Z2K  & 4.0x9.45 &3700-7450 &  2.0    \\
 82 &  2004Aug18 & 53235.64 & GHO  & 2.5x6.0  &4300-6060 &  3.1    \\
 83 &  2004Aug20 & 53237.63 & GHO  & 2.5x6.0  &3800-7100 &  2.7    \\
 84 &  2004Sep06 & 53254.63 & GHO  & 2.5x6.0  &4300-6060 &  2.7    \\
 85 &  2004Dec18 & 53357.50 & L(Sc)& 1.0x6.07 &3460-7460 &  1.6    \\
 86 &  2005Feb13 & 53415.01 & GHO  & 2.5x6.0  &3830-7090 &  3.5    \\
 87 &  2005Mar17 & 53446.89 & GHO  & 2.5x6.0  &3700-7070 &  3.0    \\
 88 &  2005Apr15 & 53475.90 & GHO  & 2.5x6.0  &4240-5940 &  2.2    \\
 89 &  2005Apr16 & 53476.79 & GHO  & 2.5x6.0  &5220-7250 &  1.8    \\
 90 &  2005May11 & 53501.85 & GHO  & 2.5x6.0  &3750-7100 &  3.1    \\
 91 &  2005May12 & 53502.83 & GHO  & 2.5x6.0  &4220-5920 &  3.0    \\
 92 &  2005May12 & 53503.48 & Z2K  & 4.0x9.45 &3700-7450 &  2.0    \\
 93 &  2005May13 & 53503.83 & GHO  & 2.5x6.0  &5580-7300 &  2.2    \\
 94 &  2005May14 & 53504.73 & SPM  & 2.5x6.0  &3880-5970 &  4.8    \\ 
 95 &  2005May15 & 53505.85 & SPM  & 2.5x6.0  &5730-7810 &  3.3    \\
 96 &  2005Jun08 & 53529.85 & GHO  & 2.5x6.0  &3720-7070 &  1.9    \\
 97 &  2005Jun09 & 53530.64 & GHO  & 2.5x6.0  &4300-5960 &  3.0    \\
 98 &  2005Jun11 & 53532.78 & GHO  & 2.5x6.0  &4250-5940 &  3.5    \\
 99 &  2005Jun16 & 53538.46 & Z2K  & 4.0x9.45 &3700-7450 &  2.5    \\
100 &  2005Jul07 & 53559.44 & Z2K  & 4.0x9.45 &3700-7450 &  5.0    \\
101 &  2005Aug26 & 53608.70 & GHO  & 2.5x6.0  &3790-7100 &  2.6    \\
102 &  2005Aug28 & 53610.64 & GHO  & 2.5x6.0  &4330-6000 &  2.8    \\
103 &  2005Aug29 & 53611.63 & GHO  & 2.5x6.0  &4320-6000 &  3.5    \\
104 &  2005Aug31 & 53613.62 & GHO  & 2.5x6.0  &4330-6000 &  2.7    \\
105 &  2005Sep08 & 53621.67 & SPM  & 2.5x6.0  &3740-5770 &  3.0    \\
106 &  2005Sep09 & 53622.67 & SPM  & 2.5x6.0  &3670-5800 &  2.8    \\
107 &  2005Oct11 & 53655.23 & Z2K  & 4.0x9.45 &3700-7450 &  3-4    \\
108 &  2005Nov09 & 53684.19 & Z2K  & 4.0x9.45 &3700-7450 &  2.7    \\
109 &  2005Dec28 & 53732.50 & L(Sc)& 1.0x6.00 &3100-7300 &  1.8    \\
110 &  2006Mar08 & 53802.96 & SPM  & 2.5x6.0  &3740-5800 &  3.5    \\
111 &  2006Mar08 & 53802.97 & GHO  & 2.5x6.0  &3740-7100 &  3.5    \\
112 &  2006Apr17 & 53842.94 & GHO  & 2.5x6.0  &3720-7090 &  3.2    \\
113 &  2006Apr18 & 53843.93 & GHO  & 2.5x6.0  &4240-5940 &  2.4    \\
114 &  2006Apr19 & 53844.86 & GHO  & 2.5x6.0  &4240-5940 &  2.0    \\
115 &  2006Aug28 & 53975.65 & GHO  & 2.5x6.0  &3600-7090 &  2.1    \\
116 &  2006Aug28 & 53976.69 & SPM  & 2.5x6.0  &3720-5790 &  3.0    \\
117 &  2006Aug29 & 53977.50 & Z2K  & 4.0x9.45 &3750-7390 &  5.0    \\
118 &  2006Aug30 & 53977.63 & GHO  & 2.5x6.0  &4130-5920 &  2.5    \\
119 &  2006Aug30 & 53978.48 & Z2K  & 4.0x9.45 &3700-7450 &  2.0    \\
120 &  2006Aug31 & 53978.63 & GHO  & 2.5x6.0  &3530-7110 &  2.1    \\
121 &  2007Mar15 & 54174.95 & SPM  & 2.5x6.0  &3740-5810 &  2.4    \\
122 &  2007May20 & 54240.85 & GHO  & 2.5x6.0  &3780-7370 &  2.0    \\
123 &  2007May23 & 54243.82 & SPM  & 2.5x6.0  &3730-5800 &  3.2    \\
124 &  2007May26 & 54246.87 & SPM  & 2.5x6.0  &3750-5820 &  3.5    \\
125 &  2007Jun08 & 54259.40 & L(Sc)& 1.0x6.07 &3100-7300 &  2.3    \\
126 &  2007Jun20 & 54271.76 & GHO  & 2.5x6.0  &4300-6120 &  3.0    \\
127 &  2007Jun21 & 54272.85 & GHO  & 2.5x6.0  &3890-7480 &  2.1    \\
128 &  2007Jul20 & 54302.38 & Z2K  & 4.0x9.45 &3700-7450 &  3.0    \\
129 &  2007Aug10 & 54322.66 & GHO  & 2.5x6.0  &3880-7470 &  3.0    \\
130 &  2007Aug11 & 54323.63 & GHO  & 2.5x6.0  &4350-6170 &  3.5    \\
131 &  2007Aug18 & 54331.31 & Z2K  & 4.0x9.45 &3700-7450 &  3.0    \\
132 &  2007Aug19 & 54332.33 & Z2K  & 4.0x9.45 &3700-7450 &  2.0    \\
133 &  2007Sep04 & 54347.63 & GHO  & 2.5x6.0  &4150-5970 &  2.5    \\
134 &  2008Jun08 & 54626.46 & Z2K  & 4.0x9.45 &3700-7450 &  2.5    \\
135 &  2009May17 & 54969.47 & Z2K  & 4.0x9.45 &3700-7450 &  2.5    \\
136 &  2009May19 & 54970.54 & Z2K  & 4.0x9.45 &3700-7450 &  3.0    \\
137 &  2009May20 & 54972.20 & L(Sc)& 1.0x6.07 &3100-7300 &  2.0    \\
138 &  2009Jul17 & 55030.33 & Z2K  & 4.0x9.45 &3700-7450 &  2.5    \\
139 &  2009Aug14 & 55058.38 & Z2K  & 4.0x9.45 &3700-7450 &  2.5    \\
140 &  2010Mar21 & 55276.51 & Z2K  & 4.0x9.45 &3700-7450 &  2.6    \\
141 &  2010Apr20 & 55306.52 & Z2K  & 4.0x9.45 &3700-7450 &  2.5    \\
142 &  2010Jun19 & 55367.42 & Z2K  & 4.0x9.45 &3700-7450 &  2.5    \\
\hline
\end{longtable}
\tablefoot{-- Col.(1): Number. Col.(2): UT date. Col.(3): Julian
date (JD). Col.(4): CODE\tablefootmark{*}. Col.(5): Projected
spectrograph entrance apertures. Col.(6): Wavelength range covered.
Col.(7): Spectral resolution. Col.(8): Mean seeing in arcsec.\\
\tablefoottext{*}{Code given according to Table~\ref{tab1}.}}
}

\subsection{Absolute calibration (scaling) of the spectra and measurements of spectra}

The standard technique of the flux calibration of  spectra (i.e.
comparison with stars of known spectral energy distribution) is not
precise enough for the study of the AGN variability, since even
under good photometric conditions, the accuracy of spectrophotometry
is usually not better than 10\%. Therefore we used standard stars
only to provide a relative flux calibration. For the absolute
calibration, the fluxes of the  narrow emission lines are adopted as
a scaling  factor of observed AGN spectra, since they are known to
remain constant on time scales of tens of years \citep{pet93}.

The scaling of the blue spectra was performed using the method of
\citet{vg92} modified by \citet{sh04}\footnote{see Appendix A in
\citet{sh04}}. We will not repeat the scaling procedure here. This
method allowed us to obtain a homogeneous set of spectra with the
same wavelength calibration and the same [O III]$\lambda$4959+5007
flux. Blue spectra were scaled using [O III]$\lambda$4959+5007. Red
spectra with a resolution of 8--10 \AA\ do not contain these [O III]
lines, but as a rule there was the blue spectra (R$\sim$8-10 \AA \
or $\sim$15 \AA) taken in the same night. Usually, the red edge of
the blue spectra and the blue edge of the red spectra overlap in an
interval of 300 \AA. Therefore, first the red spectra were rescaled
using the overlapping continuum region with the blue ones. Then, the
blue spectrum was scaled with the [O III]$\lambda$4959+5007 line. In
these cases, the scaling uncertainty was about 10\%.  Then scaling
of the red spectrum was refined using the mean flux in [O
I]$\lambda$6300 determined from a low-dispersion spectrum
(R$\sim$12--15 \AA). In Arp 102B we do not have data for [O
III]$\lambda$5007+4959 in absolute units. Therefore we used the [O
I]$\lambda$6300 in absolute units from \cite{s00}
F(6300)abs=(1.76$\pm$0.18)$\times$10$^{-14}$ erg cm$^{-2}$s$^{-1}$)
and have calculated flux for [O III] lines to be F[O
III]4959+5007=(4.9$\pm$0.49)$\times$10$^{-14}$ erg
cm$^{-2}$s$^{-1}$. For this we used the good spectra of Arp 102B,
containing both, the H$\alpha$ and H$\beta$ spectral bands
(R$\sim$12-15 \AA/px).

From the scaled spectra we determined the average flux in the blue
continuum at the observed wavelength $\sim 5225$\AA\, (or at $\sim
5100$ \AA\, in the rest frame of Arp 102B, z=0.0241), by means of
the flux averaging in the bandpass of 5200--5250\,\AA\ and in the
red continuum the observed wavelength is $\sim 6381$ \AA\, (or at
$\sim 6230$ \AA\, in the rest frame),  averaging the flux in the
wavelength range from 6356 \AA\ to 6406 \AA. In the above
wavelength intervals there are no strong absorption lines from the
host galaxy. For the determination of the observed fluxes of
H$\beta$ and H$\alpha$, it is necessary to subtract the continuum
contribution. For the continuum subtraction, a linear continuum was
constructed using windows of 20\,\AA\, width, located at 4700\,\AA\,
and 5240\,\AA\, for the H$\beta$ spectral band, and at 6380\,\AA\,
and 7000\,\AA\, for the H$\alpha$ one. After the continuum
subtraction, we defined the observed fluxes in  the lines in the
following wavelength intervals: (4845-5150)\,\AA\, for H$\beta$, and
(6500-6965)\,\AA\, for H$\alpha$ \citep[the interval is like the one
in][]{s00}.

\begin{figure}
\centering
\includegraphics[width=9cm]{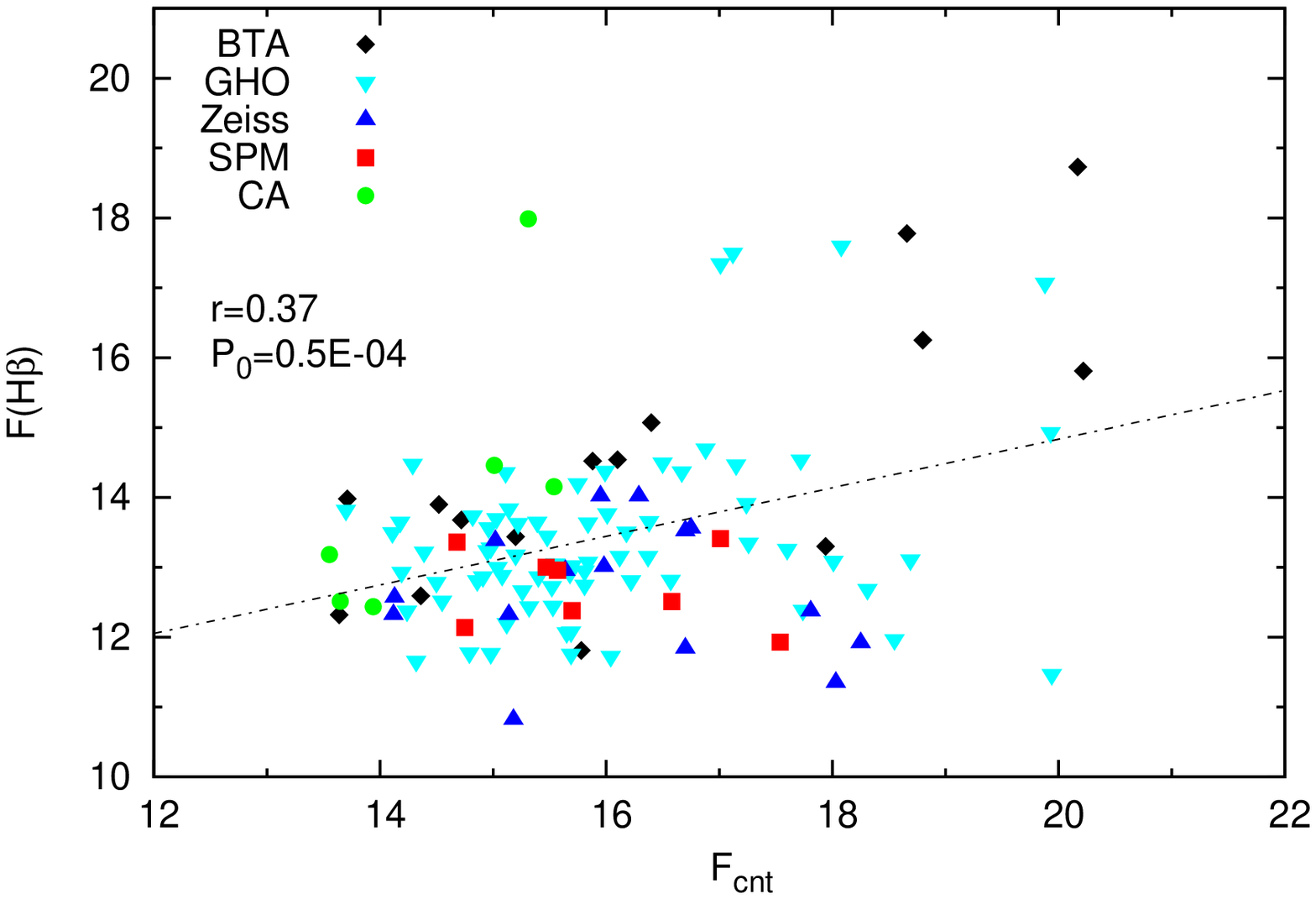}
\includegraphics[width=9cm]{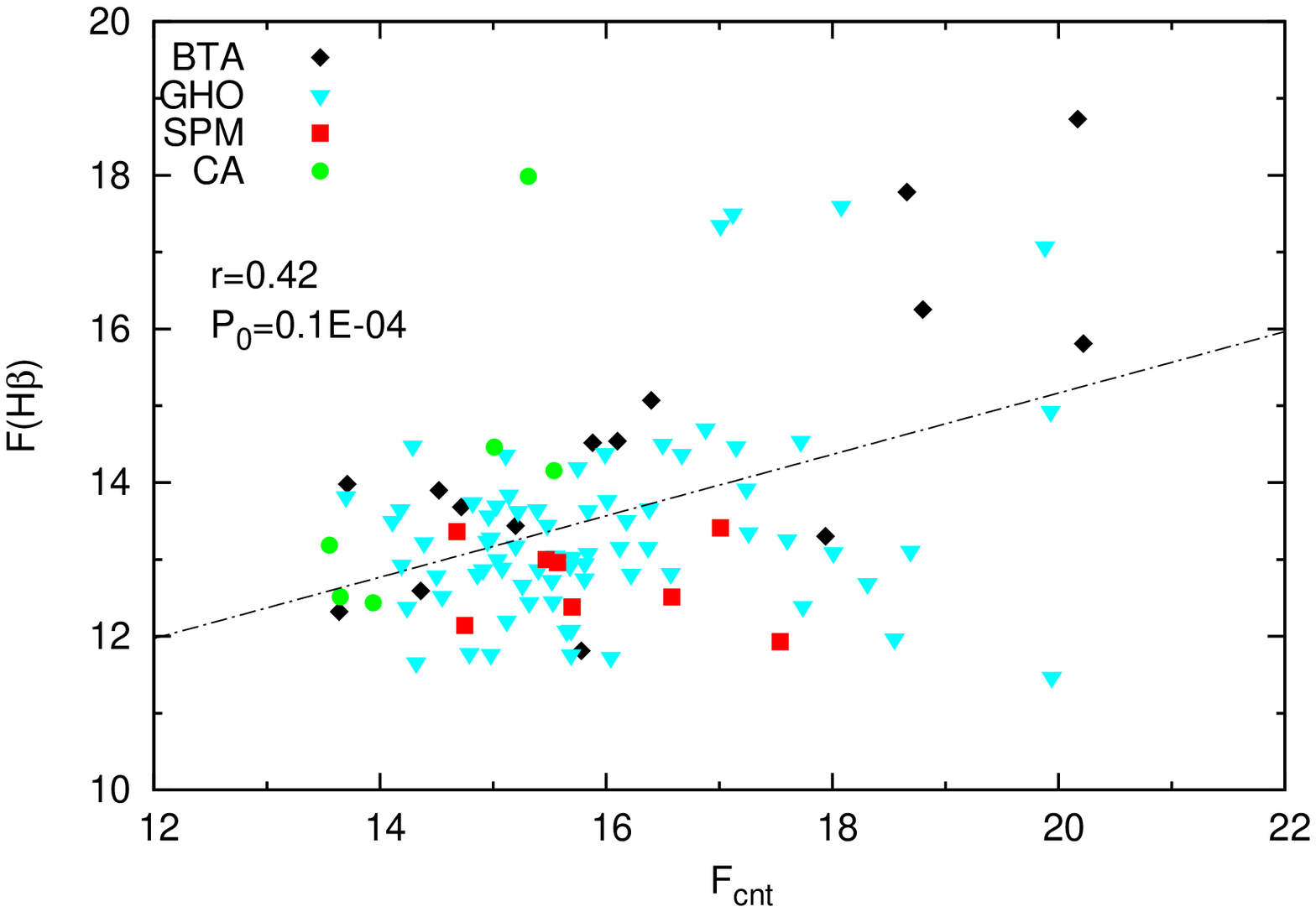}
\caption{H$\beta$ vs continuum flux. Upper plot: all observed data. Bottom: without data
observed with the Zeiss telescope. The continuum flux is in units
$10^{-16} \rm erg \ cm^{-2} s^{-1} A^{-1}$, and the line flux in
$10^{-14} \rm erg \ cm^{-2}s^{-1}$. Observations with different
telescopes are denoted with different symbols given in the upper left.
The correlation coefficient and the
corresponding p-value are also given.} \label{Hb_cnt}
\end{figure}

\begin{figure}
\centering
\includegraphics[width=9cm]{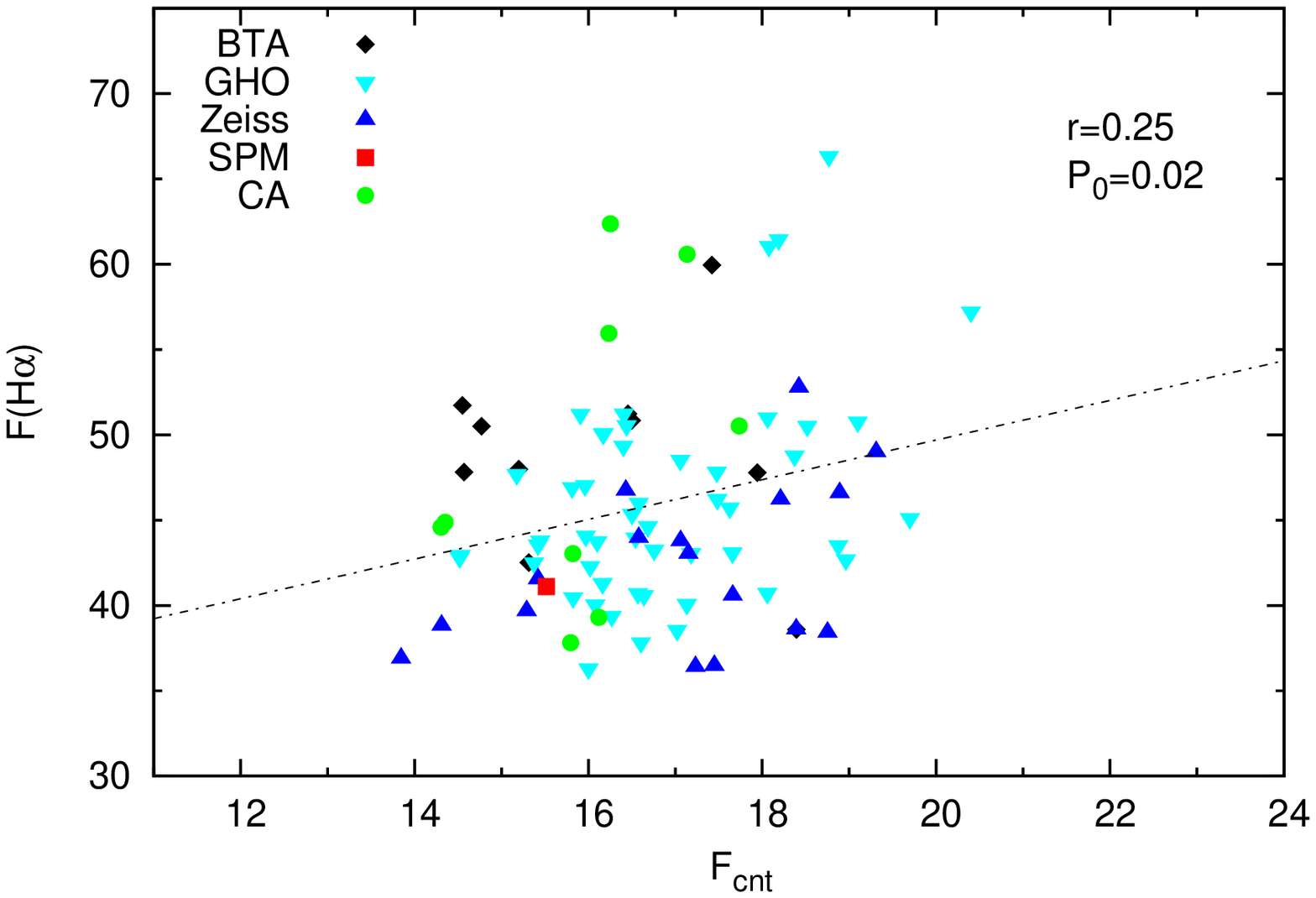}
\includegraphics[width=9cm]{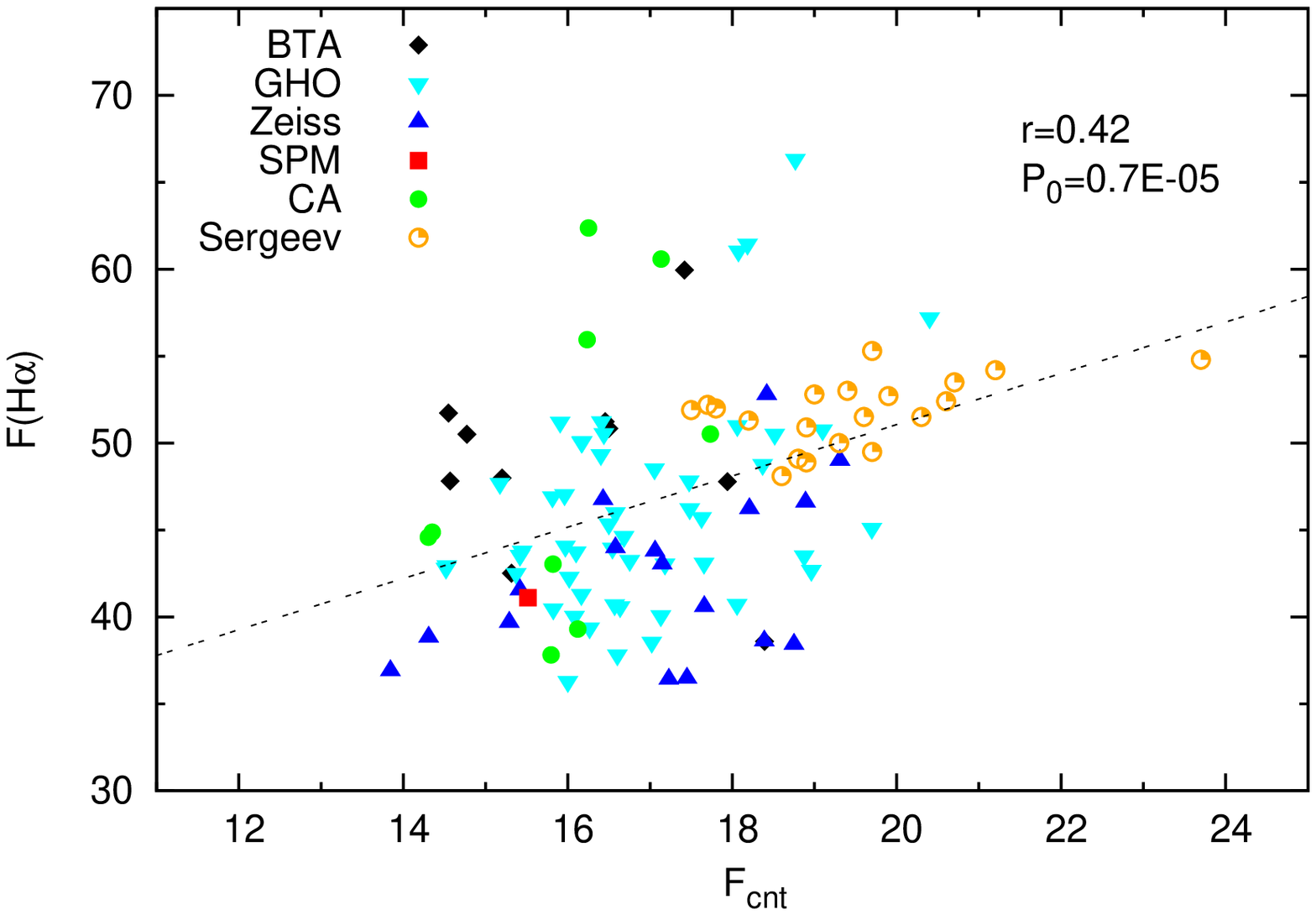}
\caption{The same as Fig. \ref{Hb_cnt} but for H$\alpha$ (upper) The
bottom panel includes also data from \cite{s00}.}
\label{Ha_cnt}
\end{figure}

\begin{table*}
\begin{center}
\caption[]{Flux scale factors for optical spectra.}\label{tab3}
\begin{tabular}{lcccc}
\hline \hline
Sample   &  Years &    Aperture  &  Scale factor & Extended source  correction\\
         &        &     (arcsec) &  ($\varphi$)  &  G(g)\tablefootmark{a}    \\
\hline
L(U,N)    & 1998--2004 &  2.0$\times$6.0   & 1.000  &  0.000 \\
L(Sc)     & 2004--2009 &  1.0$\times$6.0   & 1.019  & -0.391 \\
GHO,SPM   & 1998--2007 &  2.5$\times$6.0   & 1.000  &  0.000 \\
Z2K       & 2004--2005 &  4.0$\times$9.45  & 1.027  &  0.304 \\
Z2K       & 2006--2010 &  4.0$\times$9.45  & 1.027  &  0.633 \\
CA1       & 1987--1993 &  1.5--2.1$\times$3.5  & $\sim$ 1    & -0.200--0.000 \\
CA2       & 1992--1994 &  2.0$\times$3.5       & $\sim$ 1    &  0.000 \\
\hline
\tablefootmark{a} In units 10$^{-15} \rm erg s^{-1} cm^{-2} \AA^{-1}$
\end{tabular}
\end{center}
\end{table*}

\onllongtab{4}{
\begin{longtable}{ccccccccc}
\caption[]{\label{tab4} The measured continuum and line fluxes\tablefootmark{*}. }\\
\hline \hline
N& UT-date & MJD & F$_{\rm blue \ cnt}\pm \sigma$  & F$_{\rm blue \ cnt}^{\rm corr}\pm \sigma$\tablefootmark{**} & F(H$\beta$)$\pm \sigma$  & F(H$\alpha$)$\pm \sigma$  & F$_{\rm red \ cnt}\pm \sigma$ & F$_{\rm red \ cnt}^{\rm corr}\pm \sigma$\tablefootmark{**} \\ 
\hline1&2&3&4&5&6&7&8&9\\
\hline
\endfirsthead
\caption{Continued.}\\
\hline
N& UT-date & MJD   & F$_{\rm blue cnt}\pm \sigma$  & F$_{\rm blue \ cnt}^{\rm corr}\pm \sigma$\tablefootmark{**}  & F(H$\beta$)$\pm \sigma$  & F(H$\alpha$)$\pm \sigma$  & F$_{\rm red cnt}\pm \sigma$ & F$_{\rm red \ cnt}^{\rm corr}\pm \sigma$\tablefootmark{**} \\
\hline1&2&3&4&5&6&7&8&9\\
\hline
\endhead
\hline
\endfoot
\hline
\endlastfoot
  1 &  1987Jun28 &  46975.00 &   15.31             & 4.01         &  17.99$\pm$1.01     &  62.38$\pm$3.49    &    16.25$\pm$0.50  &  4.15$\pm$0.37  \\
  2 &  1987Jun29 &  46976.00 &   -                 & -            &  -                  &  55.95$\pm$3.13    &    16.24$\pm$0.50  &  4.14$\pm$0.37  \\
  3 &  1987Jun30 &  46977.00 &   -                 & -            &  -                  &  60.58$\pm$3.39    &    17.13$\pm$0.53  &  5.03$\pm$0.45  \\
  4 &  1988Mar08 &  47229.00 &   15.54$\pm$0.96    & 4.24$\pm$0.40&  14.16$\pm$0.99     &  43.04$\pm$4.35    &    15.82$\pm$1.42  &  3.72$\pm$0.33  \\
  5 &  1989May18 &  47665.00 &   13.65$\pm$0.85    & 2.35$\pm$0.22&  12.51$\pm$0.88     &  39.93$\pm$4.03    &    -               &    \\
  6 &  1989Oct27 &  47827.00 &   -                 & -            &  -                  &  37.82$\pm$3.82    &    15.80$\pm$1.42  &  3.70$\pm$0.33  \\
  7 &  1989Oct28 &  47828.00 &   13.94$\pm$0.86    & 2.64$\pm$0.25&  12.44$\pm$0.87     &  44.60$\pm$4.50    &    14.31$\pm$1.29  &  2.21$\pm$0.20  \\
  8 &  1992Aug30 &  48865.00 &   13.55$\pm$0.84    & 2.25$\pm$0.21&  13.18$\pm$0.92     &  44.88$\pm$4.53    &    14.35$\pm$1.29  &  2.25$\pm$0.20  \\
  9 &  1993Sep07 &  49238.00 &   15.01$\pm$0.93    & 3.71$\pm$0.35&  14.46$\pm$1.01     &  50.52$\pm$5.10    &    17.73$\pm$1.60  &  5.63$\pm$0.50  \\
 10 &  1994Sep02 &  49598.00 &   -                 & -            &  -                  &  39.30$\pm$3.97    &    16.12$\pm$1.45  &  4.02$\pm$0.36  \\
 11 &  1998May06 &  50940.34 &   18.80$\pm$  0.64  & 7.50$\pm$0.71&  16.25$\pm$  0.49   &   -                &    -               &   \\
 12 &  1998May08 &  50942.33 &   18.66$\pm$  0.63  & 7.36$\pm$0.70&  17.78$\pm$  0.53   &   -                &    -               &   \\
 13 &  1998Jun25 &  50990.29 &   20.17$\pm$  0.69  & 8.87$\pm$0.84&  18.73$\pm$  2.25   &   -                &    -               &   \\
 14 &  1998Jun26 &  50991.20 &   20.22$\pm$  0.69  & 8.92$\pm$0.85&  15.81$\pm$  1.90   &   -                &    -               &   \\
 15 &  1998Jul13 &  51008.30 &   18.08$\pm$  0.61  & 6.78$\pm$0.64&  17.60$\pm$  0.53   &   66.33$\pm$2.72   &    18.77$\pm$ 0.83 &  6.67$\pm$0.59 \\
 16 &  1998Jul16 &  51011.23 &   17.12$\pm$  0.58  & 5.82$\pm$0.55&  17.50$\pm$  0.53   &   61.46$\pm$2.52   &    18.19$\pm$ 0.80 &  6.09$\pm$0.54 \\
 17 &  1998Jul25 &  51020.26 &   17.01$\pm$  1.87  & 5.71$\pm$0.54&  17.35$\pm$  0.52   &   61.06$\pm$2.50   &    18.08$\pm$ 1.54 &  5.98$\pm$0.53 \\
 18 &  1998Jul26 &  51021.27 &   19.88$\pm$  2.19  & 8.58$\pm$0.82&  17.07$\pm$  0.51   &   57.22$\pm$2.35   &    20.40$\pm$ 1.73 &  8.30$\pm$0.74 \\
 19 &  1998Sep23 &  51079.50 &   19.93$\pm$  0.68  & 8.63$\pm$0.82&  14.93$\pm$  0.45   &   45.11$\pm$1.85   &    19.70$\pm$ 0.87 &  7.60$\pm$0.68 \\
 20 &  1999Aug19 &  51410.30 &   16.10$\pm$  0.55  & 4.80$\pm$0.46&  14.54$\pm$  0.44   &   -                &    -               &   \\
 21 &  1999Nov03 &  51486.14 &   16.40$\pm$  0.56  & 5.10$\pm$0.48&  15.07$\pm$  0.45   &   59.96$\pm$2.46   &    17.42$\pm$ 0.77 &  5.32$\pm$0.47 \\
 22 &  2000Jan26 &  51569.97 &   16.01$\pm$  0.54  & 4.71$\pm$0.45&  13.77$\pm$  0.41   &   45.36$\pm$3.45   &    16.50$\pm$ 0.73 &  4.40$\pm$0.39 \\
 23 &  2000Jan27 &  51570.96 &   -                 & -            &  -                  &   40.73$\pm$3.10   &    18.06$\pm$ 0.79 &  5.96$\pm$0.53 \\
 24 &  2000Feb25 &  51599.92 &   16.67$\pm$  0.57  & 5.37$\pm$0.51&  14.37$\pm$  0.43   &   50.77$\pm$2.08   &    19.10$\pm$ 0.84 &  7.00$\pm$0.62 \\
 25 &  2000Feb26 &  51600.88 &   17.15$\pm$  0.58  & 5.85$\pm$0.56&  14.47$\pm$  0.43   &   48.77$\pm$2.00   &    18.37$\pm$ 0.81 &  6.27$\pm$0.56 \\
 26 &  2000Apr24 &  51658.87 &   13.70$\pm$  0.47  & 2.40$\pm$0.23&  13.82$\pm$  0.41   &   47.69$\pm$1.96   &    15.17$\pm$ 0.67 &  3.07$\pm$0.27 \\
 27 &  2000Apr25 &  51659.83 &   14.39$\pm$  0.49  & 3.09$\pm$0.29&  13.22$\pm$  0.40   &   51.24$\pm$2.10   &    16.41$\pm$ 0.72 &  4.31$\pm$0.38 \\
 28 &  2000May24 &  51689.47 &   16.50$\pm$  0.56  & 5.20$\pm$0.49&  14.50$\pm$  0.44   &   47.83$\pm$1.96   &    17.48$\pm$ 0.77 &  5.38$\pm$0.48 \\
 29 &  2000May25 &  51689.54 &   15.99$\pm$  0.54  & 4.69$\pm$0.45&  14.38$\pm$  0.43   &   48.52$\pm$1.99   &    17.06$\pm$ 0.75 &  4.96$\pm$0.44 \\
 30 &  2000Jun24 &  51719.74 &   15.20$\pm$  0.52  & 3.90$\pm$0.37&  13.18$\pm$  0.40   &   51.22$\pm$2.10   &    15.91$\pm$ 0.70 &  3.81$\pm$0.34 \\
 31 &  2000Jul30 &  51756.26 &   14.52$\pm$  0.49  & 3.22$\pm$0.31&  13.90$\pm$  0.42   &   51.73$\pm$2.12   &    14.55$\pm$ 0.64 &  2.45$\pm$0.22 \\
 32 &  2000Oct17 &  51834.58 &   16.38$\pm$  0.56  & 5.08$\pm$0.48&  13.66$\pm$  0.41   &   51.01$\pm$2.09   &    18.06$\pm$ 0.79 &  5.96$\pm$0.53 \\
 33 &  2001Mar12 &  51980.61 &   18.31$\pm$  0.62  & 7.01$\pm$0.67&  12.69$\pm$  0.38   &   50.53$\pm$2.07   &    18.52$\pm$ 0.81 &  6.42$\pm$0.57 \\
 34 &  2001May11 &  52041.88 &   17.72$\pm$  0.60  & 6.42$\pm$0.61&  14.54$\pm$  0.44   &   -                &    -               &   \\
 35 &  2001May13 &  52043.94 &   -                 & -            &  -                  &   46.01$\pm$1.89   &    16.58$\pm$ 0.73 &  4.48$\pm$0.40 \\
 36 &  2001Jun13 &  52073.80 &   15.22$\pm$  0.52  & 3.92$\pm$0.37&  13.63$\pm$  0.41   &   49.35$\pm$2.02   &    16.40$\pm$ 0.72 &  4.30$\pm$0.38 \\
 37 &  2001Jul10 &  52101.46 &   -                 & -            &   -                 &   50.84$\pm$2.08   &    16.50$\pm$ 0.73 &  4.40$\pm$0.39 \\
 38 &  2001Jul11 &  52102.48 &   13.71$\pm$  0.47  & 2.41$\pm$0.23&  13.98$\pm$  0.42   &   -                &    -               &   \\
 39 &  2001Aug29 &  52151.30 &   15.88$\pm$  0.54  & 4.58$\pm$0.44&  14.52$\pm$  0.44   &   51.24$\pm$2.10   &    16.46$\pm$ 0.72 &  4.36$\pm$0.39 \\
 40 &  2001Oct08 &  52190.62 &   17.24$\pm$  0.59  & 5.94$\pm$0.56&  13.92$\pm$  0.42   &   -                &    -               &   \\
 41 &  2002Mar04 &  52337.91 &   14.96$\pm$  0.51  & 3.66$\pm$0.35&  13.57$\pm$  0.41   &   46.92$\pm$1.92   &    15.81$\pm$ 0.70 &  3.71$\pm$0.33 \\
 42 &  2002Mar05 &  52338.91 &   -                 & -            &  -                  &   50.09$\pm$2.05   &    16.18$\pm$ 0.71 &  4.08$\pm$0.36 \\
 43 &  2002Mar06 &  52339.97 &   14.29$\pm$  0.49  & 2.99$\pm$0.28&  14.48$\pm$  0.43   &   -                &    -               &   \\
 44 &  2002Mar16 &  52349.94 &   14.95$\pm$  0.51  & 3.65$\pm$0.35&  13.24$\pm$  0.40   &   -                &    -               &   \\
 45 &  2002Apr02 &  52366.85 &   15.48$\pm$  0.53  & 4.18$\pm$0.40&  13.45$\pm$  0.40   &   -                &    -               &   \\
 46 &  2002Apr03 &  52367.83 &   -                 & -            &  -                  &   43.56$\pm$1.79   &    15.42$\pm$ 1.37 &  3.32$\pm$0.30 \\
 47 &  2002Apr05 &  52369.88 &   15.84$\pm$  0.54  & 4.54$\pm$0.43&  13.64$\pm$  0.41   &   46.24$\pm$1.90   &    17.48$\pm$ 1.56 &  5.38$\pm$0.48 \\
 48 &  2002May02 &  52396.82 &   14.82$\pm$  0.50  & 3.52$\pm$0.33&  13.74$\pm$  0.41   &   -                &    -               &   \\
 49 &  2002May03 &  52397.76 &   -                 & -            &  -                  &   44.64$\pm$1.83   &    16.68$\pm$ 0.73 &  4.58$\pm$0.41 \\
 50 &  2002May04 &  52398.77 &   15.81$\pm$  0.54  & 4.51$\pm$0.43&  12.95$\pm$  0.39   &   -                &    -               &   \\
 51 &  2002Jun01 &  52426.82 &   15.53$\pm$  0.53  & 4.23$\pm$0.40&  12.45$\pm$  0.37   &   -                &    -               &   \\
 52 &  2002Jun02 &  52427.78 &   -                 & -            &  -                  &   45.72$\pm$1.87   &    17.63$\pm$ 0.78 &  5.53$\pm$0.49 \\
 53 &  2002Jun04 &  52429.77 &   16.57$\pm$  0.56  & 5.27$\pm$0.50&  12.82$\pm$  0.38   &   -                &    -               &   \\
 54 &  2002Jun24 &  52450.37 &   13.64$\pm$  0.46  & 2.34$\pm$0.22&  12.32$\pm$  0.37   &   47.82$\pm$1.96   &    14.57$\pm$ 0.64 &  2.47$\pm$0.22 \\
 55 &  2002Jul15 &  52471.42 &   14.36$\pm$  0.49  & 3.06$\pm$0.29&  12.59$\pm$  0.38   &   -                &    -               &   \\
 56 &  2002Aug15 &  52501.65 &   14.24$\pm$  0.48  & 2.94$\pm$0.28&  12.38$\pm$  0.37   &   -                &    -               &   \\
 57 &  2002Aug17 &  52503.68 &   -                 & -            &  -                  &   44.08$\pm$1.81   &    15.97$\pm$ 0.70 &  3.87$\pm$0.34 \\
 58 &  2003Mar24 &  52722.93 &   14.11$\pm$  0.48  & 2.81$\pm$0.27&  13.50$\pm$  0.41   &   -                &    -               &   \\
 59 &  2003Mar25 &  52723.92 &   15.04$\pm$  0.51  & 3.74$\pm$0.36&  13.00$\pm$  0.39   &   43.74$\pm$1.79   &    16.10$\pm$ 1.18 &  4.00$\pm$0.36 \\
 60 &  2003Mar26 &  52724.96 &   -                 & -            &  -                  &   42.96$\pm$1.76   &    14.53$\pm$ 1.06 &  2.43$\pm$0.22 \\
 61 &  2003Apr11 &  52740.86 &   -                 & -            &  -                  &   42.51$\pm$1.74   &    15.37$\pm$ 0.68 &  3.27$\pm$0.29 \\
 62 &  2003May09 &  52769.26 &   14.72$\pm$  0.50  & 3.42$\pm$0.32&  13.68$\pm$  0.41   &   -                &    -               &   \\
 63 &  2003May11 &  52771.41 &   -                 & -            &  -                  &   50.50$\pm$1.80   &    14.77$\pm$ 0.65 &  2.67$\pm$0.24 \\
 64 &  2003May22 &  52781.79 &   15.75$\pm$  0.54  & 4.45$\pm$0.42&  14.20$\pm$  0.43   &   50.55$\pm$2.07   &    16.44$\pm$ 0.72 &  4.34$\pm$0.39 \\
 65 &  2003May23 &  52782.87 &   15.11$\pm$  0.51  & 3.81$\pm$0.36&  14.36$\pm$  0.43   &   -                &    -               &   \\
 66 &  2003Jun21 &  52811.85 &   16.88$\pm$  0.57  & 5.58$\pm$0.53&  14.70$\pm$  0.44   &   -                &    -               &   \\
 67 &  2003Sep03 &  52885.65 &   18.69$\pm$  0.64  & 7.39$\pm$0.70&  13.11$\pm$  0.39   &   -                &    -               &   \\
 68 &  2003Sep17 &  52899.69 &   17.60$\pm$  0.60  & 6.30$\pm$0.60&  13.26$\pm$  0.40   &   43.53$\pm$1.78   &    18.87$\pm$ 0.83 &  6.77$\pm$0.60 \\
 69 &  2004Jan27 &  53032.00 &   15.69$\pm$  0.53  & 4.39$\pm$0.42&  11.76$\pm$  0.35   &   -                &    -               &   \\
 70 &  2004Mar02 &  53066.59 &   15.78$\pm$  0.54  & 4.48$\pm$0.43&  11.81$\pm$  0.35   &   42.52$\pm$1.74   &    15.31$\pm$ 0.67 &  3.21$\pm$0.29 \\
 71 &  2004Mar16 &  53080.94 &   15.69$\pm$  0.53  & 4.39$\pm$0.42&  12.08$\pm$  0.36   &   -                &    -               &   \\
 72 &  2004Mar18 &  53082.94 &   15.84$\pm$  0.54  & 4.54$\pm$0.43&  13.08$\pm$  0.39   &   39.38$\pm$1.61   &    16.27$\pm$ 0.72 &  4.17$\pm$0.37 \\
 73 &  2004Apr11 &  53106.94 &   15.81$\pm$  0.54  & 4.51$\pm$0.43&  12.75$\pm$  0.38   &   40.61$\pm$1.67   &    16.64$\pm$ 0.73 &  4.54$\pm$0.40 \\
 74 &  2004Apr12 &  53107.84 &   14.55$\pm$  0.49  & 3.25$\pm$0.31&  12.52$\pm$  0.38   &   -                &    -               &   \\
 75 &  2004Apr13 &  53108.92 &   -                 & -            &   -                 &   40.09$\pm$1.64   &    17.13$\pm$ 0.75 &  5.03$\pm$0.45 \\
 76 &  2004May18 &  53143.83 &   15.39$\pm$  0.52  & 4.09$\pm$0.39&  13.65$\pm$  0.41   &   40.05$\pm$1.64   &    16.08$\pm$ 0.71 &  3.98$\pm$0.35 \\
 77 &  2004May19 &  53144.82 &   15.40$\pm$  0.52  & 4.10$\pm$0.39&  12.87$\pm$  0.39   &   -                &    -               &   \\
 78 &  2004May20 &  53145.78 &   -                 & -            &  -                  &   38.57$\pm$1.58   &    17.02$\pm$ 0.75 &  4.92$\pm$0.44 \\
 79 &  2004Jun10 &  53166.80 &   16.18$\pm$  0.55  & 4.88$\pm$0.46&  13.51$\pm$  0.41   &   40.71$\pm$1.67   &    16.57$\pm$ 0.73 &  4.47$\pm$0.40 \\
 80 &  2004Jun11 &  53167.77 &   15.70$\pm$  0.53  & 4.40$\pm$0.42&  13.03$\pm$  0.39   &   -                &    -               &   \\
 81 &  2004Aug10 &  53228.32 &   14.12$\pm$  0.48  & 2.82$\pm$0.27&  12.32$\pm$  0.37   &   36.90$\pm$1.51   &    13.85$\pm$ 0.61 &  1.75$\pm$0.16 \\
 82 &  2004Aug18 &  53235.64 &   15.08$\pm$  0.51  & 3.78$\pm$0.36&  12.89$\pm$  0.39   &   -                &    -               &   \\
 83 &  2004Aug20 &  53237.63 &   15.26$\pm$  0.52  & 3.96$\pm$0.38&  12.67$\pm$  0.38   &   36.29$\pm$1.49   &    16.00$\pm$ 0.70 &  3.90$\pm$0.35 \\
 84 &  2004Sep06 &  53254.63 &   17.26$\pm$  0.59  & 5.96$\pm$0.57&  13.35$\pm$  0.40   &   -                &    -               &   \\
 85 &  2004Dec18 &  53357.50 &   17.94$\pm$  0.61  & 6.64$\pm$0.63&  13.30$\pm$  0.40   &   47.78$\pm$1.96   &    17.94$\pm$ 0.79 &  5.84$\pm$0.52 \\
 86 &  2005Feb13 &  53415.01 &   15.12$\pm$  0.51  & 3.82$\pm$0.36&  12.20$\pm$  0.37   &   43.26$\pm$1.77   &    16.76$\pm$ 0.74 &  4.66$\pm$0.42 \\
 87 &  2005Mar17 &  53446.89 &   15.55$\pm$  0.53  & 4.25$\pm$0.40&  13.05$\pm$  0.39   &   41.30$\pm$1.69   &    16.16$\pm$ 0.71 &  4.06$\pm$0.36 \\
 88 &  2005Apr15 &  53475.90 &   14.32$\pm$  0.49  & 3.02$\pm$0.29&  11.66$\pm$  0.35   &   -                &    -               &   \\
 89 &  2005Apr16 &  53476.79 &   -                 & -            &  -                  &   40.47$\pm$1.66   &    15.83$\pm$ 0.70 &  3.73$\pm$0.33 \\
 90 &  2005May11 &  53501.85 &   15.65$\pm$  0.53  & 4.35$\pm$0.41&  12.07$\pm$  0.36   &   43.06$\pm$1.77   &    17.18$\pm$ 0.76 &  5.08$\pm$0.45 \\
 91 &  2005May12 &  53502.83 &   14.79$\pm$  0.50  & 3.49$\pm$0.33&  11.78$\pm$  0.35   &   -                &    -               &   \\
 92 &  2005May12 &  53503.48 &   -                 & -            &  -                  &   43.04$\pm$1.76   &    17.15$\pm$ 0.75 &  5.05$\pm$0.45 \\
 93 &  2005May13 &  53503.83 &   -                 & -            &  -                  &   37.84$\pm$1.55   &    16.60$\pm$ 0.73 &  4.50$\pm$0.40 \\
 94 &  2005May14 &  53504.73 &   14.75$\pm$  0.50  & 3.45$\pm$0.33&  12.14$\pm$  0.36   &   -                &    -               &   \\
 95 &  2005May15 &  53505.85 &   -                 & -            &  -                  &   41.10$\pm$1.69   &    15.52$\pm$ 0.68 &  3.42$\pm$0.30 \\
 96 &  2005Jun08 &  53529.85 &   18.55$\pm$  0.63  & 7.25$\pm$0.69&  11.97$\pm$  0.36   &   42.69$\pm$1.75   &    18.96$\pm$ 0.83 &  6.86$\pm$0.61 \\
 97 &  2005Jun09 &  53530.64 &   17.74$\pm$  0.60  & 6.44$\pm$0.61&  12.39$\pm$  0.37   &   -                &    -               &   \\
 98 &  2005Jun11 &  53532.78 &   18.01$\pm$  0.61  & 6.71$\pm$0.64&  13.09$\pm$  0.39   &   -                &    -               &   \\
 99 &  2005Jun16 &  53538.46 &   17.81$\pm$  0.61  & 6.51$\pm$0.62&  12.37$\pm$  0.37   &   36.47$\pm$1.50   &    17.45$\pm$ 0.77 &  5.35$\pm$0.48 \\
100 &  2005Jul07 &  53559.44 &   18.03$\pm$  0.61  & 6.73$\pm$0.64&  11.35$\pm$  0.34   &   38.42$\pm$1.58   &    18.75$\pm$ 0.83 &  6.65$\pm$0.59 \\
101 &  2005Aug26 &  53608.70 &   15.14$\pm$  0.51  & 3.84$\pm$0.36&  13.84$\pm$  0.42   &   42.30$\pm$1.73   &    16.02$\pm$ 0.70 &  3.92$\pm$0.35 \\
102 &  2005Aug28 &  53610.64 &   16.37$\pm$  0.56  & 5.07$\pm$0.48&  13.16$\pm$  0.39   &   -                &    -               &   \\
103 &  2005Aug29 &  53611.63 &   15.52$\pm$  0.53  & 4.22$\pm$0.40&  12.73$\pm$  0.38   &   -                &    -               &   \\
104 &  2005Aug31 &  53613.62 &   15.32$\pm$  0.52  & 4.02$\pm$0.38&  12.44$\pm$  0.37   &   -                &    -               &   \\
105 &  2005Sep08 &  53621.67 &   15.70$\pm$  0.53  & 4.40$\pm$0.42&  12.38$\pm$  0.37   &   -                &    -               &   \\
106 &  2005Sep09 &  53622.67 &   15.57$\pm$  0.53  & 4.27$\pm$0.41&  12.96$\pm$  0.39   &   -                &    -               &   \\
107 &  2005Oct11 &  53655.23 &   -                 & -            &  -                  &   46.61$\pm$1.91   &    18.89$\pm$ 0.83 &  6.79$\pm$0.60 \\
108 &  2005Nov09 &  53684.19 &   -                 & -            &  -                  &   52.79$\pm$2.16   &    18.42$\pm$ 0.81 &  6.32$\pm$0.56 \\
109 &  2005Dec28 &  53732.50 &   -                 & -            &  12.79$\pm$  0.38   &   48.46$\pm$1.99   &    -               &   \\
110 &  2006Mar08 &  53802.96 &   17.54$\pm$  0.60  & 6.24$\pm$0.59&  11.93$\pm$  0.36   &   -                &    -               &   \\
111 &  2006Mar08 &  53802.97 &   19.94$\pm$  0.68  & 8.64$\pm$0.82&  11.47$\pm$  0.34   &   -                &    -               &   \\
112 &  2006Apr17 &  53842.94 &   14.98$\pm$  0.51  & 3.68$\pm$0.35&  11.77$\pm$  0.35   &   43.97$\pm$1.80   &    16.54$\pm$ 0.73 &  4.44$\pm$0.40 \\
113 &  2006Apr18 &  53843.93 &   16.04$\pm$  0.55  & 4.74$\pm$0.45&  11.73$\pm$  0.35   &   -                &    -               &   \\
114 &  2006Apr19 &  53844.86 &   15.68$\pm$  0.53  & 4.38$\pm$0.42&  12.92$\pm$  0.39   &   -                &    -               &   \\
115 &  2006Aug28 &  53975.65 &   16.22$\pm$  0.55  & 4.92$\pm$0.47&  12.81$\pm$  0.38   &   43.11$\pm$1.77   &    17.66$\pm$ 0.78 &  5.56$\pm$0.50 \\
116 &  2006Aug28 &  53976.69 &   14.68$\pm$  0.50  & 3.38$\pm$0.32&  13.36$\pm$  0.40   &   -                &    -               &   \\
117 &  2006Aug29 &  53977.50 &   15.95$\pm$  0.54  & 4.65$\pm$0.44&  14.02$\pm$  0.42   &   46.75$\pm$1.92   &    16.43$\pm$ 0.72 &  4.33$\pm$0.39 \\
118 &  2006Aug30 &  53977.63 &   15.03$\pm$  0.51  & 3.73$\pm$0.35&  13.70$\pm$  0.41   &   -                &    -               &   \\
119 &  2006Aug30 &  53978.48 &   15.02$\pm$  0.51  & 3.72$\pm$0.35&  13.38$\pm$  0.40   &   41.55$\pm$1.70   &    15.42$\pm$ 0.68 &  3.32$\pm$0.30 \\
120 &  2006Aug31 &  53978.63 &   14.18$\pm$  0.48  & 2.88$\pm$0.27&  13.65$\pm$  0.41   &   43.79$\pm$1.80   &    15.45$\pm$ 0.68 &  3.35$\pm$0.30 \\
121 &  2007Mar15 &  54174.95 &   15.47$\pm$  0.53  & 4.17$\pm$0.40&  13.00$\pm$  0.39   &   -                &    -               &   \\
122 &  2007May20 &  54240.85 &   14.98$\pm$  1.08  & 3.68$\pm$0.35&  13.28$\pm$  0.40   &   47.04$\pm$1.93   &    15.96$\pm$ 0.70 &  3.86$\pm$0.34 \\
123 &  2007May23 &  54243.82 &   16.58$\pm$  1.19  & 5.28$\pm$0.50&  12.51$\pm$  0.38   &   -                &    -               &   \\
124 &  2007May26 &  54246.87 &   17.01$\pm$  0.58  & 5.71$\pm$0.54&  13.41$\pm$  0.40   &   -                &    -               &   \\
125 &  2007Jun08 &  54259.40 &   15.20$\pm$  0.52  & 3.90$\pm$0.37&  13.44$\pm$  0.40   &   50.07$\pm$2.05   &    15.20$\pm$ 0.67 &  3.10$\pm$0.28 \\
126 &  2007Jun20 &  54271.76 &   16.12$\pm$  0.55  & 4.82$\pm$0.46&  13.16$\pm$  0.39   &   -                &    -               &   \\
127 &  2007Jun21 &  54272.85 &   14.91$\pm$  0.51  & 3.61$\pm$0.34&  12.87$\pm$  0.39   &   47.99$\pm$1.97   &    16.17$\pm$ 0.71 &  4.07$\pm$0.36 \\
128 &  2007Jul20 &  54302.38 &   15.64$\pm$  0.53  & 4.34$\pm$0.41&  12.95$\pm$  0.39   &   46.22$\pm$1.90   &    18.21$\pm$ 0.80 &  6.11$\pm$0.54 \\
129 &  2007Aug10 &  54322.66 &   14.19$\pm$  0.48  & 2.89$\pm$0.27&  12.93$\pm$  0.39   &   42.82$\pm$1.76   &    14.52$\pm$ 0.64 &  2.42$\pm$0.22 \\
130 &  2007Aug11 &  54323.63 &   14.50$\pm$  0.49  & 3.20$\pm$0.30&  12.79$\pm$  0.38   &   -                &    -               &   \\
131 &  2007Aug18 &  54331.31 &   16.29$\pm$  0.55  & 4.99$\pm$0.47&  14.02$\pm$  0.42   &   45.58$\pm$1.87   &    14.84$\pm$ 0.65 &  2.74$\pm$0.24 \\
132 &  2007Aug19 &  54332.33 &   16.70$\pm$  0.57  & 5.40$\pm$0.51&  13.52$\pm$  0.41   &   43.98$\pm$1.80   &    16.58$\pm$ 0.73 &  4.48$\pm$0.40 \\
133 &  2007Sep04 &  54347.63 &   14.86$\pm$  0.51  & 3.56$\pm$0.34&  12.81$\pm$  0.38   &   -                &    -               &   \\
134 &  2008Jun08 &  54626.46 &   15.98$\pm$  0.54  & 4.68$\pm$0.44&  13.01$\pm$  0.39   &   43.78$\pm$1.79   &    17.06$\pm$ 0.75 &  4.96$\pm$0.44 \\
135 &  2009May17 &  54969.47 &   14.13$\pm$  0.48  & 2.83$\pm$0.27&  12.57$\pm$  0.38   &   39.68$\pm$1.63   &    15.29$\pm$ 1.57 &  3.19$\pm$0.28 \\
136 &  2009May19 &  54970.54 &   15.14$\pm$  0.51  & 3.84$\pm$0.36&  12.32$\pm$  0.37   &   38.61$\pm$1.58   &    18.39$\pm$ 1.89 &  6.29$\pm$0.56 \\
137 &  2009May20 &  54972.20 &   -                 & -            &  11.44$\pm$  0.34   &   38.59$\pm$1.58   &    18.39$\pm$ 1.89 &  6.29$\pm$0.56 \\
138 &  2009Jul17 &  55030.33 &   -                 & -            &  -                  &   38.83$\pm$1.59   &    14.31$\pm$ 0.63 &  2.21$\pm$0.20 \\
139 &  2009Aug14 &  55058.38 &   16.70$\pm$  0.57  & 5.40$\pm$0.51&  11.84$\pm$  0.36   &   49.00$\pm$2.01   &    19.31$\pm$ 0.85 &  7.21$\pm$0.64 \\
140 &  2010Mar21 &  55276.51 &   18.25$\pm$  0.62  & 6.95$\pm$0.66&  11.92$\pm$  0.36   &   -                &    -               &   \\
141 &  2010Apr20 &  55306.52 &   16.75$\pm$  0.57  & 5.45$\pm$0.52&  13.56$\pm$  0.41   &   40.59$\pm$1.66   &    17.66$\pm$ 0.78 &  5.56$\pm$0.50 \\
142 &  2010Jun19 &  55367.42 &   15.18$\pm$  0.52  & 3.88$\pm$0.37&  10.82$\pm$  0.32   &   36.41$\pm$1.49   &    17.23$\pm$ 0.76 &  5.13$\pm$0.46 \\
\hline
\end{longtable}
\tablefoottext{*}{Continuum flux is given in $10^{-16} \rm erg \, cm^{-2} s^{-1} \AA^{-1}$ and
 line flux in $ 10^{-14} \rm erg \, cm^{-2} s^{-1} $.}\\
\tablefoottext{**}{The fluxes with the mark "cor" are corrected for the host-galaxy contribution. }
}

\begin{table*}
\begin{center}
\caption[]{Estimates of the errors for line and line-segment fluxes.}\label{tab5}
\begin{tabular}{lcccc}
\hline \hline
 Line              &  \multicolumn{2}{c}{Spectral region}  & $\sigma \pm$e  {  (host-galaxy corrected)} & $V_r$ region \\
                   &   [\AA] (obs) & [\AA] (rest)          &  [\%]             & [km s$^{-1}$] \\
 \hline
 cont 5100         &   5200--5250  & 5077--5126  &  3.4$\pm$2.6   (9.5$\pm$6.2) &  -   \\
 cont 6200         &   6356--6406  & 6206--6255  &  4.4$\pm$3.1   (8.9$\pm$6.0) &  -   \\
H$\alpha$ - total  &   6500--6965  & 6347--6801  &  4.1$\pm$2.8   (4.1$\pm$2.8) &  (-9875;+10865)  \\
 H$\beta$ - total  &   4845--5150  & 4731--5028  &  3.0$\pm$2.3   (3.2$\pm$2.8) &  (-8074;+10291)  \\
H$\alpha$ - blue   &   6540--6660  & 6386--6503  &  3.7$\pm$2.8     &  (-8090; -2730)  \\
H$\alpha$ - core   &   6660--6760  & 6503--6600  &  3.3$\pm$2.2     &  (-2730; +1741)  \\
H$\alpha$ - red 1  &   6760--6860  & 6600--6698  &  4.5$\pm$2.2     &  (+1739; +6197)  \\
H$\alpha$ - red 2  &   6920--6940  & 6757--6776  & 13.1$\pm$15.3    &  (+8875; +9768)  \\
H$\beta$ - blue    &   4854--4937  & 4739--4820  &  5.8$\pm$4.2     &  (-7532; -2531)  \\
H$\beta$ - core    &   4937--5021  & 4820--4902  &  4.3$\pm$3.6     &  (-2531; +2531)  \\
H$\beta$ - red     &   5021--5104  & 4902--4984  &  4.3$\pm$3.2     &  (+2531; +7532)  \\
\hline
\end{tabular}
\end{center}
\end{table*}

\subsection{Unification of the spectral data}

In order to investigate the long term spectral variability of an
AGN, it is necessary to make a consistent, uniformed data set. Since
observations were carried out with four different instruments, one
has to correct the line and continuum fluxes for aperture effects
\citep{pc83}. As in our previous papers
\citep{sh01,sh04,sh08,sh10,sh12} we determined  a point-source
correction factor ($\varphi$) and correction for contribution host
galaxy (G) using the the following expressions \citep[see][]{pet95}:
$$F({\rm H}\beta)_{\rm true} =\varphi \cdot F({\rm H}\beta)_{\rm obs},$$
where $F({\rm H}\beta)_{\rm obs}$ is the observed H$\beta$ flux;
$F({\rm H}\beta)_{\rm true}$ is the aperture corrected H$\beta$ flux. And
$$ F({\rm cnt})_{\rm true} = \varphi \cdot F({\rm cont})_{\rm obs} - G(g),$$
where $F({\rm cnt})_{\rm obs}$ is the continuum flux at observed wavelength;
$G(g)$ is an aperture-dependent correction factor to account for the
host galaxy contribution.

The GHO observing scheme (Table \ref{tab1}), which correspond to a
projected aperture ($2.5\arcsec\times6.0\arcsec$) of the 2.1 m
telescope, was adopted as a standard with $\varphi= 1.0$ and
$G(g)=0.0$, since with the GHO we collected the largest number of
observed spectra of Arp102B.

The correction factors $\varphi$ and $G(g)$ are determined
empirically by comparing pairs of simultaneous observations from
each of given telescope data sets to that of the standard data set
(as it was done in AGN Watch \citep[as e.g.][]{pet94,pet99,pet02}.
Due to the small number of CA spectra (1987--1994) it was impossible
to determine the correction factors as described above. The CA
apertures for 7 spectra were close to our standard (slit width
$2.0\arcsec -2.1\arcsec$) so we set $\varphi$=1.000 and
$G(g)$=0.000, while for only 3 spectra - JD2446976 (1987Jun29),
JD2446977 (1987Jun30) and JD2447229 (1988Mar08) the slit width was
different ($1.5\arcsec$). In these last cases we expect that the
corrections are the average between the aperture $1.0\arcsec$ and
$2.0\arcsec$ (Table \ref{tab3}). However, in any case, to CA fluxes
should be taken with caution. In practice, intervals which we
defined as "nearly simultaneous" are typically of 1-3 days.
Therefore, the variability on short time scales ($<$ 3 days) is
suppressed. The point-source correction factors  and $G(g)$ values
for different samples are listed in Table \ref{tab3}. Using these
factors, we re-calibrated the observed fluxes of H$\alpha$,
H$\beta$,  and the blue and red continuum to a common scale
corresponding to our standard aperture $2.5\arcsec \times 6.0
\arcsec$ (Table \ref{tab4}). {  Also we determined the observed 
fluxes in the H$\alpha,\beta$ line-segments (core and wings -- 
the wavelengths  intervals are given in Table \ref{tab5}). Then we 
re-calibrated the H$\alpha,\beta$ line-segment fluxes  to a 
common scale that corresponds to our standard aperture $2.5\arcsec \times 6.0
\arcsec$ (see Tables \ref{tab_seg_ha}-\ref{tab_seg_hb}).}
The mean errors (uncertainties) in the {  observed}
continuum fluxes and in the {  observed fluxes} of emission lines
and their segments are given in Table \ref{tab5}.

\onllongtab{6}{
\begin{longtable}{cccccc}
\caption[]{\label{tab_seg_ha} The measured line segment fluxes of H$\alpha$
in units of $ 10^{-14} \rm erg \, cm^{-2} s^{-1} $.}\\
\hline \hline
N& MJD    & F(H$\alpha$)blue &  F(H$\alpha$)red1 & F(H$\alpha$)red2 & F(H$\alpha$)core  \\
\hline
\endfirsthead
\caption{Continued.}\\
\hline
N& MJD    & F(H$\alpha$)blue &  F(H$\alpha$)red1 & F(H$\alpha$)red2 & F(H$\alpha$)core  \\
\hline
\endhead
\hline
\endfoot
\hline
\endlastfoot
  1 & 46975  &  14.794  &  11.854   &  0.887    &  24.248    \\
  2 & 46976  &  12.816  &  10.648   &  0.920    &  21.660    \\
  3 & 46977  &  14.383  &  11.946   &  0.933    &  23.243    \\
  4 & 47229  &  10.668  &  7.403    &  0.388    &  17.304    \\
  5 & 47827  &  8.576   &  6.254    &  0.557    &  15.528   \\
  6 & 47828  &  10.540  &  7.367    &  0.674    &  17.832    \\
  7 & 48865  &  9.416   &  7.215    &  0.804    &  18.006   \\
  8 & 49238  &  11.721  &  7.845    &  0.660    &  21.903    \\
  9 & 49598  &  7.755   &  7.066    &  0.611    &  16.159   \\
 10 & 51008  &  18.692  &  13.213   &  0.427    &  25.368    \\
 11 & 51011  &  16.101  &  12.064   &  0.509    &  24.633    \\
 12 & 51020  &  16.001  &  11.986   &  0.501    &  24.474    \\
 13 & 51021  &  15.167  &  11.007   &  0.480    &  23.163    \\
 14 & 51079  &  12.188  &  8.036    &  0.305    &  19.490    \\
 15 & 51486  &  15.873  &  12.697   &  0.511    &  26.352    \\
 16 & 51569  &  11.241  &  8.229    &  0.316    &  18.947    \\
 17 & 51570  &  10.084  &  7.563    &  0.265    &  17.343    \\
 18 & 51599  &  12.826  &  9.327    &  0.396    &  20.861    \\
 19 & 51600  &  12.494  &  8.875    &  0.424    &  19.998    \\
 20 & 51658  &  10.760  &  8.735    &  0.625    &  20.116    \\
 21 & 51659  &  11.664  &  9.490    &  0.608    &  21.453    \\
 22 & 51689  &  11.013  &  8.831    &  0.592    &  19.831    \\
 23 & 51689  &  11.473  &  8.863    &  0.624    &  19.811    \\
 24 & 51719  &  11.581  &  9.522    &  0.585    &  21.452    \\
 25 & 51756  &  11.902  &  9.751    &  0.711    &  20.953    \\
 26 & 51834  &  12.018  &  9.429    &  0.434    &  21.915    \\
 27 & 51980  &  12.791  &  9.116    &  0.397    &  21.041    \\
 28 & 52043  &  12.377  &  8.447    &  0.329    &  18.740    \\
 29 & 52073  &  12.480  &  8.734    &  0.385    &  20.522    \\
 30 & 52101  &  11.915  &  10.391   &  0.336    &  20.367    \\
 31 & 52151  &  12.999  &  9.062    &  0.481    &  21.449    \\
 32 & 52337  &  11.167  &  7.784    &  0.484    &  20.412    \\
 33 & 52338  &  12.265  &  8.361    &  0.478    &  21.373    \\
 34 & 52367  &  10.505  &  7.073    &  0.381    &  18.847    \\
 35 & 52369  &  10.629  &  7.557    &  0.400    &  20.547    \\
 36 & 52397  &  10.459  &  6.999    &  0.291    &  19.883    \\
 37 & 52427  &  10.387  &  7.288    &  0.409    &  19.972    \\
 38 & 52450  &  10.695  &  7.786    &  0.388    &  21.101    \\
 39 & 52503  &  10.389  &  7.438    &  0.335    &  18.909    \\
 40 & 52723  &  9.797   &  7.071    &  0.446    &  19.382   \\
 41 & 52724  &  9.976   &  6.695    &  0.481    &  18.911   \\
 42 & 52740  &  10.030  &  6.696    &  0.414    &  18.758    \\
 43 & 52771  &  12.170  &  8.838    &  0.450    &  21.080    \\
 44 & 52781  &  12.294  &  8.721    &  0.594    &  20.522    \\
 45 & 52899  &  9.718   &  7.618    &  0.373    &  19.244   \\
 46 & 53066  &  8.878   &  8.050    &  0.500    &  18.185   \\
 47 & 53082  &  8.427   &  6.584    &  0.323    &  17.824   \\
 48 & 53106  &  8.926   &  6.375    &  0.441    &  18.256   \\
 49 & 53108  &  9.314   &  6.356    &  0.395    &  18.167   \\
 50 & 53143  &  9.198   &  6.260    &  0.318    &  17.877   \\
 51 & 53145  &  9.368   &  5.688    &  0.294    &  17.187   \\
 52 & 53166  &  9.559   &  6.793    &  0.435    &  17.345   \\
 53 & 53228  &  8.095   &  5.892    &  0.238    &  16.072   \\
 54 & 53237  &  8.556   &  5.635    &  0.239    &  16.262   \\
 55 & 53357  &  9.366   &  8.181    &  0.569    &  20.268   \\
 56 & 53415  &  9.133   &  7.882    &  0.306    &  18.973   \\
 57 & 53446  &  8.594   &  7.016    &  0.392    &  18.415   \\
 58 & 53475  &  8.508   &  6.912    &  0.302    &  18.228   \\
 59 & 53501  &  8.943   &  7.329    &  0.255    &  19.345   \\
 60 & 53503  &  8.035   &  6.506    &  0.254    &  16.976   \\
 61 & 53505  &  8.018   &  7.249    &  0.365    &  18.888   \\
 62 & 53529  &  9.004   &  7.252    &  0.301    &  19.172   \\
 63 & 53538  &  7.446   &  5.731    &  0.218    &  16.732   \\
 64 & 53559  &  7.867   &  6.649    &  0.181    &  16.976   \\
 65 & 53608  &  9.126   &  7.366    &  0.307    &  18.751   \\
 66 & 53655  &  9.604   &  8.188    &  0.197    &  20.401   \\
 67 & 53684  &  12.131  &  10.553   &  0.464    &  24.961    \\
 68 & 53842  &  10.396  &  6.379    &  0.553    &  19.629    \\
 69 & 53975  &  10.951  &  6.679    &  0.170    &  19.055    \\
 70 & 53977  &  11.265  &  7.314    &  0.425    &  19.991    \\
 71 & 53978  &  10.228  &  6.468    &  0.397    &  18.383    \\
 72 & 53978  &  10.805  &  6.745    &  0.342    &  19.226    \\
 73 & 54240  &  12.045  &  6.348    &  0.340    &  21.030    \\
 74 & 54259  &  11.982  &  7.419    &  0.445    &  20.837    \\
 75 & 54272  &  12.635  &  7.060    &  0.300    &  20.877    \\
 76 & 54302  &  11.924  &  7.121    &  0.149    &  19.526    \\
 77 & 54322  &  10.771  &  6.534    &  0.233    &  18.902    \\
 78 & 54331  &  11.053  &  7.132    &  0.286    &  19.579    \\
 79 & 54332  &  10.558  &  6.830    &  0.193    &  19.593    \\
 80 & 54626  &  9.193   &  6.560    &  0.189    &  20.469   \\
 81 & 54969  &  7.713   &  5.902    &  0.294    &  18.841   \\
 82 & 54970  &  7.765   &  5.742    &  0.471    &  17.646   \\
 83 & 54972  &  7.100   &  5.595    &  0.462    &  17.120   \\
 84 & 55030  &  7.379   &  5.871    &  0.102    &  17.097   \\
 85 & 55058  &  9.795   &  7.485    &  0.469    &  22.373   \\
 86 & 55306  &  8.972   &  5.514    &  0.379    &  18.180   \\
 87 & 55367  &  7.419   &  4.419    &  0.272    &  17.829   \\
\hline
\end{longtable}
}

\onllongtab{7}{
\begin{longtable}{ccccc}
\caption[]{\label{tab_seg_hb} The measured line segment fluxes of H$\beta$
in units of $ 10^{-14} \rm erg \, cm^{-2} s^{-1} $.}\\
\hline \hline
N& MJD    & F(H$\beta$)blue  & F(H$\beta$)core & F(H$\beta$)red \\
\hline
\endfirsthead
\caption{Continued.}\\
\hline
N& MJD    & F(H$\beta$)blue  & F(H$\beta$)core & F(H$\beta$)red \\
\hline
\endhead
\hline
\endfoot
\hline
\endlastfoot
  1 & 46975 &  4.455 & 4.793 & 4.581   \\
  2 & 47229 &  3.188 & 3.676 & 3.465   \\
  3 & 47665 &  2.537 & 3.088 & 3.008   \\
  4 & 47828 &  2.358 & 3.052 & 3.065   \\
  5 & 48865 &  2.462 & 3.314 & 3.581   \\
  6 & 49238 &  3.164 & 3.749 & 3.536   \\
  7 & 50940 &  3.930 & 4.123 & 4.338   \\
  8 & 50942 &  4.308 & 4.654 & 4.744   \\
  9 & 50990 &  5.374 & 4.691 & 5.157   \\
 10 & 50991 &  4.430 & 3.739 & 4.424   \\
 11 & 51008 &  4.648 & 4.644 & 4.741   \\
 12 & 51011 &  4.128 & 4.842 & 4.681   \\
 13 & 51020 &  4.096 & 4.801 & 4.636   \\
 14 & 51021 &  4.043 & 4.705 & 4.579   \\
 15 & 51079 &  3.578 & 3.827 & 3.874   \\
 16 & 51410 &  3.251 & 3.575 & 3.845   \\
 17 & 51486 &  3.139 & 3.742 & 4.278   \\
 18 & 51569 &  2.942 & 3.371 & 3.862   \\
 19 & 51599 &  2.953 & 3.441 & 4.116   \\
 20 & 51600 &  3.150 & 3.528 & 4.048   \\
 21 & 51658 &  2.527 & 3.424 & 3.856   \\
 22 & 51659 &  2.419 & 3.286 & 3.663   \\
 23 & 51689 &  2.809 & 3.554 & 3.952   \\
 24 & 51689 &  2.953 & 3.526 & 3.914   \\
 25 & 51719 &  2.375 & 3.469 & 3.504   \\
 26 & 51756 &  2.729 & 3.470 & 3.677   \\
 27 & 51834 &  2.589 & 3.506 & 3.756   \\
 28 & 51980 &  2.445 & 3.141 & 3.519   \\
 29 & 52041 &  3.188 & 3.624 & 4.040   \\
 30 & 52073 &  2.779 & 3.378 & 3.624   \\
 31 & 52102 &  2.793 & 3.558 & 3.725   \\
 32 & 52151 &  3.308 & 3.765 & 3.648   \\
 33 & 52190 &  2.861 & 3.648 & 3.439   \\
 34 & 52337 &  2.632 & 3.531 & 3.570   \\
 35 & 52339 &  3.397 & 3.975 & 3.382   \\
 36 & 52349 &  2.767 & 3.618 & 3.118   \\
 37 & 52366 &  3.026 & 3.477 & 3.089   \\
 38 & 52369 &  2.663 & 3.439 & 3.582   \\
 39 & 52396 &  3.107 & 3.735 & 3.185   \\
 40 & 52398 &  2.815 & 3.364 & 3.002   \\
 41 & 52426 &  2.663 & 3.166 & 2.891   \\
 42 & 52429 &  2.901 & 3.308 & 2.861   \\
 43 & 52450 &  2.354 & 3.233 & 3.049   \\
 44 & 52471 &  2.421 & 3.181 & 3.212   \\
 45 & 52501 &  2.612 & 3.282 & 2.858   \\
 46 & 52722 &  2.618 & 3.539 & 3.519   \\
 47 & 52723 &  2.488 & 3.324 & 3.446   \\
 48 & 52769 &  2.760 & 3.416 & 3.592   \\
 49 & 52781 &  2.953 & 3.491 & 3.845   \\
 50 & 52782 &  3.112 & 3.505 & 3.798   \\
 51 & 52811 &  3.427 & 3.602 & 3.766   \\
 52 & 52885 &  2.589 & 3.373 & 3.399   \\
 53 & 52899 &  2.545 & 3.342 & 3.482   \\
 54 & 53031 &  2.426 & 3.400 & 2.647   \\
 55 & 53066 &  1.922 & 3.098 & 3.030   \\
 56 & 53080 &  2.216 & 3.352 & 2.998   \\
 57 & 53082 &  2.243 & 3.448 & 3.508   \\
 58 & 53106 &  2.102 & 3.286 & 3.367   \\
 59 & 53107 &  2.312 & 3.470 & 3.181   \\
 60 & 53143 &  2.674 & 3.497 & 3.479   \\
 61 & 53144 &  2.572 & 3.177 & 3.302   \\
 62 & 53166 &  2.659 & 3.343 & 3.563   \\
 63 & 53167 &  2.826 & 3.107 & 3.391   \\
 64 & 53228 &  2.847 & 3.031 & 2.965   \\
 65 & 53235 &  2.688 & 3.182 & 3.320   \\
 66 & 53237 &  2.518 & 3.140 & 3.289   \\
 67 & 53254 &  2.773 & 3.277 & 3.546   \\
 68 & 53357 &  2.465 & 3.395 & 3.446   \\
 69 & 53415 &  2.143 & 3.104 & 3.298   \\
 70 & 53446 &  2.323 & 3.306 & 3.570   \\
 71 & 53475 &  2.323 & 3.122 & 2.732   \\
 72 & 53501 &  2.004 & 3.010 & 3.222   \\
 73 & 53502 &  1.741 & 2.965 & 3.411   \\
 74 & 53051 &  2.221 & 3.027 & 3.251   \\
 75 & 53529 &  2.105 & 2.980 & 3.186   \\
 76 & 53530 &  2.226 & 2.932 & 3.496   \\
 77 & 53532 &  2.560 & 3.240 & 3.410   \\
 78 & 53538 &  2.304 & 3.040 & 3.240   \\
 79 & 53559 &  1.941 & 2.789 & 2.962   \\
 80 & 53608 &  2.545 & 3.547 & 3.692   \\
 81 & 53610 &  2.537 & 3.421 & 3.296   \\
 82 & 53611 &  2.635 & 3.399 & 3.043   \\
 83 & 53613 &  2.352 & 3.299 & 2.994   \\
 84 & 53621 &  2.415 & 3.129 & 3.315   \\
 85 & 53622 &  2.580 & 3.278 & 3.416   \\
 86 & 53732 &  2.493 & 3.137 & 3.338   \\
 87 & 53802 &  2.367 & 2.894 & 3.022   \\
 88 & 53802 &  2.263 & 3.067 & 2.627   \\
 89 & 53842 &  2.052 & 2.855 & 3.169   \\
 90 & 53843 &  2.251 & 3.097 & 2.774   \\
 91 & 53844 &  2.586 & 3.296 & 3.212   \\
 92 & 53975 &  2.865 & 3.263 & 3.159   \\
 93 & 53976 &  2.971 & 3.387 & 3.353   \\ 
 94 & 53977 &  3.527 & 3.011 & 3.338   \\
 95 & 53977 &  3.304 & 3.656 & 3.045   \\
 96 & 53978 &  3.019 & 3.196 & 3.305   \\
 97 & 53978 &  3.021 & 3.735 & 3.171   \\
 98 & 54174 &  2.776 & 3.365 & 3.172   \\
 99 & 54240 &  2.836 & 3.563 & 3.148   \\
100 & 54243 &  2.762 & 3.014 & 2.996   \\
101 & 54246 &  2.935 & 3.310 & 3.361   \\
102 & 54259 &  2.985 & 3.459 & 2.985   \\
103 & 54271 &  3.162 & 3.280 & 2.856   \\
104 & 54272 &  2.941 & 3.177 & 3.177   \\
105 & 54302 &  2.439 & 2.650 & 3.206   \\
106 & 54322 &  2.655 & 3.370 & 3.262   \\
107 & 54323 &  2.690 & 3.277 & 3.018   \\
108 & 54331 &  3.139 & 3.246 & 3.627   \\
109 & 54332 &  3.004 & 3.254 & 3.436   \\
110 & 54347 &  2.662 & 3.382 & 2.998   \\
111 & 54626 &  2.714 & 3.095 & 3.209   \\
112 & 54969 &  2.261 & 3.182 & 3.231   \\
113 & 54970 &  2.281 & 3.150 & 3.153   \\
114 & 54972 &  1.983 & 2.780 & 2.912   \\
115 & 55058 &  2.249 & 2.828 & 3.006   \\
116 & 55276 &  2.276 & 3.073 & 3.199   \\
117 & 55306 &  2.577 & 3.353 & 3.642   \\
118 & 55367 &  2.134 & 2.460 & 2.498   \\
\hline
\end{longtable}
}

\subsection{  The host-galaxy contribution to the spectra of Arp 102B}

{  In the optical spectra of Arp 102B we observe a substantial contribution of
the starlight (e.g. a strong Mg Ib stellar absorption 
line at 5176\AA\ and a strong Na I D line at 5893\AA\ etc.).
The starlight effectively dilutes the non-stellar agn-continuum and
leads to a smaller apparent variability amplitude. Also, the starlight 
affects the profiles and fluxes of the broad Balmer lines.
Therefore, we need to estimate the starlight contribution (i.e. host-galaxy
contribution) to the fluxes of the observed continua, the H$\alpha$ and
H$\beta$ emission lines.

For this we used spectra of Arp 102B and NGC 4339 (E0 galaxy, as Arp 102B), obtained 
on Mar 25, 2003 (JD52723.92)  at 2.1 m GHO's telescope (Mexico) with the aperture 
2.5$\arcsec$ $\times$6.0$\arcsec$ and resolution $\sim$12\AA\ and under the 
same weather conditions (good transparency and seeing $\sim$2.5$\arcsec$).
We scaled the spectra of Arp 102B and NGC 4339 to z=0 and, changing the contribution 
of the galaxy NGC 4339 to Arp 102B in region Mg Ib (for the blue part of the spectrum), 
we received that the best-fitting (when Mg Ib is completely removed) is 
for the 75$\pm$3\% of the host-galaxy contribution to the continuum at $\sim$5100\AA. 
In the spectrum of Arp 102B the emission line at $\sim$5200\AA\
corresponding to N II 5198\AA\ remains, while the absorption line Na I D (at $\sim$5893\AA) 
is fully removed and only a weak emission line He I 5876 remains (Fig. \ref{host}). 
As seen from Fig. \ref{host} (bottom)  the agn-continuum of Arp 102B has  
approximately flat form and it is  $\sim$25\% relatively to
the observed continuum at 5100\AA\ and 6200\AA.  
For the observed continuum of Arp 102B we took the blue and red continuum fluxes
for the date Mar 25, 2003 (JD52723.92) from Table \ref{tab4} and obtained the galaxy
contribution to the observed continuum in absolute units,
which is 75\% of the observed continuum (see Table \ref{host-gal}). Note here, that, depending
on the activity of the AGN, the contribution of the host galaxy continuum to the total observed 
continuum is between $\sim$ 60\% and $\sim$ 80\% (see Table \ref{tab4}).

Then we estimated the host-galaxy contribution to the H$\alpha$ and H$\beta$ emission line fluxes.
We measured H$\alpha$, H$\beta$ line fluxes in the spectrum of Mar 25, 2003 (JD52723.92),
after removing the spectrum of the NGC 4339 galaxy as described above (see Fig. \ref{host}, 
bottom spectrum).  The linear continuum in the blue (near H$\beta$) and red (near H$\alpha$)
regions were constructed in the same way as described in \S 2.2. The H$\alpha$, H$\beta$ line 
fluxes are defined in the same wavelength intervals as in \S 2.2.
In Table \ref{host-gal} we give for the blue and red continua, H$\alpha$, and H$\beta$ emission lines 
fluxes what is the observed flux, host-galaxy contribution flux,
and agn-fluxes corrected for the host-galaxy contribution. As it can be seen from  
Table \ref{host-gal} the contribution of the host-galaxy to H$\alpha$ and H$\beta$ observed 
fluxes is $\sim$4\% and 10\%, respectively. Then we determined  the host-galaxy corrected 
fluxes of all data of the blue and red continua, the H$\alpha$ and H$\beta$ emission lines, 
by subtracting from the observed flux the host-galaxy flux from Table \ref{host-gal}
in absolute units. This is possible because the observed fluxes from 
Table \ref{tab4} were brought (i.e. unified) to the same aperture
(2.5$\arcsec$ $\times$ 6.0$\arcsec$) in \S 2.3 and the galaxy contribution is also estimated 
for the same aperture.
As in \S 2.2  we used the corrected fluxes for receiving the mean 
errors (uncertainties) in the corrected continuum fluxes,  and the H$\alpha$
and H$\beta$ emission line fluxes, by comparing fluxes in the intervals of 0--3 days. 
The mean errors in the corrected continuum and emission line total fluxes are given 
in Table \ref{tab5} in brackets. From Table \ref{tab5} it is clear that the errors
in the corrected continuum fluxes (agn-continuum) are $\sim$2-3 times larger 
then in the observed continuum fluxes, but the errors in the corrected line fluxes  
are close to those in the observed fluxes. The host-galaxy corrected fluxes in the blue and 
red continuum,  H$\alpha$, and H$\beta$ emission lines and their errors are also given 
in Tables \ref{tab4} and \ref{tab5}.}

\begin{table*}
\begin{center}
\caption[]{  Estimate of the host-galaxy starlight contribution to the continuum 
(in units $10^{-16} \rm erg \, cm^{-2} s^{-1} \AA^{-1}$) and line fluxes (in units
$10^{-14} \rm erg \, cm^{-2} s^{-1}$) for the spectrum taken on Mar 25, 2003.}\label{host-gal}
\begin{tabular}{lcccccc}
\hline \hline
 2003 Mar 25 & F(5225\AA) & F(6381\AA) & F(H$\beta$) & F(H$\alpha$) & F(H$\beta$)gal/F(H$\beta$)obs & F(H$\alpha$)gal/F(H$\alpha$)obs \\                                                                        
\hline
  Observed     &  15.04   &   16.10 &  13.00 &   43.74  &       &   \\           
  Host-galaxy  &  11.28   &   12.08 &   1.35 &    1.73  & 10\%  & 4\%   \\     
  AGN          &  3.76    &    4.02 &  11.65 &   42.01  &       &  \\ 	       
\hline
\end{tabular}
\end{center}
\end{table*}

\subsection{  The narrow emission line contribution}

{  In order to estimate the narrow line contributions to the total (and line-segment) line fluxes, 
we used two spectra of Arp 102B obtained with a different spectral resolution on: 2003Mar25, 
JD2452724 (R$\sim$14\AA) and 2006Aug31, JD2453978(R$\sim$10\AA). We estimated and subtracted the broad 
H$\alpha$ and H$\beta$ components using the spline method. The estimated contributions of
the narrow H$\beta$ and [OIII]4959,5007 lines to the observed total H$\beta$ flux as well as
the narrow H$\alpha$+[NII]6584 and [SII]6717,6731 lines to the total H$\alpha$ flux are given in Table \ref{narrow}   
(in absulute units and percentage). 
 The mean errors in  the H$\alpha$ and H$\beta$ line fluxes  are $\sim$1.5-1.8  times larger in the corected 
then in the observed ones.  Additionally, we estimated the narrow line flux contributions  to 
the H$\alpha$ and H$\beta$ core and H$\beta$ red wing (Table \ref{narrow}). Note that the 
core and the red wing of the H$\beta$ line are contaminated with only the narrow H$\beta$ and 
[OIII]4959 lines, respectively ([OIII]5007 is out of the red wing of H$\beta$, see Table \ref{tab5}). 
The core of the H$\alpha$ line is contaminated with the narrow H$\alpha$+[NII]6584 lines
(the [SII]6717,6731 doublet is also out of the red wing of the H$\alpha$ line). 
In Table \ref{narrow} we give the contribution (in \%)  of the narrow lines to the core 
H$\alpha$ and H$\beta$, and the red wing of H$\beta$ relative to the corresponding 
observed mean flux obtained from the spectra taken on 2003Mar25 and 2006Aug31.}

\begin{table*}
\begin{center}
\caption[]{  Contributions the of narrow lines in absolute units to 
the total, wing and core flux of the H$\alpha$ and H$\beta$ lines. (in
units of $10^{-14} \rm erg \ cm^{-2}s^{-1}$). }\label{narrow}
\begin{tabular}{lccccc}
\hline \hline
Line fluxes  &  H$\beta$nar & [OIII]4959\tablefootmark{**} & [OIII]5007 & (H$\alpha$+NII)nar & [SII]6717,6731 \\ 
\hline                                                                       
mean F(nar)  &1.27$\pm$0.02 & 1.07$\pm$0.17 &  3.12$\pm$0.20   & 10.86$\pm$0.34&  2.66$\pm$0.30 \\
F(nar)/F(tot)&   9.5\%      &     8.0\%     &     24\%         &    24.8\%     &     6.1\%      \\
F(nar)/F(core)\tablefootmark{*}& 35.8\% &   &                  &    56.3\%     &                \\
F(nar)/F(red wing)\tablefootmark{*}&& 32.3\%&                  &               &                \\
\hline
\end{tabular}
\end{center}
\tablefoottext{*}{F(nar)/F(core) and F(nar)/F(red wing) - the ratios of the narrow line flux to 
the mean core flux or mean red wing (in \%).  F(core) - mean flux for H$\beta$ core and H$\alpha$ 
core obtained from the spectra taken on 2003Mar25 and 2006Aug31. F(red wing) - mean flux for 
the red H$\beta$ wing btained from the same spectra.}\\
\tablefoottext{**}{The ratio of the [OIII] lines is 5007/4959=2.9 \citep[see][]{dim07}.}
\end{table*}

\section{Data analysis}

We measured and analyzed variations in the continuum and lines using
total of 118 spectra covering the H$\beta$ wavelength region, and 90
spectra covering the H$\alpha$ line. Also, we considered the
variability in the line segments (blue, red, and central segment) of
these lines (see Tables \ref{tab_seg_ha} -- \ref{tab_seg_hb},
available electronically only), where the wavelength ranges of line
segments are given in Table \ref{tab5}.  In Fig. \ref{arp} we
give two examples of total optical spectra taken with the GHO
telescope in July 1998 (when the object was in higher state) and Aug
2006 (when the object was in the lower state of activity). In Fig.
\ref{lc} we present the light curves of the H$\alpha$ and H$\beta$
lines and the corresponding blue (in the rest-frame 5100\AA \ for
H$\beta$) and red (in the rest-frame 6200\AA \ for H$\alpha$)
continuum. {  The dashed lines on (1st and 4th) panels  in Fig. \ref{lc}
present the contributions of the starlight-continuum
of the host galaxy to the blue and red continua.}
 The trend of a high intensity in lines from  1998 is also
seen in the continuum, but it is not the case in 1987. The
variability in lines as well as in the continuum is not so high
(Table \ref{tab6}), i.e. there are several flare-like peaks.
The line and continuum variations are not prominent in the
monitoring period (Table \ref{tab6}). 

We calculated mean {  observed (obs) and corrected (cor) for the host-galaxy 
contribution} fluxes of H$\alpha$, H$\beta$ and continuum in
different periods of the monitoring period  and results are given in
Table \ref{tab6b}. Three things can be noted from Table \ref{tab6b}:

a) during the monitoring period different mean {  observed} red continuum fluxes
(at 6200\AA \ in rest frame) are always larger (at $\sim$5-7\%) than
blue (at 5100\AA \ in rest frame) ones, that is  caused by
the host-galaxy contribution, {  since (as noted in \S 2.4)
the corrected blue and red contina (i.e. agn-continuum) have a nearly 
flat shape}; 

b) in 1987 and 1998, the H$\beta$ and H$\alpha$ different 
mean {  observed and corrected for the host-galaxy contribution (obs and cor in Table \ref{tab6b})} 
fluxes are larger for $\sim$(32-35)\% (H$\beta$) and $\sim$(38-39)\% (H$\alpha$) comparing with those in
the period (1988--1994) for 1987, and period (1999--2010) for 1998;
{   i.e. the variation of the mean line fluxes is independent on the 
contribution of the host-galaxy.}

c) different mean red and blue continuum fluxes declined in
1988--1994 in comparison with the one observed in 1987 for only
$\sim$(6--7)\%, and in (1999--2010) in comparison with the one
observed in 1998 for (13--19)\% (Table \ref{tab6b}). It is
interesting to note that changes in different mean fluxes of lines
between 1987 and (1988--1994) and between 1998 and (1999--2010) are
significantly larger than in the {  observed} continuum fluxes (obs
 in Table \ref{tab6b}). 
{  However, the corrected mean red and blue 
continuum fluxes (i.e. agn-continuum) decreased  $\sim$1.3 times in
1988--1994  with  respect to the one observed in 1987,  and  $\sim$1.53--1.69 times 
in 1999--2010  with respect the 
one in 1998 (see cor in Table \ref{tab6b}). 
But the mean observed and corrected broad line fluxes  have almost 
the same variability amplitude in above considered periods ($\sim$1.35), that is not depending 
on the host-galaxy contribution.} 

{   d) The mean broad H$\alpha,\beta$ line fluxes which are corrected for the narrow line contributions
are given in Table \ref{tab6b} (see: cor-line). Obviously, there is a change of the mean line fluxes 
(decrease of $\sim$1.46--1.63 times in 1999--2010 with respect to the one observed in 1998) that is 
slightly smaller than in the blue-red continuum flux changes, but  it is significantly
higher than in the case where the 
narrow line contributions are not taken into account}.

\begin{table*}
\begin{center}
\caption[]{Parameters of the continuum and line variabilities. The
continuum flux is in units $10^{-16} \rm erg \ cm^{-2} s^{-1}
\AA^{-1}$, and the line and line-segment fluxes are in
$10^{-14} \rm erg \ cm^{-2}s^{-1}$. }\label{tab6}
\begin{tabular}{lccccc}
\hline \hline
 Feature     &   N  &  $F$(mean) & $\sigma$($F$) & $R$(max/min) & $F$(var) \\
 1           &   2  &   3        &  4         & 5             & 6         \\
 \hline
 cont 5100    & 110 &    16.04    & {  1.50}   &  1.48    &  {  0.085} \\
 cont 6200    & 79  &    16.79    & {  1.40}   &  1.47    &  {  0.065} \\
 H$\alpha$ - total & 80  &    45.48    & {  5.91}   &  1.83    &  {  0.123} \\
 H$\beta$ - total  & 112 &    13.37    & {  1.39}   &  1.73    &  {  0.098} \\
\hline
H$\alpha$ - blue  &  78   &  10.64   &   2.18   &  2.63   & 0.202     \\
H$\alpha$ - core  &  78   &  19.73   &   2.05   &  1.64   & 0.098     \\
H$\alpha$ - red 1 &  78   &  7.73    &   1.70   &  2.99   & 0.215     \\
H$\alpha$ - red 2 &  78   &  0.38    &   0.13   &  6.94   & 0.296     \\
H$\beta$ - blue   &  112  &  2.77    &   0.59   &  3.09   &  0.203    \\
H$\beta$ - core   &  112  &  3.39    &   0.41   &  1.97   &  0.114    \\
H$\beta$ - red    &  112  &  3.43    &   0.48   &  2.06   &  0.133    \\
\hline
+CA data\\
\hline
cont 5100          & 115 &    15.97    & {  1.51}   &  1.49    &  {  0.086} \\
cont 6200          & 88  &    16.71    & {  1.39}   &  1.47    &  {  0.064} \\
H$\alpha$ - total  & 90  &    45.75    & {  6.31}   &  1.83    &  {  0.128} \\
H$\beta$ - total   & 118 &    13.41    & {  1.43}   &  1.73    &  {  0.099} \\
\hline
{  Host-galaxy corrected data}\\
\hline
 cont 5100         & 116 &     4.66    & {  1.51}   &  3.96    &  {  0.308} \\
 cont 6200         & 88  &     4.61    & {  1.39}   &  4.74    &  {  0.287} \\
 H$\alpha$ - total & 88  &    44.14    & {  6.35}   &  1.87    &  {  0.138} \\
 H$\beta$ - total  & 116 &    12.08    & {  1.43}   &  1.84    &  {  0.114} \\ 
\hline
{  Narrow-lines subtracted data}\\
\hline
 H$\alpha$ - total & 88  &  30.62   & {  6.35}   &  2.43   &  {  0.203} \\
 H$\beta$ - total  & 116 &  6.62    & {  1.43}   &  2.97   &  {  0.213} \\ 
 H$\alpha$ - core  & 87  &  8.87    & 2.05   &  2.97   &  0.229ß \\
 H$\beta$ - core   & 118 &  2.13    & 0.43   &  3.00   &  0.195 \\
 H$\beta$ - red    & 118 &  2.37    & 0.48   &  2.86   &  0.199 \\ 
\hline
\end{tabular}
\tablefoot{-- Col.(1): Analyzed feature of the spectrum. Col.(2):
Total number of spectra. Col.(3): Mean flux. Col.(4): Standard
deviation. Col.(5): Ratio of the maximal to minimal flux . Col.(6):
Variation amplitude (see text). \\}
\end{center}
\end{table*}

\begin{table*}
\begin{center}
\caption[]{  Possible flares in the pure AGN-continuum (blue and red),
and the corresponding changes of the H$\beta$ and H$\alpha$ fluxes.}\label{flares}
\begin{tabular}{lccccccccc}
\hline \hline
N & UT-Date   &  cnt-agn5100 & Amplitude & cnt-agn6200 & Amplitude & F(H$\beta$)agn & Var & F(H$\alpha$)agn & Var\\
1 &  2        &   3          & 4         & 5           & 6         &  7             &  8   &  9              & 10\\
\hline                                                                   
\hline                                                                   
1 & 1989Oct27 &      &         &   3.70 & 35.6\% &       &      & 36.09 & 12\%\\
  & 1989Oct28 &      &         &   2.21 &        &       &      & 42.87 &     \\
\hline                                                                   
2 & 1998Jul25 & 5.71 & 28\%    &   5.98 &  23\%  & 16.0  & 1.2\%& 59.33 & 4.7\%\\
  & 1998Jul26 & 8.58 &         &   8.30 &        & 15.72 &      & 55.49 &       \\
\hline                                                                   
3 & 2000Apr24 & 2.40 & 17.7\%  &   3.07 &  23.8\%& 12.47 & 3.5\%& 45.96 & 5.3\%\\
  & 2000Apr25 & 3.09 &         &   4.31 &        & 11.87 &      & 49.51 &     \\
\hline                                                                   
4 & 2002Apr02 & 4.18 & 5.8\%   &        &        & 12.1  & 1.1\%&       &     \\
  & 2002Apr03 &      &         &   3.32 &  33.4\%&       &      & 41.83 & 4.4\%\\
  & 2002Apr05 & 4.54 &         &   5.38 &        & 12.29 &      & 44.51 &        \\
\hline                                                                   
5 & 2003Mar24 & 2.81 & 20\%    &        &        & 12.15 & 3.0\%&       &      \\
  & 2003Mar25 & 3.74 &         &   4.0  &  34.5\%& 11.65 &      & 42.01 & 1.3\%\\
  & 2003Mar26 &      &         &   2.43 &        &       &      & 41.23 &      \\
\hline                                                                   
6 & 2004Apr11 & 4.51 & 23\%    &   4.54 &  7.2\% & 11.4  & 1.4\%& 38.88 & 1.0\%\\
  & 2004Apr12 & 3.25 &         &        &        & 11.17 &      &       &    \\
  & 2004Apr13 &      &         &   5.03 &        &       &      & 38.36 &    \\
\hline                                                                   
7 & 2006Aug28 & 3.38 & 19.8\%  &   5.56 &  25.5\%& 12.01 & 3.3\%& 41.38 & 5.2\%\\
  & 2006Aug28 & 4.92 &         &   4.33 &        & 11.46 &      &       &   \\
  & 2006Aug29 & 4.65 &         &        &        & 12.67 &      & 45.02 &    \\
  & 2006Aug30 & 3.72 &         &   3.32 &        & 12.03 &      & 39.82 &    \\
  & 2006Aug30 & 3.73 &         &        &        & 12.35 &      &       &   \\
  & 2006Aug31 & 2.88 &         &   3.35 &        & 12.30 &      &  42.06&     \\
\hline                                                                   
8 & 2009May17 & 2.83 & 21.4\%  &  3.19  & 34\%  &  11.22 & 1.6\% & 37.95&  1.7\%\\
  & 2009May19 & 3.84 &         &  6.29  &      &  10.97  &     & 36.88  &   \\
  & 2009May20 &      &         &  6.29  &      &         &     & 36.86  &   \\
\hline
\end{tabular}
\tablefoot{-- Col.(1): Number of the possible flare. Col.(2):
Date. Cols.(3 and 5): Pure AGN blue (at 5100$\AA$) and red (at 6200$\AA$) continuum flux,
corrected for the host-galaxy contribution, in units $10^{-16} \rm erg \ cm^{-2} s^{-1} \AA^{-1}$. 
Col.(4 and 6): Flare amplitude in \%. Cols.(7 and 9): H$\beta$ and H$\alpha$ fluxes corrected 
for the host-galaxy contribution, in units $10^{-14} \rm erg \ cm^{-2} s^{-1}$. 
Cols.(8 and 10): The change in the line flux in \%. \\ }
\end{center}
\end{table*}

\begin{table*}
\begin{center}
\caption[]{Mean {  observed (obs), host-galaxy corrected (cor)
and narrow-lines subtracted (cor-line)} 
fluxes and standard deviations of H$\alpha$, H$\beta$ and continuum in 
different monitoring periods. The continuum flux is in units 
$10^{-16} \rm erg \ cm^{-2} s^{-1} \AA^{-1}$, and the line flux in
$10^{-14} \rm erg \ cm^{-2}s^{-1}$. }\label{tab6b}
\begin{tabular}{lcccccc}
\hline \hline
\multicolumn{2}{c}{UT-date} & JD period &  $F_{cnt}$(5100)$\pm \sigma$ & $F$(H$\beta$)$\pm \sigma$ & $F$(H$\alpha$)$\pm \sigma$ & $F_{cnt}$(6200)$\pm \sigma$\\
\multicolumn{2}{c}{1}  & 2        &   3        &  4         & 5             & 6         \\
\hline
1987 & obs &  46976  & 15.31\tablefootmark{*} & 17.99\tablefootmark{*} & 59.64$\pm$3.31(5.6\%)& 16.54$\pm$0.51(3.1\%)\\
     & cor &         &  4.01\tablefootmark{*} & 16.64\tablefootmark{*} & 57.91$\pm$3.32(5.7\%)&  4.44$\pm$0.51(11.5\%)\\
     & cor-line &    &  -                     & 11.18\tablefootmark{*} & 44.39$\pm$3.32(7.5\%)&  -\\     
\\
1988--1994 &  obs &  48413  &  14.34$\pm$0.89(6.2\%) & 13.35$\pm$0.93(6.9\%)& 42.87$\pm$4.33(10.1\%)& 15.69$\pm$1.27(8.1\%)\\
           &  cor &         &   3.04$\pm$0.89(29.2\%)& 12.00$\pm$0.93(7.7\%)& 41.63$\pm$4.53(10.9\%)&  3.59$\pm$1.27(35.4\%)\\
           & cor-line &     &  -                     &  7.31$\pm$2.07(28.3\%)& 28.11$\pm$4.53(16.1\%) &  -\\                  
\hline
1987/(1988--1994) &  obs &     &  1.07   & 1.35  & 1.39  & 1.05 \\
                  &  cor &     &  1.32   & 1.39  & 1.39  & 1.24 \\
                  &  cor-line& &  -      & 1.53  & 1.58  & -    \\
 \hline
1998  & obs& 50940-51021 & 18.74$\pm$1.29(6.9\%) & 17.26$\pm$0.91(5.3\%)& 61.52$\pm$3.73(6.1\%) & 18.86$\pm$1.07(5.7\%)\\
      & cor&             &  7.57$\pm$1.27(16.7\%)& 15.65$\pm$1.15(7.4\%)& 56.51$\pm$8.02(14.2\%)&  6.93$\pm$1.00(14.4\%) \\
      & cor-line &       &  -                    & 10.19$\pm$1.15(11.3\%)&42.99$\pm$8.02(18.7\%) &  -\\       
\\
1999--2010 &  obs& 51410-55367  &15.79$\pm$1.24(7.8\%) & 13.05$\pm$0.86(6.6\%)& 44.63$\pm$4.70(10.5\%)& 16.64$\pm$1.29(7.8\%)\\
           &  cor&              & 4.49$\pm$1.24(27.5\%)& 11.72$\pm$0.85(7.3\%)& 42.95$\pm$4.76(11.1\%)&  4.54$\pm$1.29(28.5\%) \\
           & cor-line &         &  -                   & 6.26$\pm$0.85(13.6\%)& 29.43$\pm$4.76(16.2\%) &  -\\             
\hline
1998/(1999--2010) &  obs&         & 1.19             & 1.32           &  1.38 & 1.13 \\
                  &  cor&         & 1.69             & 1.34           &  1.32 & 1.53 \\
                  &  cor-line&    & -                & 1.63           &  1.46 & -    \\
\hline
\end{tabular}
\tablefoot{-- Col.(1): Observed period. Col.(2): Julian date period
in units of 2400000+. Col.(3-6): Mean flux and standard deviations
of blue continuum, H$\beta$, H$\alpha$, and red continuum, the error
in percentages is given in the brackets. The middle and last row give the ratio of
the mean fluxes in year 1987 and period 1988--1994, and in year 1998 
and period 1999--2010, respectively.\\}
\end{center}
\tablefoottext{*}{  For year 1987 it was not possible to derive a standard deviation for the blue continuum
and H$\beta$ line flux.}
\end{table*}

\subsection{Variability of the emission lines and  continuum}

To estimate an amount of the variability in different line segments,
we used the method given by \cite{ob98} and defined  several
parameters characterizing the variability of the continuum, total
line, and line-segments fluxes (Table \ref{tab6}). There, N is the
number of spectra, $F$ denotes the mean flux over the whole
observing period and $\sigma(F)$ is the standard deviation, and
$R$(max/min) is the ratio of the maximal to minimal flux in the
monitoring period. The parameter $F$(var) is an inferred
(uncertainty-corrected) estimate of the variation amplitude with
respect to the mean flux, defined as:
$$ F({\rm var})= [\sqrt{\sigma(F)^2 -e^2}]/F({\rm mean}) $$
$e^2$ being the mean square value of the individual measurement
uncertainty for N observations, i.e. $e^2=\frac{1}{N}\sum_i^N
e(i)^2$ \citep{ob98}.  As it can be seen from Table \ref{tab6}, 
the indicator of variability $F$(var) is not high ({   $\sim$10-12\%} for
{  the observed} H$\alpha$, H$\beta$, {   $\sim$9\%} for the continuum at 5100\AA, and 
$\sim$7\% for the continuum at 6200\AA). The blue wing of H$\alpha$ and H$\beta$,
and H$\alpha$-red1 vary more ($\sim$20\%) than corresponding line cores
($\sim$11\%) and red wing of H$\beta$ ($\sim$13\%). 
{   But, the relative variation amplitude F(var) of the continuum fluxes changed more much ($\sim$30\%) when we 
removed the contribution of the host-galaxy (i.e. the corrected or agn-continuum), while F(var)
of the H$\alpha$ and H$\beta$ line fluxes remaine almost unchanged (see: host-galaxy corrected data in Table \ref{tab6}).  
In the corrected blue and red continuum light-curves there are some possible flare-like events with an amplitude of up to 
30\% lasting for a few (2-3) days (see Table \ref{flares}), while in the corrected H$\alpha$ and H$\beta$ 
line light-curves there are no observed flare-like events at the corresponding epochs.

Note here that the narrow line contamination of the H$\alpha$ and H$\beta$ broad lines and their 
line-segments can affect the variation amplitude F(var). This contaminations may cause the measured small variation 
in the H$\alpha$ and H$\beta$ core and in the H$\beta$ red wing. 
We corrected the H$\alpha$ and H$\beta$ line fluxes and their segments for the narrow line contribution 
(Table \ref{tab6}, narrow line subtracted data) and obtained that the indicator of variability  F(var) 
is $\sim$20\% in the corrected H$\alpha$ and H$\beta$ line and line-segment fluxes.}

{  Also we should note} that the part of the H$\alpha$ red wings (H$\alpha$-red2 in
Tables \ref{tab5}-\ref{tab6}) is very weak and its contribution to the H$\alpha$
flux is negligible. We use H$\alpha$-red2 flux only for investigation of
variations in the red to blue line-segment flux ratio (see \S 3.2.1).

\begin{figure}
\centering
\includegraphics[width=9cm]{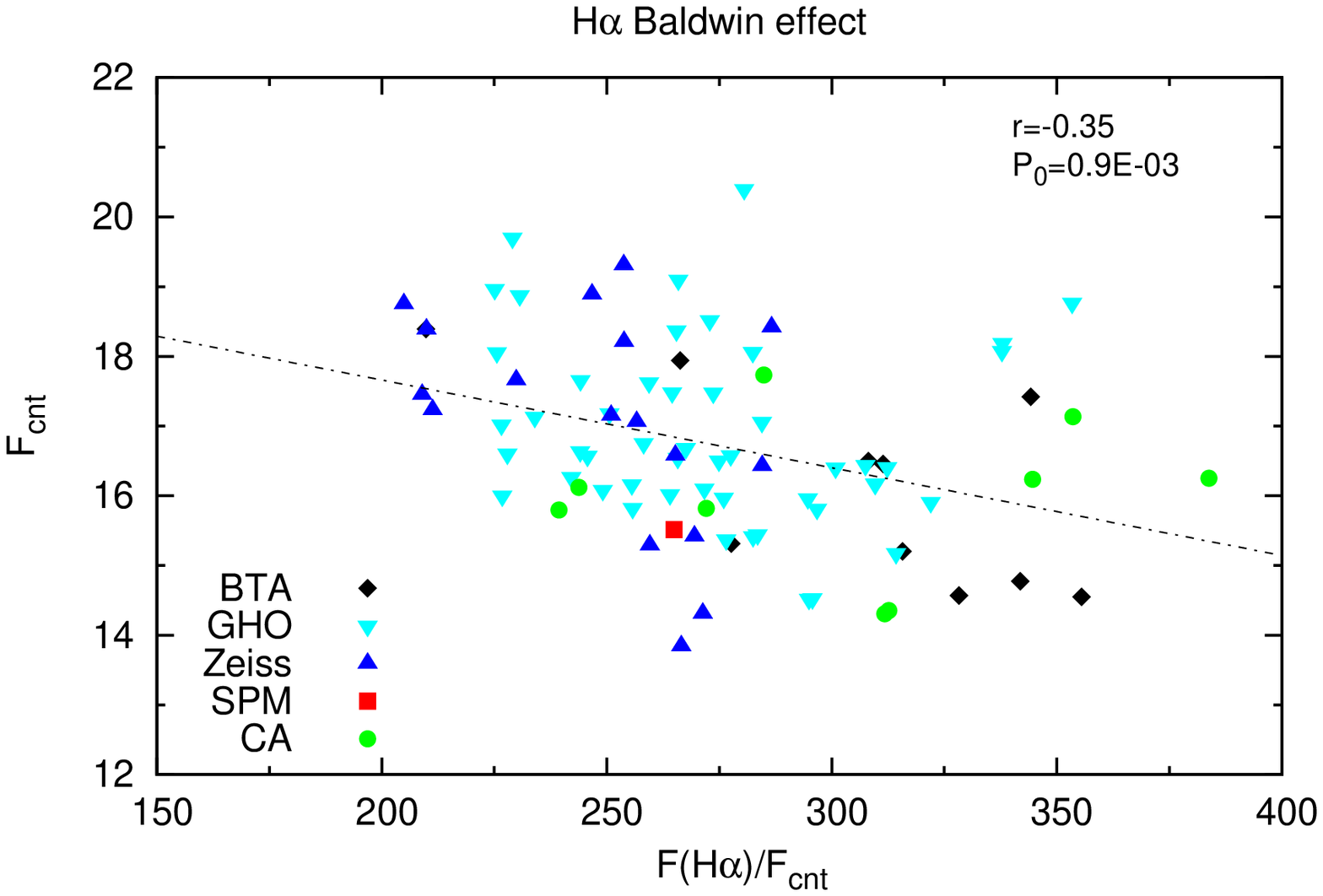}
\includegraphics[width=9cm]{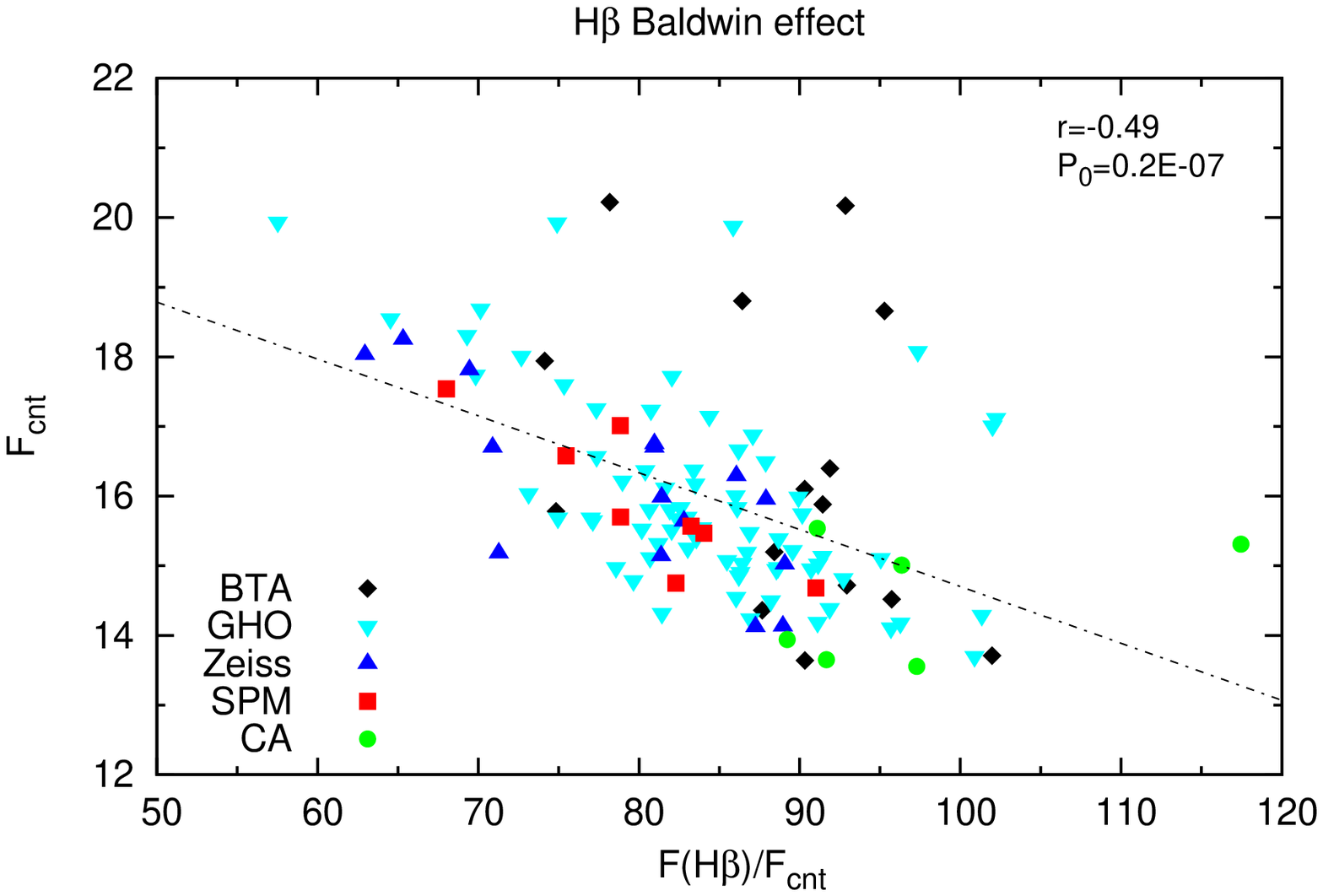}
\caption{Baldwin effect in the H$\alpha$ (upper) and H$\beta$ line
(bottom). The continuum flux is in units $10^{-16} \rm erg \ cm^{-2}
s^{-1} \AA^{-1}$. Observations with different telescopes are denoted
with different symbols given in the bottom left. The correlation
coefficient and the corresponding p-value are also given.}
\label{BE}
\end{figure}

In Figs. \ref{Hb_cnt} and \ref{Ha_cnt} we plot the H$\beta$ and
H$\alpha$ line fluxes as a function of the continuum. Since the
observations with Zeiss should be taken with caution, we present the
line {\rm vs.} continuum  flux with and without data obtained with
Zeiss (see Tables 1 and 2, code Z2K). As it can be seen from
figures, there is a relatively week correlation between the line and
continuum fluxes, r=0.42 for H$\beta$ and 0.42 for H$\alpha$
(without Zeiss data). Such small correlation between the line and
the continuum flux also indicates that beside the continuum central
source there may be present other effects in photoionization.

On the other hand in Fig. \ref{BE} we plot the intrinsic Baldwin
effect, and as it can be seen there is an anticorrelation between
the continuum flux and equivalent widths of the H$\beta$ ($r=-0.49,
P_0=0.6\cdot10^{-07}$) and H$\alpha$ lines ($r=-0.35,
P_0=0.9\cdot10^{-03}$), i.e. that there is the intrinsic Baldwin in
the  H$\beta$ and  H$\alpha$ lines, similar as it is observed in
another (single peaked) AGNs \citep[see e.g.][]{gp03}.

\begin{figure*}
\centering
\includegraphics[width=8cm]{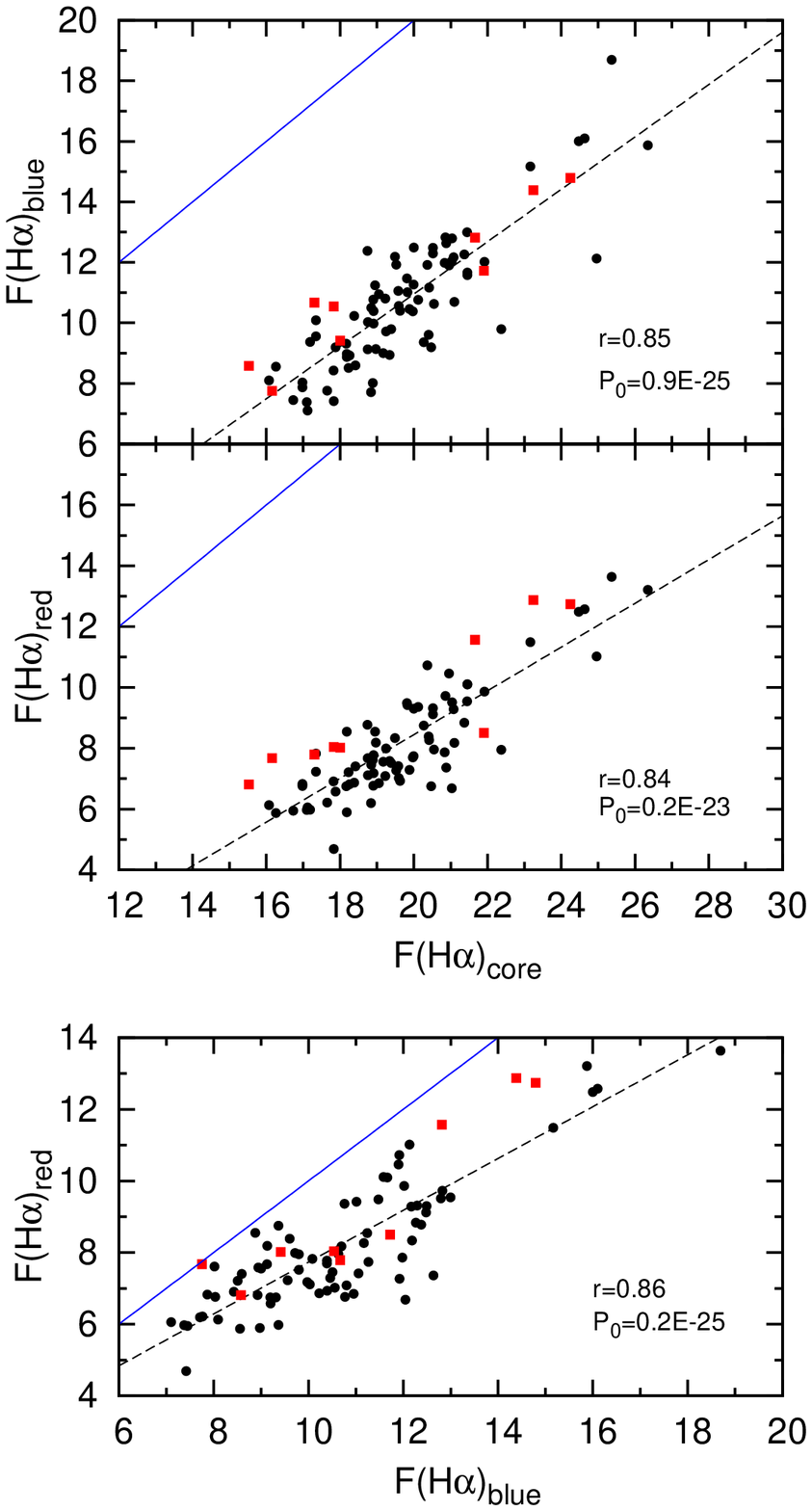}
\includegraphics[width=8cm]{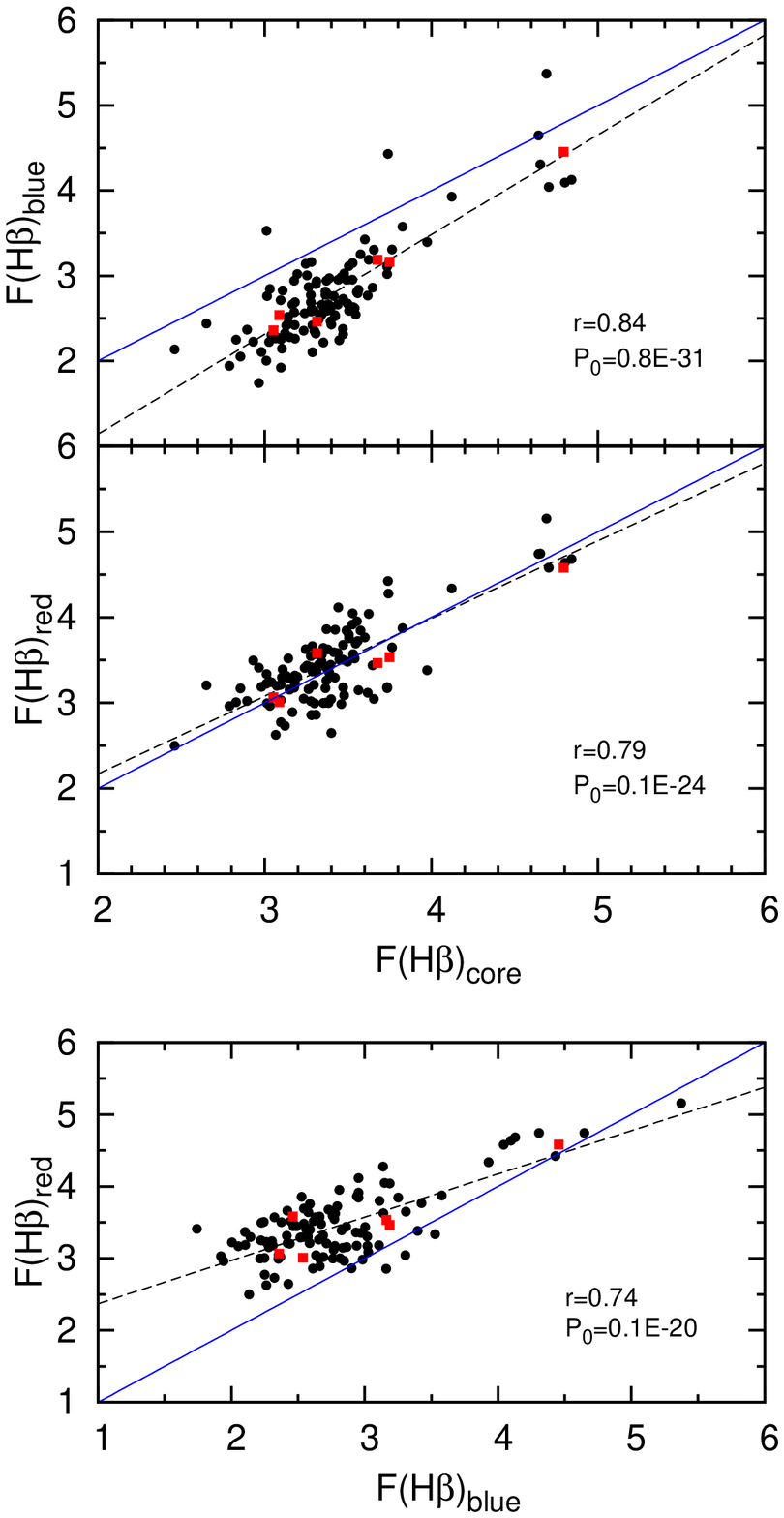}
\caption{H$\alpha$ and H$\beta$ line-wing fluxes (blue, red) vs.
line-core flux (upper panels), and red vs. blue-wing (bottom panel).
The line-segment fluxes are in units $10^{-14} \rm erg \
cm^{-2}s^{-1}$. The correlation coefficient and the corresponding
p-value are given in the bottom right corner. The CA-data are denoted with squares.
{  The dashed line gives the linear best-fitting of the data, while the solid line 
markes the linear function with the slope equals to 1.}}
\label{wings}
\end{figure*}

We defined the observed fluxes in the blue and red wings, and in the
core of the H$\alpha$ and H$\beta$ lines {  (Tables \ref{tab_seg_ha}-\ref{tab_seg_hb})} 
in the  wavelength intervals as they are given in Table \ref{tab5}. In 
Fig. \ref{wings} we plot the H$\alpha$ and H$\beta$ line-wing fluxes (blue, red) vs.
line-core flux (upper panels), and red vs. blue-wing (bottom panel).
{   The dashed line in Fig. \ref{wings} represents the best fit, while the
solid line represents the expected slope in the case that the different parts 
of the line profile vary proportionally to each other, i.e. the slope of the 
best fit is 1.}

As it can be seen  in {  Fig. \ref{wings}} the correlations in variation between the blue/red
wings and central component are high, as well as between the red and
blue wings ($r\sim0.8$). {  However, the slopes of the best-fit lines are not consistent with 1, except
in the case of  H$\beta$ red wing vs. core, where the best fit is very close to 1.}

\begin{figure*}
\centering
\includegraphics[width=8cm]{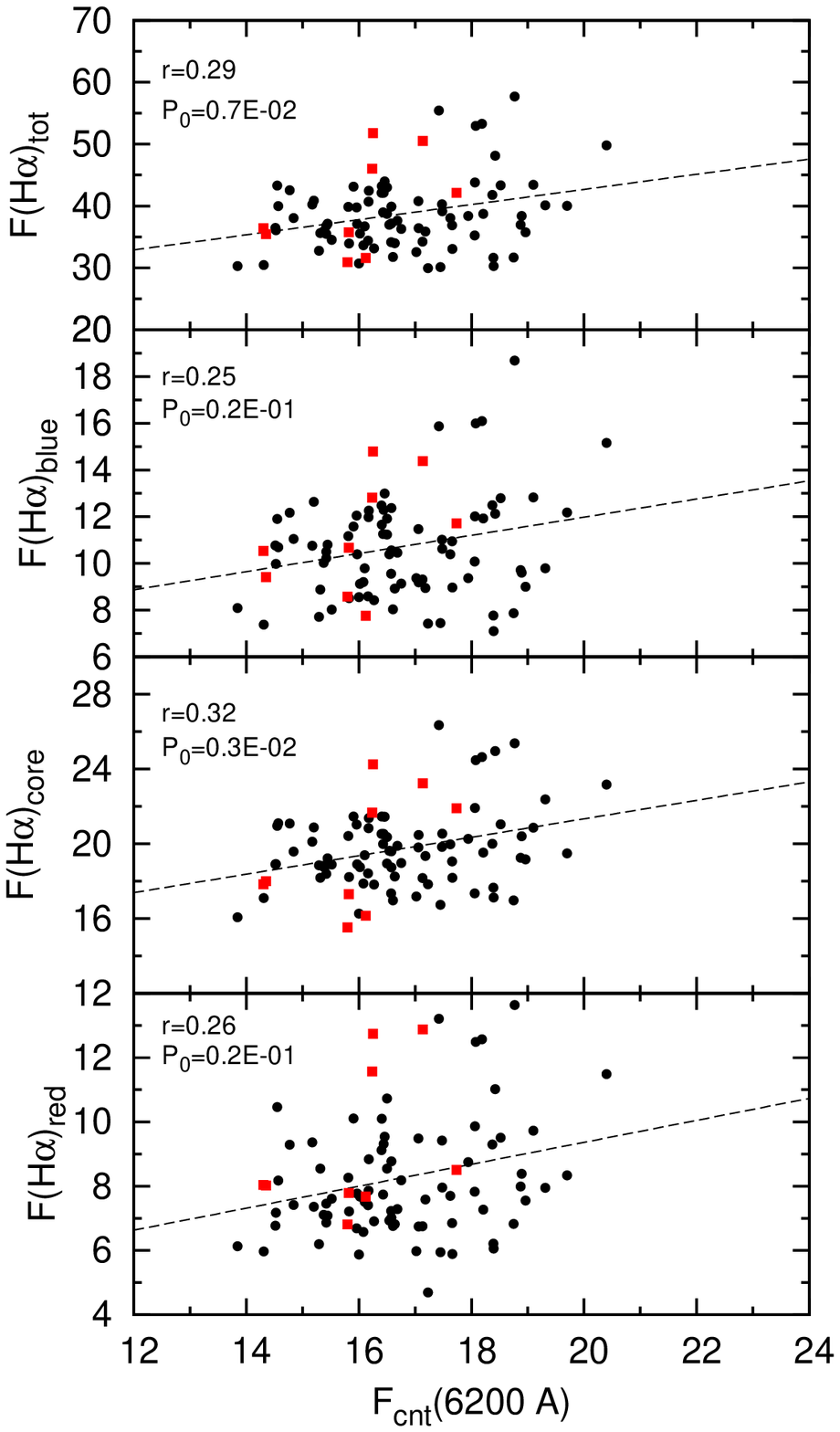}
\includegraphics[width=8cm]{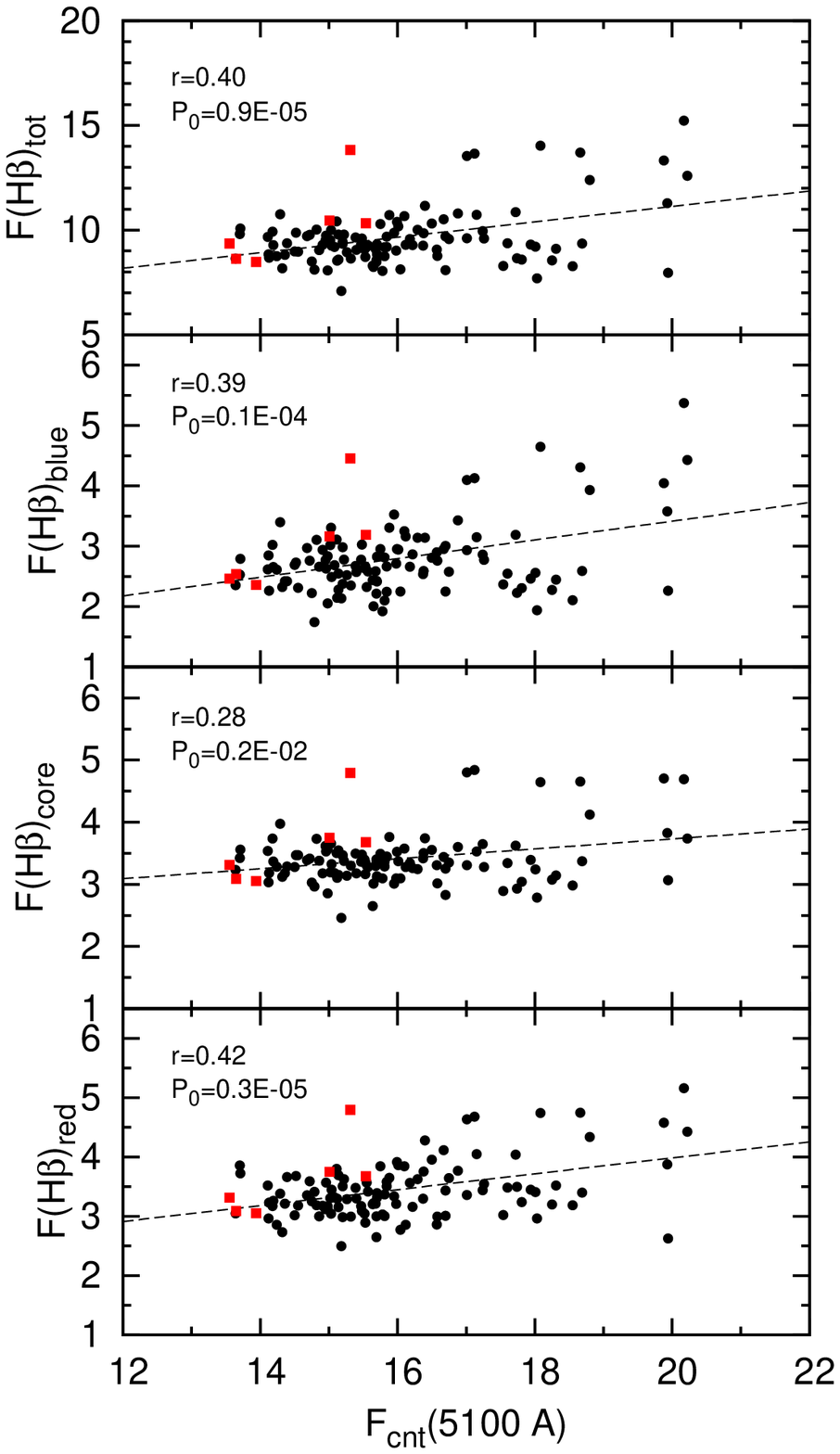}
\caption{H$\alpha$ and H$\beta$ line and line-segment fluxes (blue,
core and red) vs. continuum flux at 6300 and 5100, respectively. The
continuum flux is in units $10^{-16} \rm erg \ cm^{-2} s^{-1}
A^{-1}$, and the line-segment fluxes are in $10^{-14} \rm erg \
cm^{-2}s^{-1}$.The correlation coefficient and the corresponding
p-value are given in the upper left corner. The CA-data are denoted with squares.}
\label{wings_c}
\end{figure*}

\begin{figure*}
\centering
\includegraphics[width=8cm]{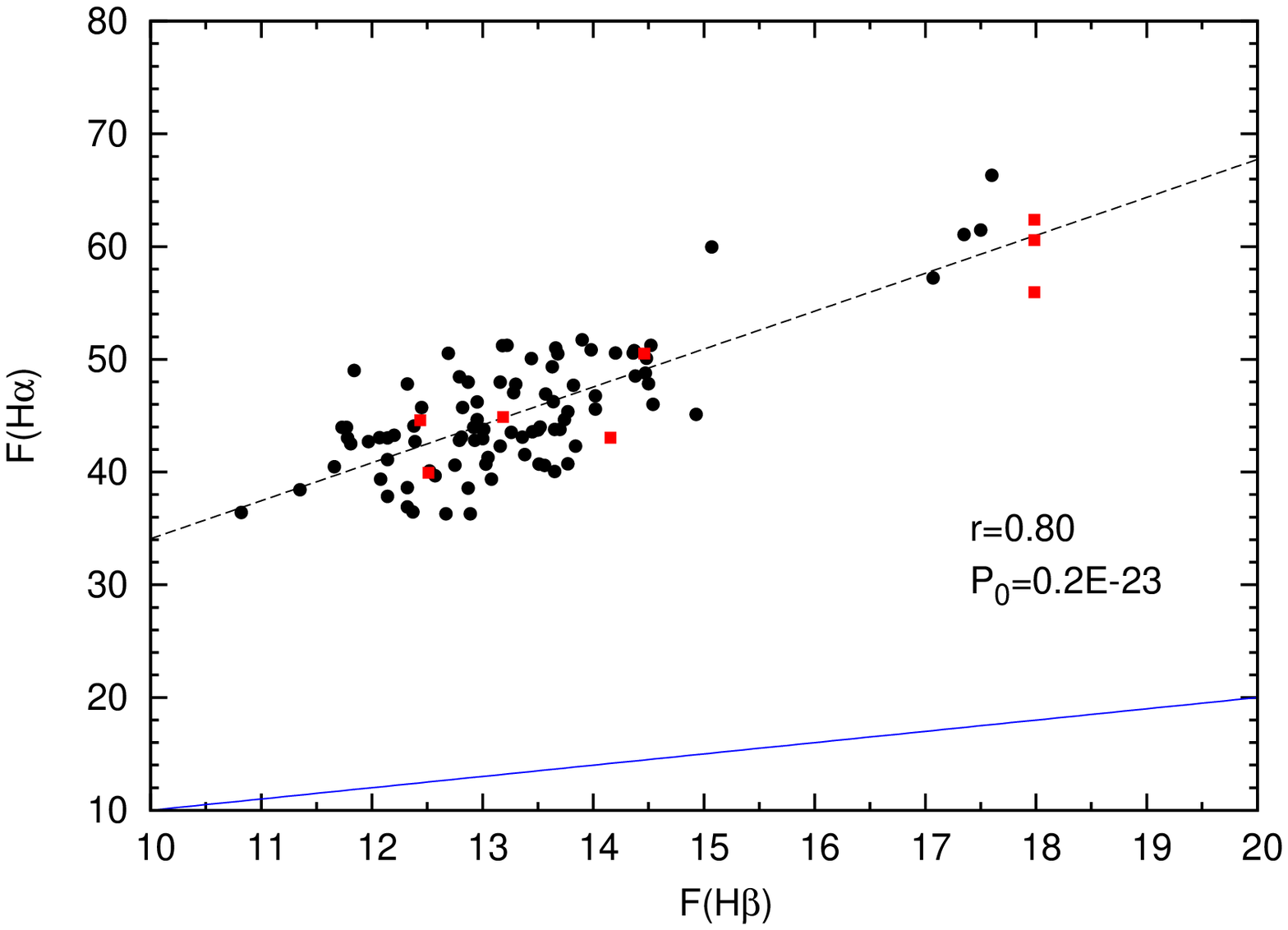}
\includegraphics[width=8cm]{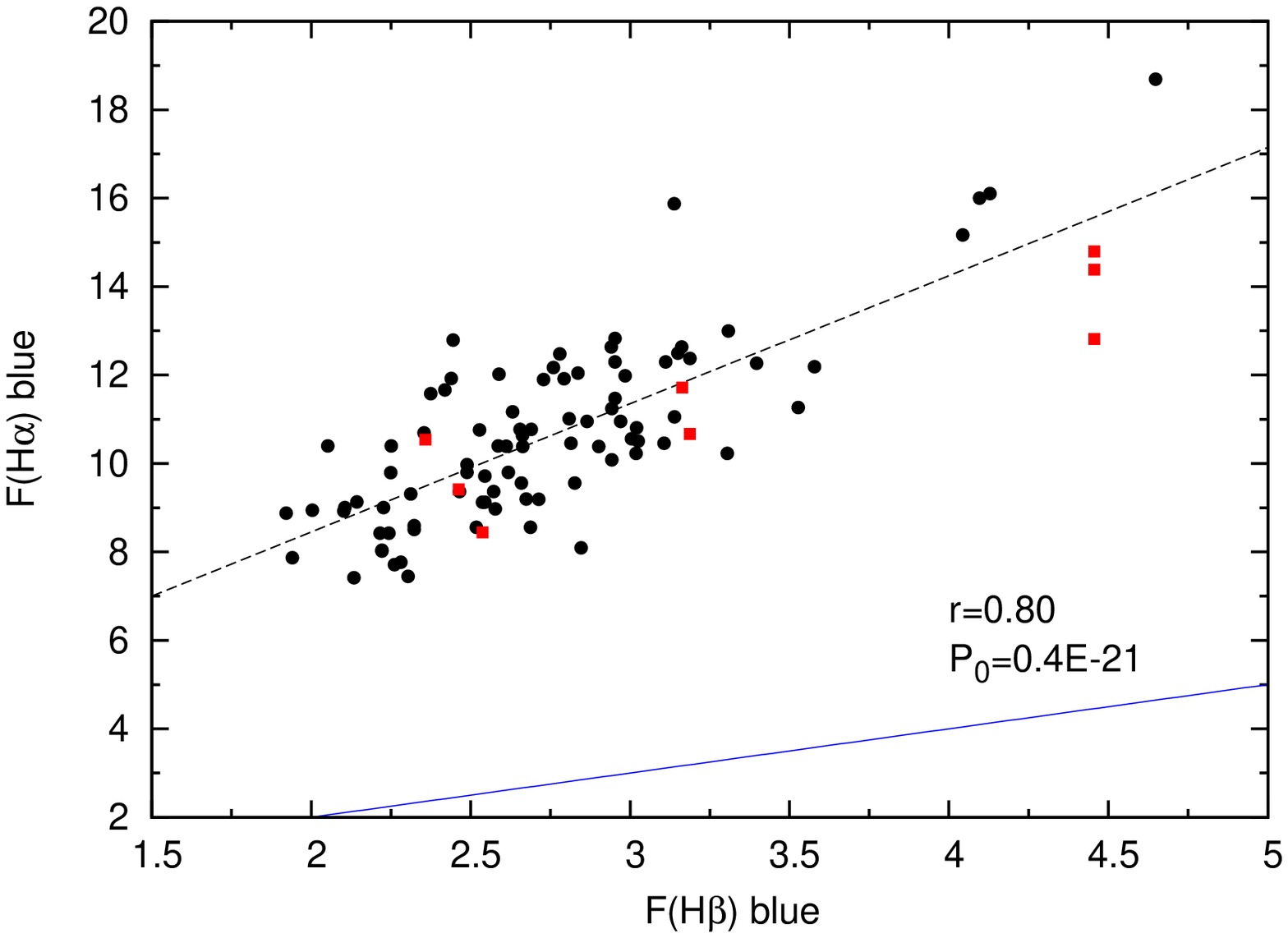}
\includegraphics[width=8cm]{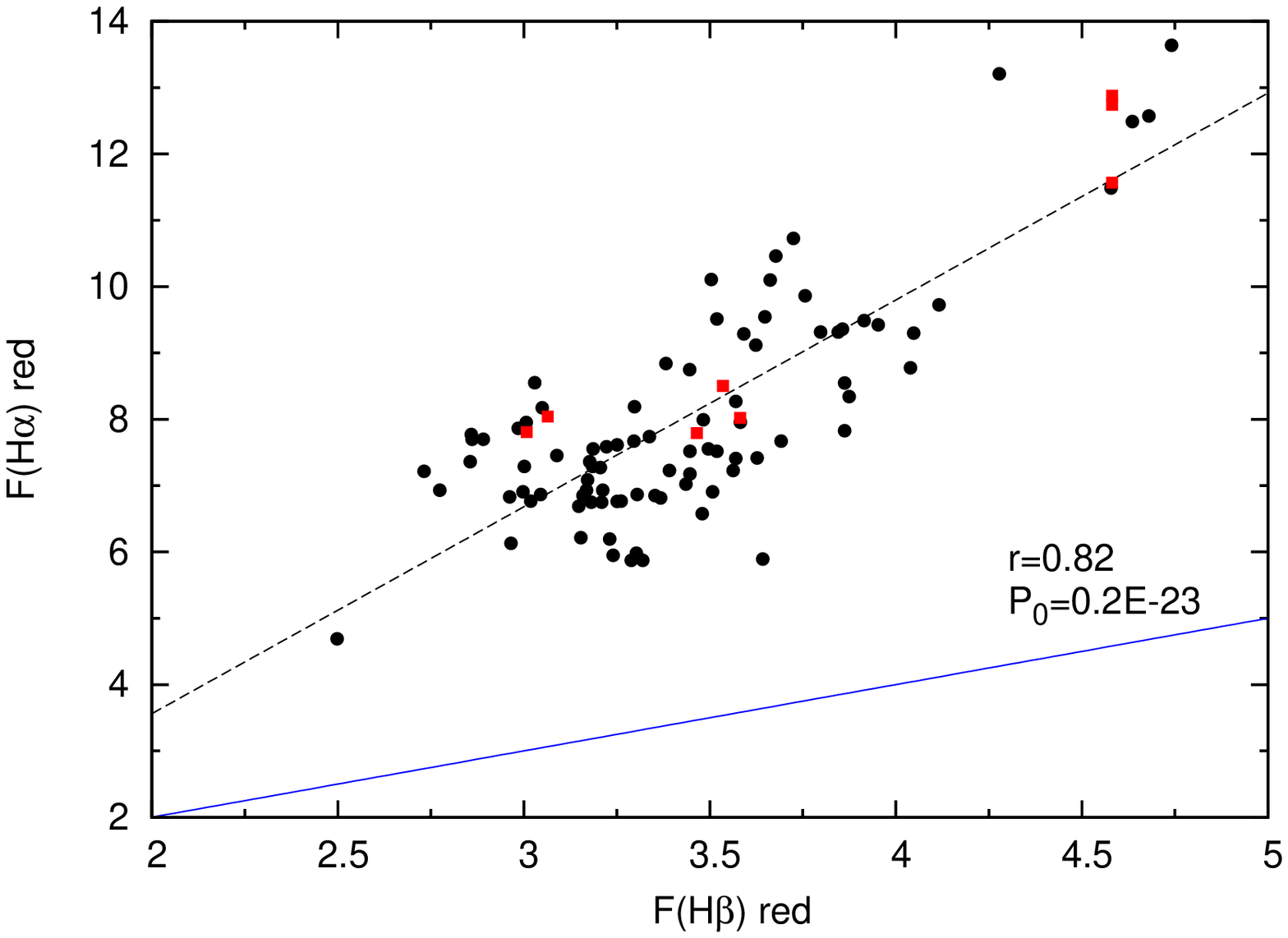}
\includegraphics[width=8cm]{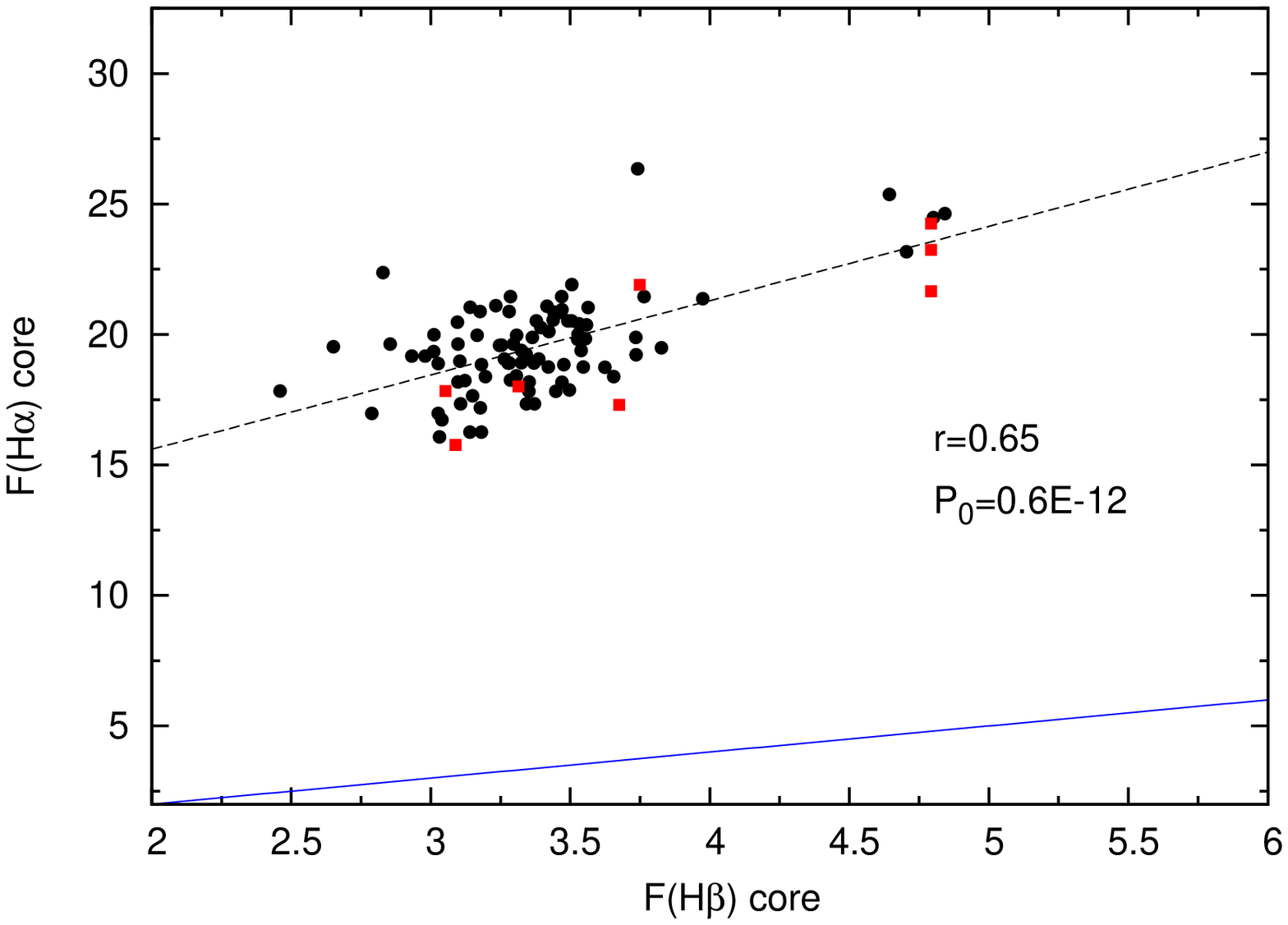}
\caption{H$\alpha$ vs. H$\beta$ total line and line segment fluxes (blue, red, core).
The correlation coefficient and the corresponding p-value are given in the bottom right corner.
The CA-data are denoted with squares. {  The dashed line gives the linear best-fitting of the data, while the solid line 
markes the linear function with the slope equals to 1.}} \label{hahb}
\end{figure*}

{  Although the correlations between the H$\alpha$
vs. H$\beta$ total line and line segment fluxes  (Fig. \ref{hahb})
are very good ($r\sim0.8$), the best fits are far away from the slope of 1. Moreover,
 in the case of the line cores,
F(H$\alpha)$core  vs. F(H$\beta$)core,  the correlation
coefficient is smaller ($r\sim0.65$).} On the other hand,
the correlation between the line segment and continuum flux is very
low, i.e. almost absent and statistically not important in the case
of H$\alpha$ (see Fig. \ref{wings_c}). The situation is slightly
better in the case of the H$\beta$ line.

\begin{table*}
\begin{center}
\caption[]{Sampling characteristics and cross-correlation analysis of H$\beta$.}\label{ccf}
\begin{tabular}{lccc}
\hline \hline
Light curve  & N  & Lag ZDCF & ZDCF   \\
\hline
cnt vs H${\alpha}$                       & 79  & $14.94_{-13.81}^{15.66}$ & $0.19_{-0.14}^{0.14}$ \\
cnt vs H${\alpha}$ (bad Zeiss discarded) & 60  & $16.29_{-14.34}^{14.30}$ & $0.28_{-0.16}^{0.15}$ \\
cnt vs H${\beta}$                        & 110 & $20.61_{-18.71}^{54.33}$ & $0.31_{-0.09}^{0.09}$ \\
cnt vs H${\beta}$ (only Mexico points)   & 80  & $10.75_{-9.76}^{19.31} $ & $0.15_{-0.13}^{0.13}$ \\
cnt vs H${\beta}$ (without Zeiss)        & 95  & $16.85_{-14.86}^{20.13}$ & $0.34_{-0.11}^{0.10}$ \\
\hline
\end{tabular}
\tablefoot{-- Col.(1): Analyzed light curves. Col.(2): Number of used spectra.
Col.(3): Lag calculated using ZDCF method. Col.(4): Cross correlation coefficient
calculated using ZDCF.\\}
\end{center}
\end{table*}


\subsubsection{CCF analysis}

As it can be seen, the light curves shown in Fig. \ref{lc} are
complex, with a number of peaks, and  the observed fluxes show  only
modest indications for variations, which is indicated by F(var)
parameter in Table \ref{tab6}. In spite of the small correlation
between the line and continuum fluxes, we apply on our data-samples
the Z-transformed DCF method, called ZDCF \citep[see][]{Al13}. We
did several calculations, using different number of data, first of
all, we discarded bad Zeiss spectra, also we calculated CCF using
only spectra observed with two telescopes from Mexico. The results
of the cross-correlation analysis are given in Table \ref{ccf}.

It could be seen that the continuum and both line emission light
curves H${\alpha}$ and H${\beta}$ (without bad Zeiss spectra) varies
similarly.  As it can be seen from Table \ref{ccf}, the lags for
H$\beta$, for different number of points, are between 20 and 29
days, and for H$\alpha$ between 17 and 29 days. The cross
correlations are small, but the ZDCF coefficient is larger in the
case of H${\beta}$.

\begin{figure*}
\centering
\includegraphics[width=8cm, angle=90]{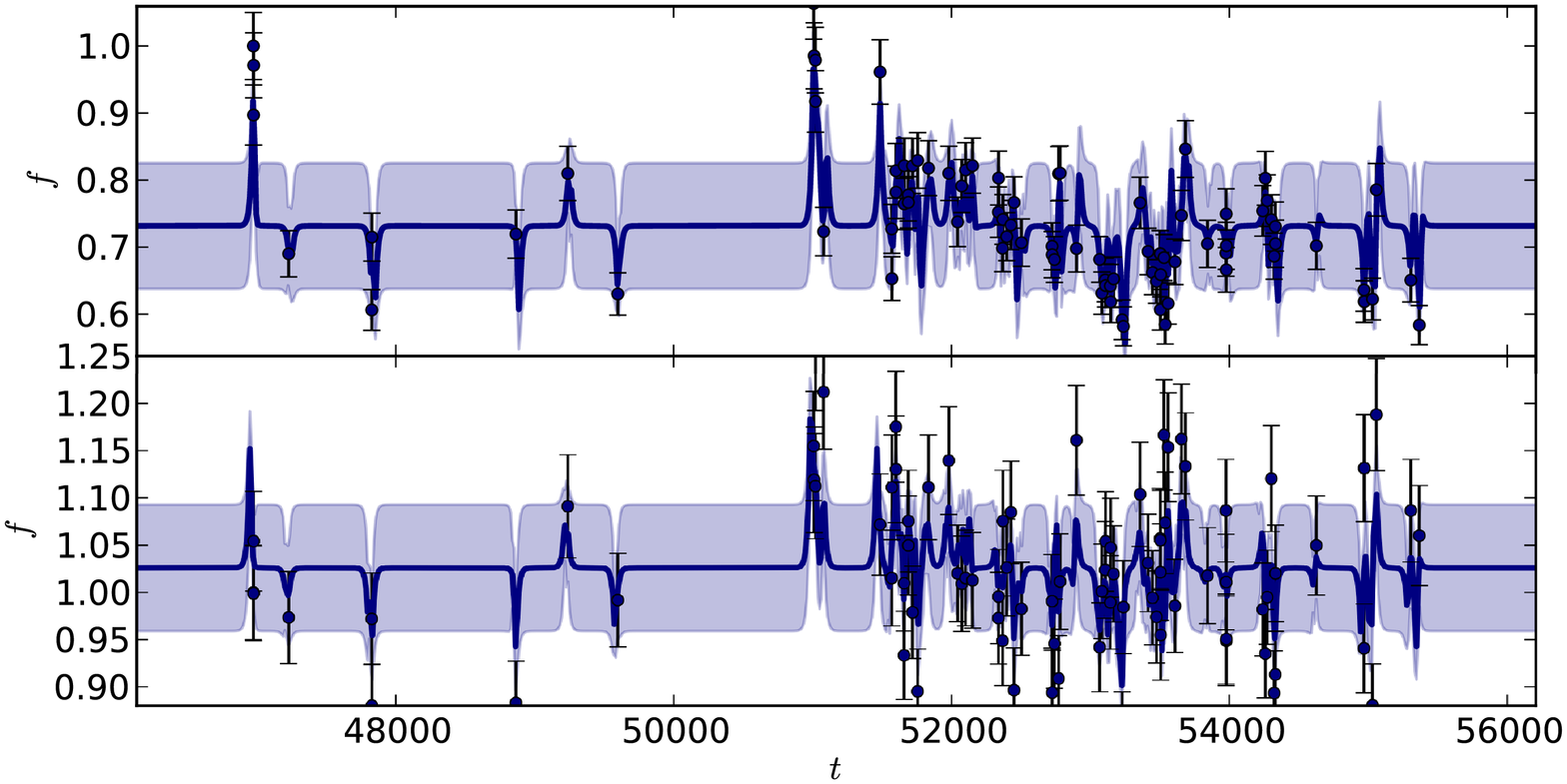}
\includegraphics[width=8cm, angle=90]{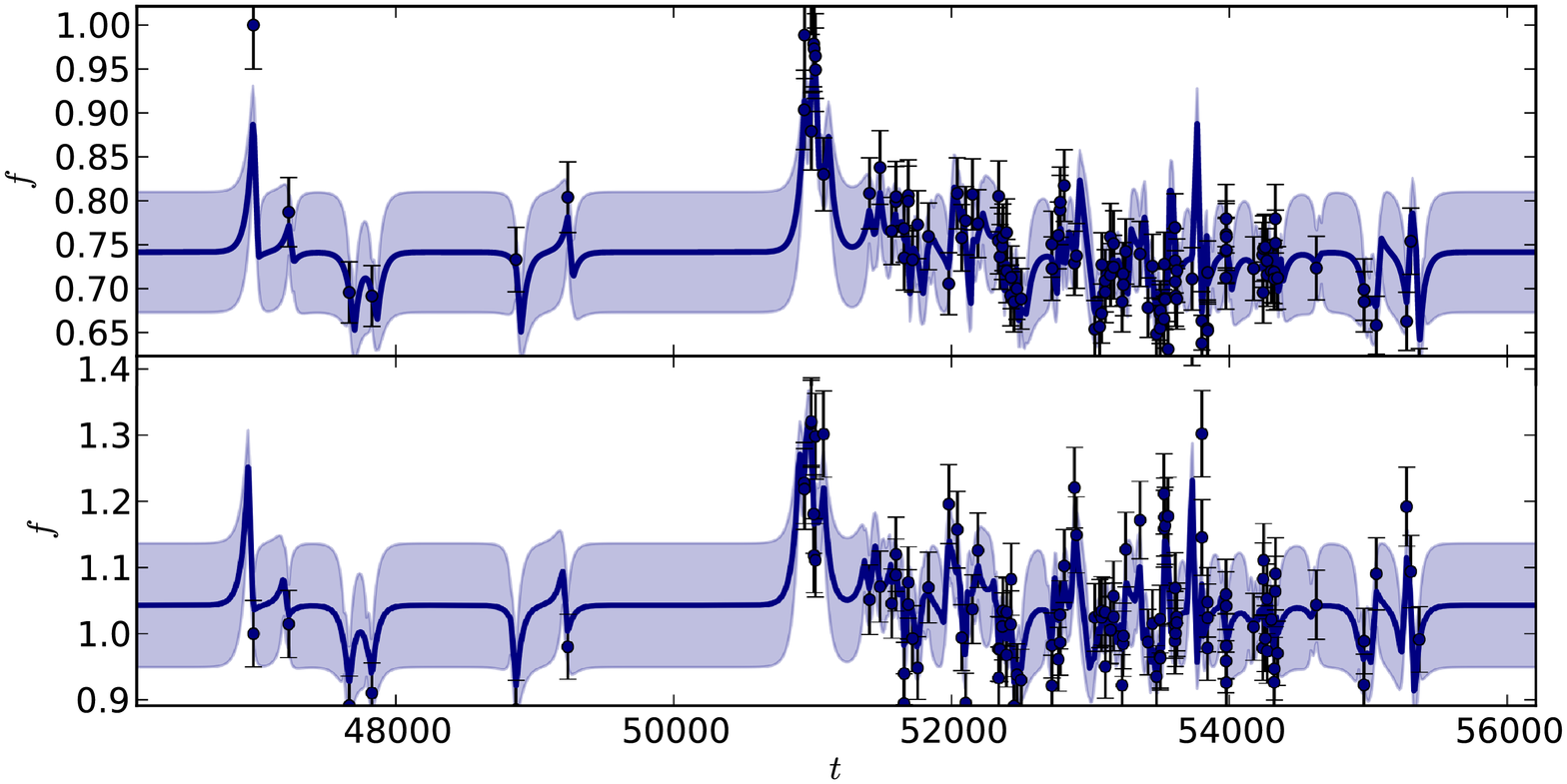}
\caption{Models of H{$\alpha$} (up) and H$\beta$ line (down), where
the upper panel gives the line and the bottom panel continuum light
curve. On all plots full line is the model of the expected mean light
curve, while dots represent the observed data. The blue band shows the
expected spread of light curves around the mean consistent with the data.
The x-axis gives modified Julian Date (MJD) while the y-axis gives
normalized line/continuum fluxes.}
\label{javelincnthb}
\end{figure*}

Additionally,  we apply the method of \cite{zu11} for the lag
estimation of the H$\beta$ line (Fig. \ref{javelincnthb}). We also
applied this method on the H$\alpha$ line, however, it did not
produce valid results. In these calculations, the first step is to
build a continuum model to determine the Zu model parameters of the
continuum light curve. The continuum light curve is generated from
the  model with the time scale of 100 days and variability amplitude
of $\sigma=2$. The posterior distribution of the two  parameters of
the continuum variability $(\tau_d, \sigma )$ are calculated from Zu
model using 40000 MCMC (Markov Chain Monte Carlo) method
burn-in iterations. In order to measure the
lag between the continuum and the H$\beta$ light curve, Zu model
then interpolates the continuum light curve based on the posteriors
$(\tau_d, \sigma )$ derived, and then shifts, smooths, and scales
continuum light curve to compare to the observed H$\beta$ light
curve. After doing this 10000 times in an MCMC run, we derived the
posterior distribution of the lag, the tophat width w, and the scale
factor s of the emission line, along with updated posteriors for the
timescale $\tau_d$ and the amplitude $\sigma$ of the continuum. The
model gives for lag between continuum and H$\beta$ line
$36.95_{19.72}^{47.53}$ days and for continuum and H{$\alpha$} line
the lag is $22.86_{16.84}^{30.20}$ days   which is approximately
within 3$\sigma$ distance from lag values between continuum and
emission lines obtained by classical methods given in Table
\ref{ccf}. In order to know what the best fitting parameters from
the last MCMC run look like, we give in Fig. \ref{javelincnthb}  a
comparison of the best-fitting light curves and the observed ones.
It could be seen that the observation between MJD 46975 and 50991
are closer to the best-fitting light curve in the case of
H{$\alpha$} than in the case of the H{$\beta$} line.

\begin{figure*}
\centering
\includegraphics[width=9cm]{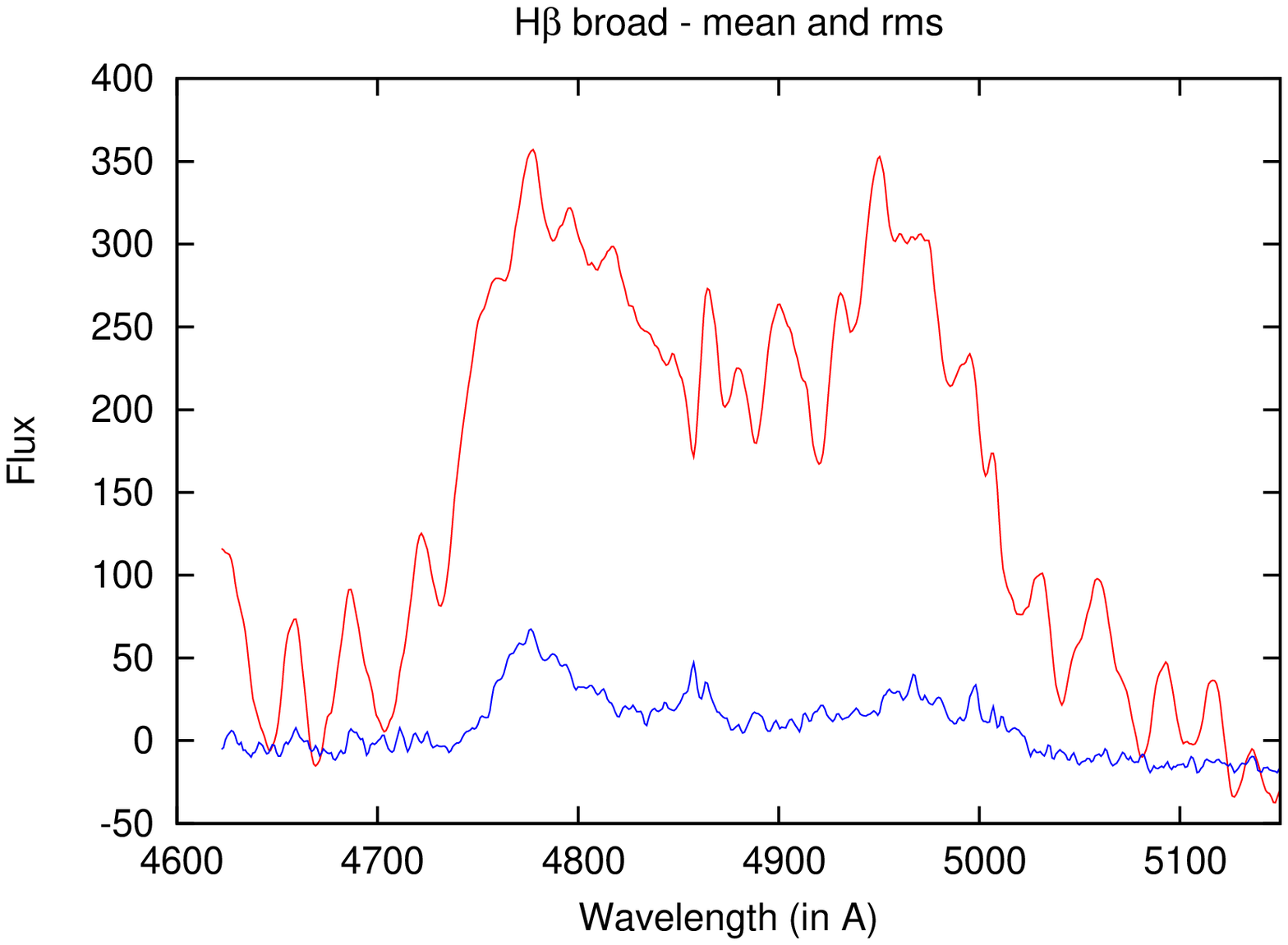}
\includegraphics[width=9cm]{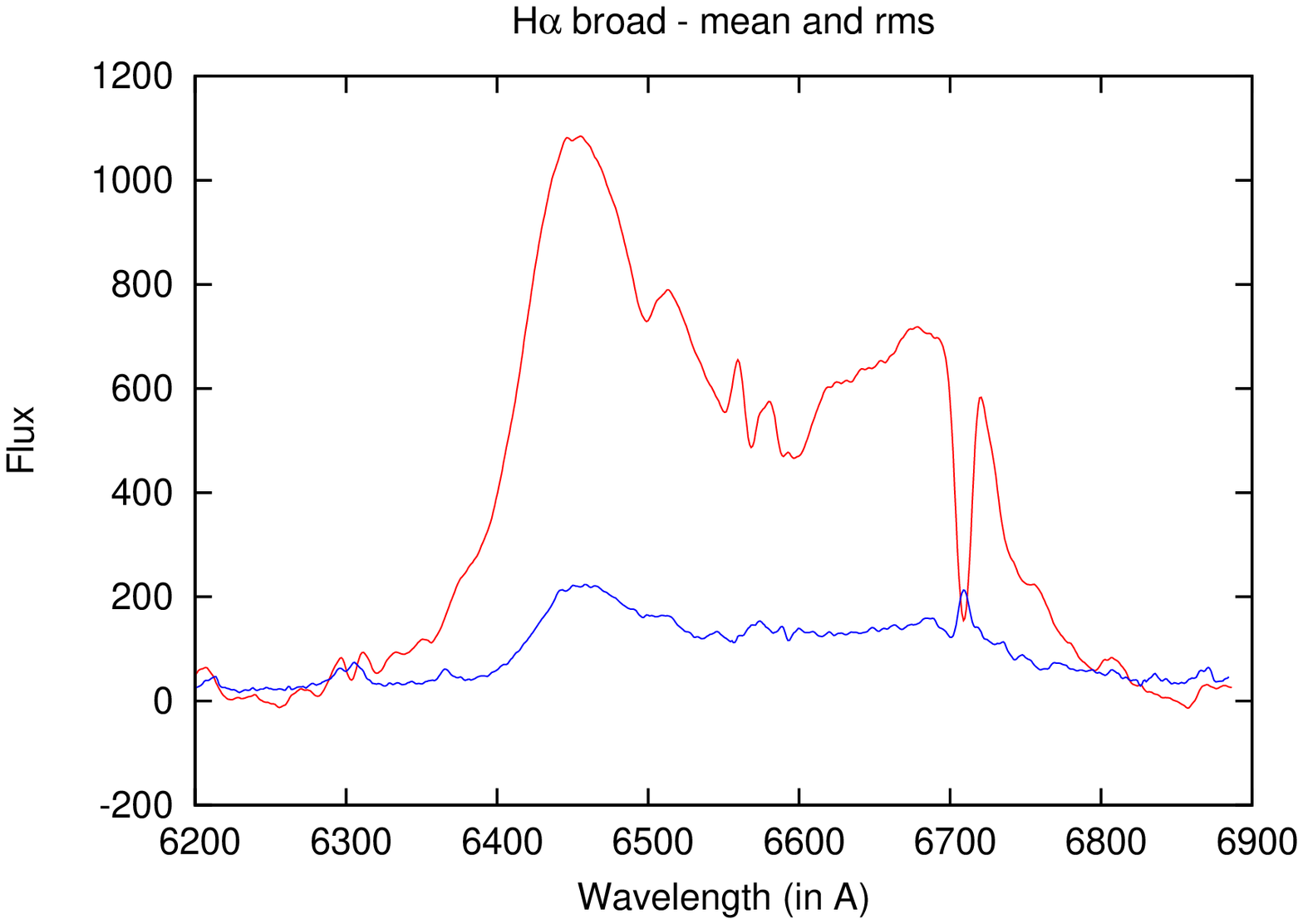}
\includegraphics[width=9cm]{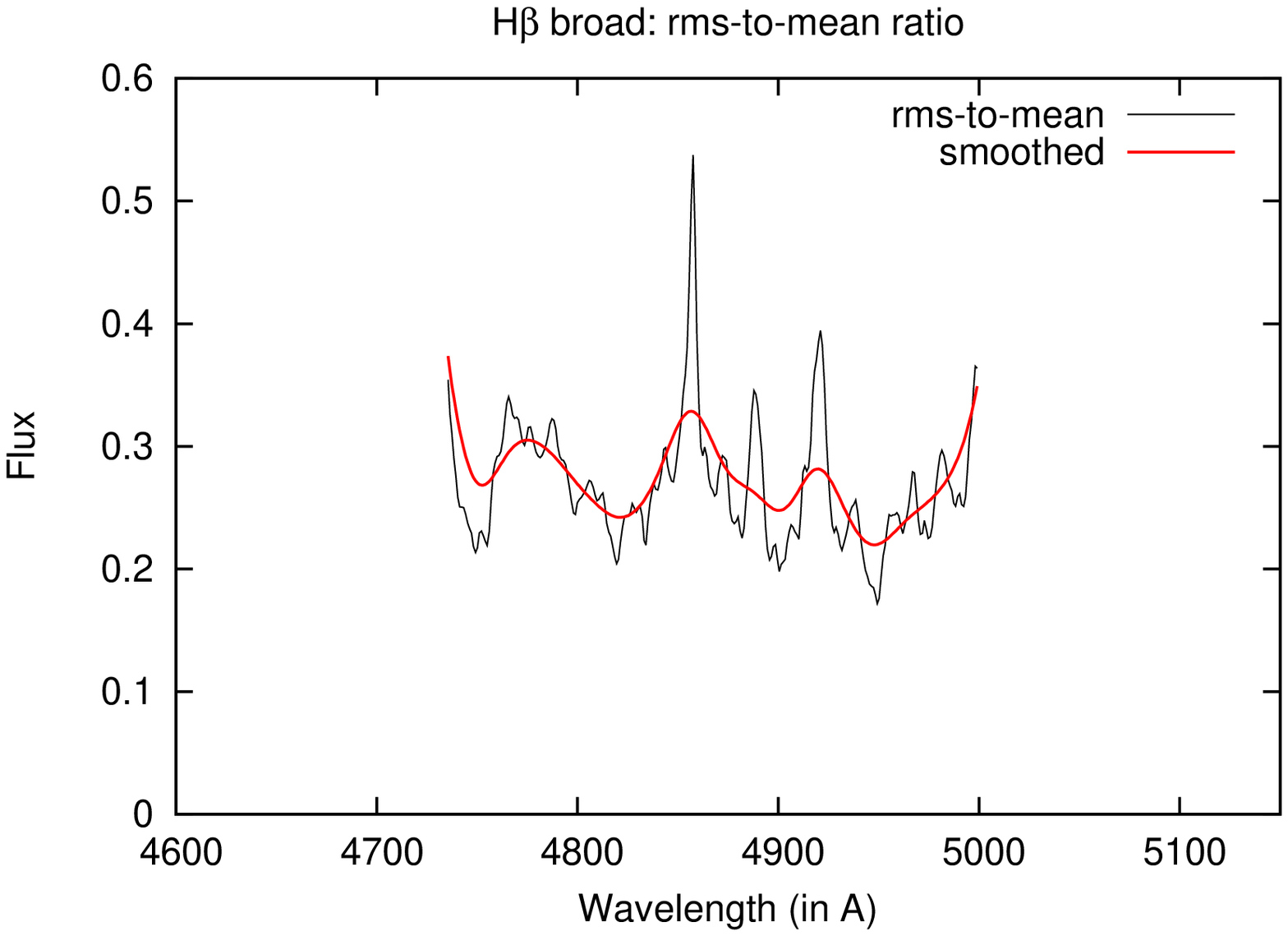}
\includegraphics[width=9cm]{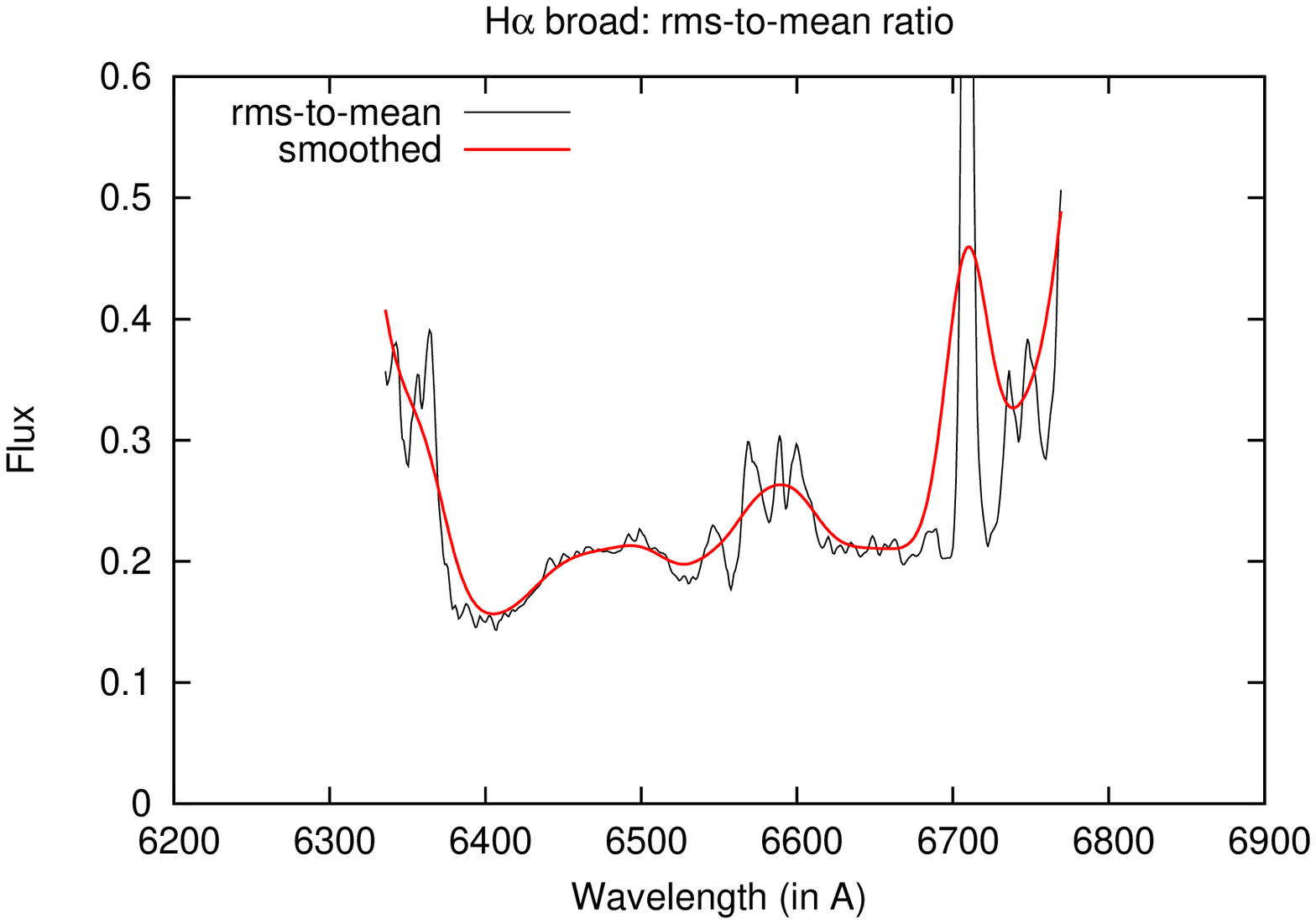}
\caption{{  Top panels give the mean and rms profiles, while the bottom panels give the ratio of the rms-to-mean flux} 
of the H$\beta$ (left) and H$\alpha$ (right) lines.} \label{mp}
\end{figure*}

Finally, we applied the interpolation cross-correlation function
method (ICCF) method \citep[][]{bk99} to cross-correlate the flux of
the continuum with H$\beta,\alpha$ flux. Also in this case, the
error-bars in lags were large and there is an indication for a lag
of 20 days, that is in agreement with previous two methods.
Therefore, for the black hole mass estimation we will accept a time
lag for H$\beta$ of 20 days.

\subsection{Changes in the broad line profiles}

We are going to discuss and model the line shapes variability in
Paper II, here we will give some characteristics of the line profile
and peaks variations, since Arp 102B is a prototype of double peaked
emitters.

During the monitoring period, the broad H$\beta$ and H$\alpha$ lines
have double-peaked profiles. In Fig. \ref{mp} we present the mean
profile accross the line profile 
H$\beta$ and H$\alpha$ profiles and their rms profile {  (top panels) and 
normalized rms on the mean profile across the line profile (bottom panels).} The FWHM of
the mean and rms profiles are: H$\alpha$ mean 14,320 km s$^{-1}$ and
rms 14,450 km s$^{-1}$, and H$\beta$ mean 15,900 km s$^{-1}$ (15,840
km s$^{-1}$) and rms 14,870 (or 16,080 if not corrected for the
underlying continuum). The distance between the two peaks is around
11,000 km s$^{-1}$, the blue peak is located around -5,000 km
s$^{-1}$ and red around 6,000 km s$^{-1}$ from the line center. Such
big distances between the peaks indicate a fast rotating disk, that
is probably close to the black hole.
As it can be seen from Fig. \ref{mp}, the changes in the line
profile have also two peaked rms, that indicates that the changes in
the broad line profile are in both red and blue peak in both lines,
but changes in the blue wing are significantly bigger, than in the
red one. Note here, that there is one, central, peak in the rms,
that may be caused by a central component \citep[see
e.g.][]{pop04,bo06,bo09}.

\subsubsection{Red to blue peak ratio}

As it was reported in \cite{n97} and \cite{s00} there is the
variation in the red-to-blue flux ratio of H$\alpha$ that has a periodical
characteristic. The observed wavelength intervals of H$\alpha$ are defined in
such a way that the red wing excludes the narrow forbidden lines of
[N II]$\lambda$6584 and [S II] $\lambda\lambda$6717,6731 (H$\alpha$
- red1 and H$\alpha$ - red2 from Table \ref{tab5}). The blue and red
wavelength intervals (see Table \ref{tab5}) correspond to intervals
from \cite{n97} in case of H$\alpha$. We measure the line-segment flux ratio for H$\alpha$ 
and H$\beta$ (Tables \ref{tab_seg_ha} -- \ref{tab_seg_hb}, available electronically only) 
and apply the so called Lomb-Scargle periodogram \citep[][]{l76,sc82} 
to find possible periodical variations in this ratio.

\begin{figure*}
\centering
\includegraphics[width=9cm]{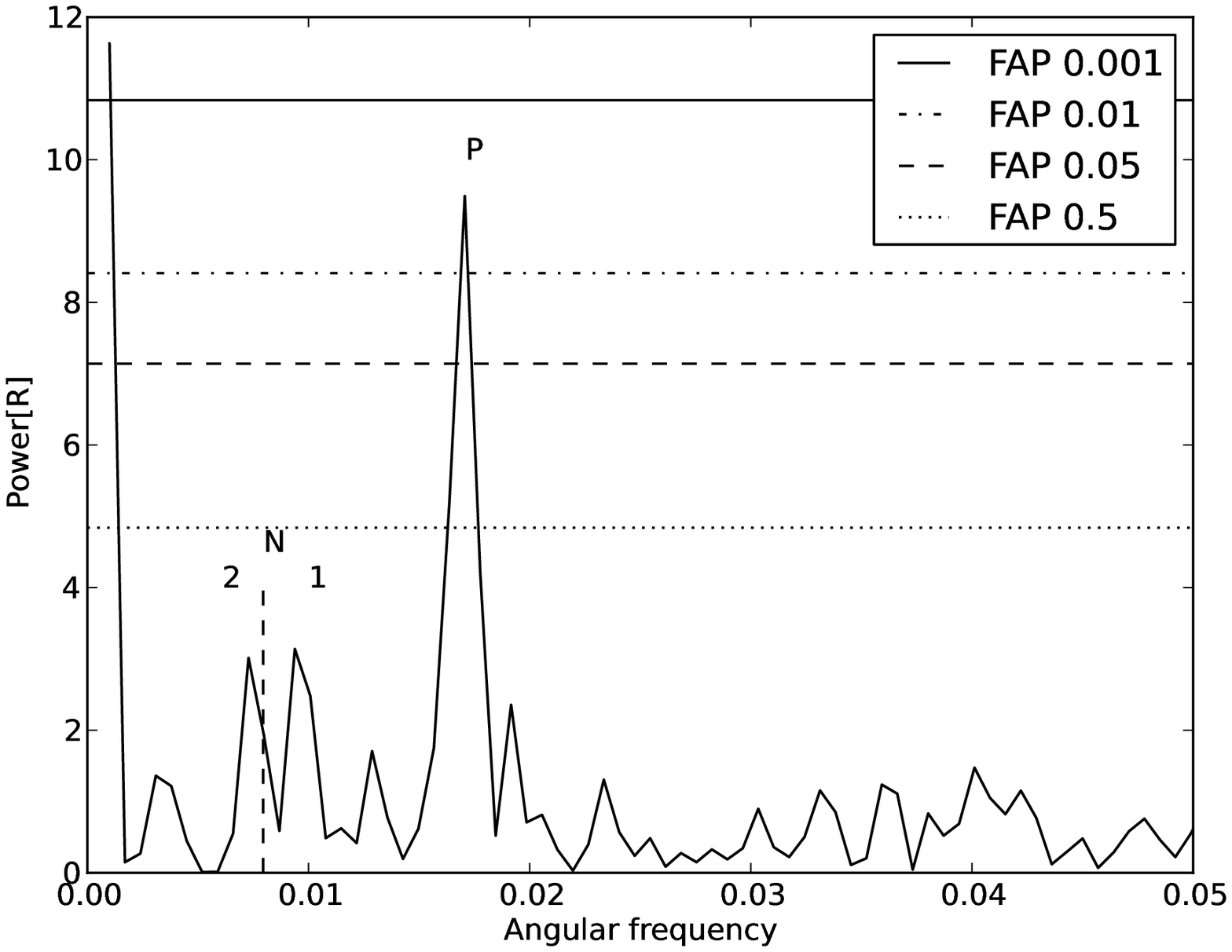}
\includegraphics[width=9cm]{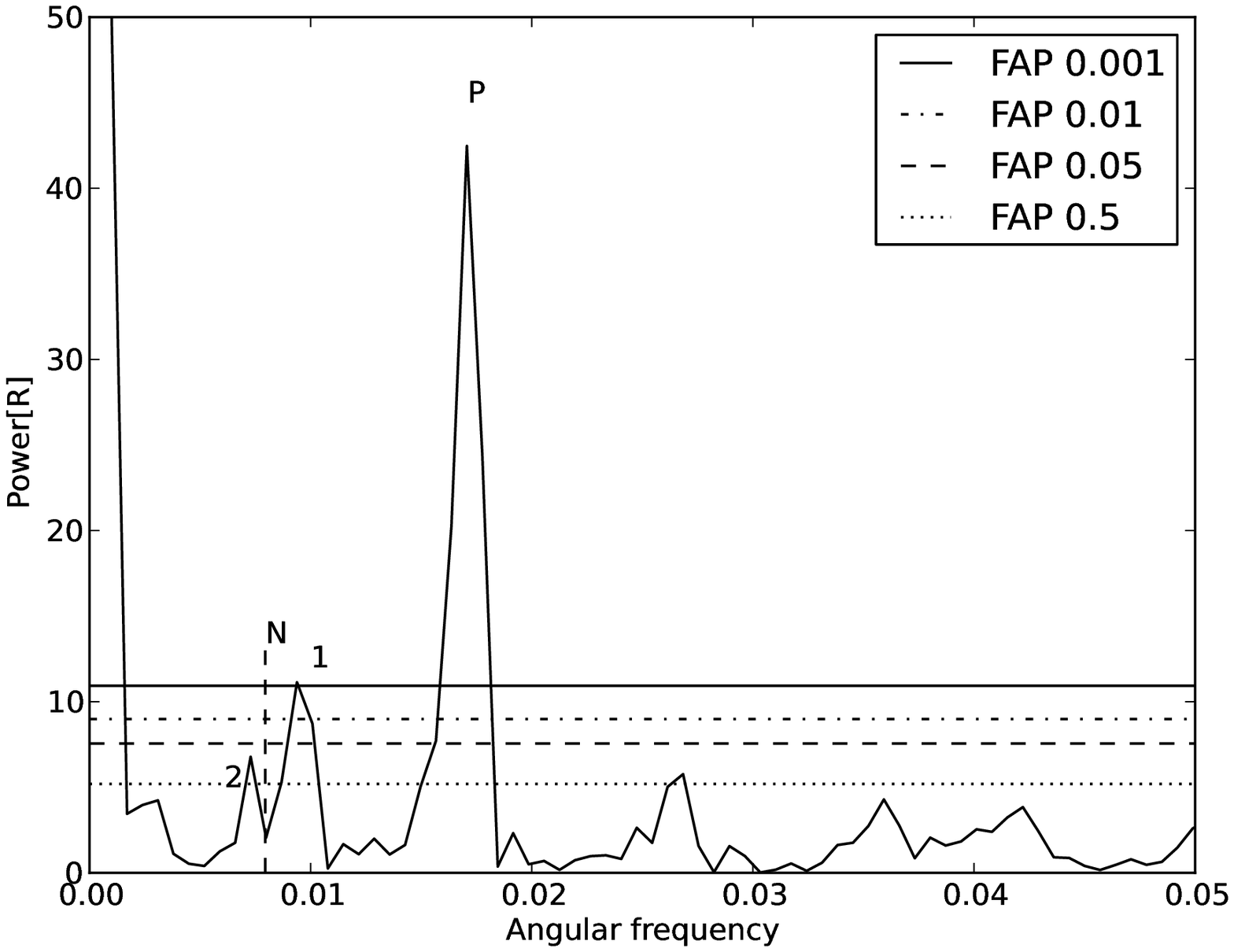}
\caption{Left: Lomb Scargle periodograms of the ratio of the red-to-blue 
line-segment fluxes $R$ of H$\alpha$ (left) and H$\beta$ (right) lines 
(false-alarm probability (FAP) lines are also indicated).} \label{lomb}
\end{figure*}

 Fig. \ref{lomb} gives the Lomb Scargle periodograms of 
the ratio of the red-to-blue line-segment fluxes ($R$=$F$(red)/$F$(blue)) of 
the H$\alpha$ and H$\beta$ lines.
As it can be seen in Fig. \ref{lomb} there are three peaks, 
where one peak is clearly distingueshed and higher than 0.01 (99\%) of 
false-alarm probability (FAP).\footnote{The false-alarm probability (FPA) 
describes the probability that at least one out of M independent power values 
in a prescribed search band of a power spectrum computed from a white-noise time 
series is expected to be as large as or larger than a given value. 
Note that the low FAP values indicate a
high degree of significance in the associated periodic signal.} 
This peak (denoted as P-peak) at angular frequency of $\sim$0.017 
in case of both lines, corresponds to the period of $\sim$370 days. The
angular frequency of 0.00795 corresponding to the period of
$\sim$790 days found by \cite{n97} and \cite{ge07} (denoted with N), seems to be located 
between two other peaks of lower significance (denoted as 1- and 2-peak in 
Fig. \ref{lomb}): the 1-peak is at angular frequency of $\sim$0.00968 for 
H$\alpha$ (0.0094 for H$\beta$) corresponding to the period of
$\sim$650 days (670 days), and the second peak is at angular frequency of
$\sim$0.0074 (0.0077) corresponding to the period of $\sim$850 days (815 days).

\section{Discussion}

{  Here we discuss our results, but before that, we should point out that the
 host-galaxy of Arp 102b is an E-galaxy with 
strong stellar absorption lines (Mg Ib,Na ID, CaII H,K etc.) typical for this type of galaxies.
The host galaxy contininuum can strongly affect the observed spectra,
 therefore we 
determined the host-galaxy contribution
to the blue and red fluxes of the observed continuum as well as to the H$\alpha$, H$\beta$ emission line fluxes
(given in 
absolute units, see Table \ref{host-gal}). To estimate the host galaxy continuum contribution,
we  used Arp 102b and NGC 4339 (E0) spectra  observed in the same night
 under the same  good weather conditions (see \S 2.4).  
To subtract the 
NGC 4339 galaxy spectrum from  Arp 102B  one, we use Mg Ib, and found that the best 
fit (when the Mg1b and Na ID are completely removed from the composite spectra) 
is with a 75$\pm$3\% of the host-galaxy
contribution to the total Arp 102B continuum at $\sim$5100 \AA\ in the considered spectrum. However, depending on 
AGN activity phase, the host galaxy contribution to the continuum at  $\sim$5100 \AA\ was between $\sim$ 60\% and
$\sim$ 80\% in the monitoring period.

Additonally,
 we estimated the narrow line contributions to the total and line-segment fluxes of H$\alpha$ and H$\beta$ 
  (see Table \ref{narrow}). Consequently, further in the discussion we will point out the parts 
  where these contributions are improtant.} 

\subsection{Variation of the broad lines and continuum}

As it can be seen in Fig. \ref{lc} the broad line and continuum
fluxes change for 10-30\% during the monitoring period (1987--2010).
Some flares with amplitude of $\sim$(10-20)\% are observed, and
these are especially strong in H$\alpha$ and continuum. {  The flare like
variation of Arp 102B was noted in \cite{ge07}. 

Actually, in 1987 (from  
Jun28 -- JD2446975 to Jun30 -- JD2446977), and in 1998 (from May6 -- JD2450940 to Jul26 -- 2451021) 
there has been a rise in the observed flux of the H$\alpha$ and H$\beta$ emission lines 
of 30-40\%, in the form of a flare-like event with duration of about 80 days in 1998, but in the 
continuum such strong oserved flux changes were not observed.} In Table
\ref{tab6b} we obtain that the different {  observed} mean line fluxes between
1987 and 1988--1994, and also between 1998 and 1999--2010 changed
significantly, more than the continuum fluxes 32--35\% in H$\beta$
and $\sim$(38--39)\% in H$\alpha$, compared to 6--13\%  and 7--19\%
in the red and blue continuum, respectively.

{  However, after  the host-galaxy
contribution subtraction we obtained that in the same (above noted) periods 
the relative amplitude of the variability of the mean corrected continuum (i.e. agn-continuum) 
is significantly larger: 1987/(1988--1994)$\sim$1.3 and 1998/(1999--2010)$\sim$1.6
than it was in the host galaxy+agn continuum: 1987/(1988--1994)$\sim$1.07 and 
1998/(1999--2010)$\sim$1.19--1.13. These variations are comparable with relative line flux variations.

A similar result was obtained for the parameter F(var) - the variation amplitude 
of fluxes with respect to the mean flux (Table \ref{tab6}, host-galaxy corrected data).
As seen from Table \ref{tab6},  F(var) in the corrected continuum fluxes (in agn-continuum)
has a significantly larger continuum variability amplitude, than the observed
total continuum (for continuum at 5100A: 31\% against $\sim$9\% in non-corrected spectra;
 for continuum at 6200A: 29\% against $\sim$7\% in non-corrected spectra).
On the other hand,  F(var) in the H$\alpha$ and H$\beta$ lines, after removing the contribution of the host-galaxy,
remaineds almost unchanged (see host galaxy corrected data in Table \ref{tab6}).  
In the corrected blue and red continuum there are some possible flares of amplitude up to 30\%
with a duration of a few (2-3) days (see Table \ref{flares}), but these flares are not detected in 
the corrected H$\alpha$
and H$\beta$ line fluxes.

As noted above, we observed strong changes of mean  fluxes in the corrected  continuum in almost 
constant changes mean flux in the broad lines i.e. greatly reduced mean continuum flux (in 1.3 times to in 
1988--1994 and 1.6 times in 1999--2010) while changes in the broad line fluxes remaine at the same level 
(1.3 times).
The amplitude of variability of the continuum and line fluxes stays closer each other, when we
corrected the line fluxes for the narrow line contributions, but  a small correlation between the agn-continuum  
and line fluxes remains (see Fig. \ref{hahb}).

Additionally, we should note  that small  amount of the optical continuum emission may not be 
too effective in the broad line reproduction. As e.g. } Chen et al. (1989) have shown that 
the gravitational energy liberated in
a standard accretion disk is not much larger than the observed line
emissivity in Arp 102B, so an extra source of heating possibly may be required.
They hypothesize that ion-supported tori might account for the
unusual properties of Arp 102B. {  On the other hand  the amplitude of variability
of the optical continuum is comparable with the line one. Additionally,
 there may be situation that the amplitude of variability and intensity
of the far UV and X-ray contina is even strong for driving of the line variability and intensity. Also, it is 
well known \citep[see e.g.][]{vi13} that the variability of the high-energy bands (as X-ray) often is
 more pronounced that the variability of the UV/optical band. But the question of the weak correlation between
 the continuum and line fluxes remains.}

The most interesting result is the almost lack of correlation
between the continuum and broad lines (r$\sim$0.31, see Fig.
\ref{Hb_cnt} and Fig. \ref{Ha_cnt}) We obtained delays between
H$\alpha$ and H$\beta$ emission line fluxes and continuum flux {  (lags
$\sim$15-20 days)}, see Table \ref{ccf}, using different CCF methods.
However the errors in lags are huge due to the fact that the flux
variability in both lines and continuum during the monitoring period
was small ($\sim$10-20\%, see Table \ref{tab6}).  Note here that
in the case of AGNs with very broad lines, the lag calculation is
difficult \citep[see e.g. the cases of NGC7603 and Mrk926
in][respectively]{ko00,ko10}. The weak correlations between the
line and continuum fluxes and non-confidently  lags-data indicate
also a possible another sources of photoionization in the BLR except
the central source (e.g. collisions of orbiting discrete clouds in
disk \citep{s00})

Additionally, similar as in some single peaked AGNs  \citep[see
e.g.][]{gp03}, the H$\beta$ and H$\alpha$ lines show intrinsic
Baldwin effect, or significant  anticorrelation between the
continuum flux and equivalent widths of broad lines (see Fig.
\ref{BE}). The anticorrelation tends to be stronger in H$\beta$ (and
statistically more significant) than in the H$\alpha$ line.

\subsection{Black hole mass in Arp 102B}

Mostly, reverberation calculations have been applied on the cases
which give black hole masses within the range $10^7 - 10^9
M\odot$. Up to now, emission-line lags have been measured for a number
of AGNs by using the cross correlation between the continuum and emission-line
light curves.

{  Using the virial theorem, the mass of black hole ($M_{BH}$) is \citep{pet98,Wa99}:

$$M_{GRAV}=f{\Delta V_{FWHM}\cdot R_{BLR}\over G},$$
where $\Delta V_{FWHM}$ is the  orbital velocity at that radius $R_{BLR}$ of the BLR, and it is estimated by the width
 of the emission line (specifically, the variable part of the line); the $f$ is the factor that depends on the geometry of
the BLR. Using a sample disk model for the Arp 102B BLR, one can use the relation givne in \cite{on04}:
 $$ M_{BH}=f{\Delta V_{FWHM}\cdot R_{BLR}\over G}\cdot {\sin(i)\over{2\ln(2)}},$$
where $i$ is the inclination of the disk. Here we take $f=5.5$ as estimated by  
\cite{on04} and $i\sim 30^\circ$ as given in \cite{ch89}.

The velocity dispersion of the disk can be estimated as  \citep[see][]{la09}
$$\Delta v={FWHN(H\beta)\over{8\sin(i)}},$$ 
where for our measurements $v_{\rm FWHM}({\rm rms})\approx 15000$\,km\,s$^{-1}$
(Fig \ref{mp}b) and  $i\sim 30^\circ$ gives $\Delta v=3750\rm \,km\,s^{-1}$. 
Taking into account that our estimate is $R_{BLR}\approx 20$ light days (Table \ref{ccf}),
we obtaine a mass of the Arp 102B black hole to be $\sim 1.1\cdot 10^8 M_\odot$.

There are other estimates of the BH mass in Arp 102B;}
using the  hot spot model for the explanation of the variations
of the line profile,  \cite{n97} estimated the black hole mass in Arp 102B to
be $2.2 \times 10^8 M_{\odot}$, while from the rotational clouds model, \cite{s00}
estimated the central body mass of $3.5 \times 10^8 M_{\odot}$.

{  It is interesting to compare the obtained BH masses mentioned above with  $M-\sigma*$ relation. 
Recently, the   $M-\sigma*$ relation was given by \cite{gu09} as:
$$\log(M_{BH}/M_\odot) = \alpha + \beta\log(\sigma*/200 {\rm km\ s^{-1}})$$
where $\alpha = (8.12 \pm 0.08$ and $\beta= 4.24 \pm 0.41$ are constants, and $\sigma*$ is the stellar disspersion.
The disperssion velocity for Arp 102B was given by \cite{ba02} as $\sigma_*=188 \pm 8\
\rm km \ s^{-1}$ that gives an estimated Arp 102B black hole mass as $M_{BH}\sim 1.01\cdot 10^8 M\odot$. 

Comparing the data obtained from our estimates, as well as estimates of  \cite{n97} and  \cite{s00}, the 
$M-\sigma*$ relation gives a smaller mass for Arp 102B than estimated by \cite{n97} and  \cite{s00} 
(two and three times, respectively). However, comparing our results and that obtained from  $M-\sigma_*$ relation
the agreement is very good $M_{rev}/M_{M-\sigma*}\sim 1.1$.

The black hole masses of AGNs with double-peaked lines obtained with  $M-\sigma_*$ relation seem to disagree 
systematically with the virial masses obtained from one epoch measurements \citep[see][]{wu04,le06}. 
But it seems that the revereberation method
gives better agreement, that is the case in another double-peaked AGN, 3C390.3, namely \cite{di12} found that 
in the case of double-peaked line emitter 3C390.3 the reverberation based mass of the BH also agrees with  $M-\sigma_*$ relations.}

\subsection{H$\alpha$ and H$\beta$ flux variability}

We explored changes in the total line fluxes of H$\alpha$ and
H$\beta$ as well as in the line segments (see Table \ref{tab5}).
There are  changes in the line profiles (blue and red double peaks,
see \S3.4) and their rms. The observed line parts in both lines  are varying around $\sim$20\% (see Table
\ref{tab6}). {  The correlations between the line segments are significant, however the slopes of the best fit
are not are not consistent with one, only in the case of the H$\beta$ red wing vs. the H$\beta$ line core
the slope follows of 1 (see Fig. \ref{wings}).} 

As seen from Fig. \ref{wings_c} there
is only a weak correlation between fluxes of different segments of
both lines and continuum flux (r$\sim$0.3--0.4). However, we find a
very good correlation (r$\sim$0.8) between the H$\alpha$ and
H$\beta$ (for total line and for line-segment fluxes, see Fig.
\ref{hahb}), and between the fluxes of the blue, red, and core
segments of the  H$\alpha$ and H$\beta$ (Fig. \ref{wings}), that is
in agreement with the disk geometry.

On the other hand, we found an indication of periodical changes in
the peak ratio (i.e. red-to-blue line-segment flux ratio), with the
period of around 370 days, that is smaller than the period obtained
by \cite{n97} {  and  \cite{ge07}}. However, we found two peaks in the periodogram very
close to the period obtained by \cite{n97} (with lower significance,
especially for the H$\alpha$ line where the peaks are below 50\% of
the false-alarm probability line). \citet{n97} pointed out  that the determined  period
seems to be an orbital motion through the evolution of the  ratio of
red to blue fluxes (or, alternatively, through the evolution of
parameter $\theta$, the azimuthal angular extent of the hot spot, in
the models). {  Also \cite{ge07} interpreted the observed oscillation as two
different bright spots orbiting at different radii in the disk at different times, but the authors
mentoined the problem with this model, since the two bright
spots rotation periods did not yield consistent values of the black hole mass. However, such a scenario --
two different bright spots orbiting at different radii -- successfully explains the line shape variability in
the case of 3C390.3 \citep[see][]{jov10}.}

Indeed, the only proposed source of variability in an
AGN, that would cause simple periodical sinusoidal variation in this
ratio with little apparent decay (in amplitude or frequency) for
nearly two complete cycles, is orbital in nature. But, the problem
here is that there are indications for three possible rotation
period \citep[two with lower significance close to period given
in][and one that is almost two times smaller]{n97}.

{  Alternativelly, it is
possible that the rotation of the bright spot in the accretion disk is not pure Keplerian; 
the bright spot, as e.g. may be affected  by a wave in the disk that rotates at a 
different speed \citep[see][]{le10}. 
This may be an explanation for multi-periodical oscillations seen in the spectra of Arp 102B, but this we
will consider in more details in the forthcoming paper.}

\section{Conclusion}

Here we present a long term (1987--2010) spectroscopical
observations of Arp 102B in the optical band, an AGN with prominent
double-peaked broad line profiles. We investigated the continuum and line variations during this period and from our 
investigations we can outline the following conclusions:

i) {  We found a significant contribution the host-galaxy continuum (between $\sim$ 60\% and $\sim$ 80\% in the monitoring
period) to the 
total observed continuum in Arp 102B.  The corrected (agn) continuum flux 
of Arp 102b approximately has a flat shape and it contributes around $\sim$25\%  to the total observed (host+agn)
continuum (Fig. \ref{host}, bottom). The flux variation of the agn-continuum  
has a significantly larger amplitude than it can be seen in  the observed total
 continuum (Table \ref{tab6}, F(var)). However,  the H$\alpha$ and H$\beta$ line fluxes
 are not sensitive to the host-galaxy correction, i.e.
  the variability amplitude remains almost unchanged (Table \ref{tab6}). 
As noted in \S 3, the corrected AGN continuum shows some possible flare-like events, i.e. 
an increase in the flux up to $>30$ \% within a few (2-3) days
(see Table \ref{flares}), but correspond flare-like events were not observed in
the broad H$\alpha$ and H$\beta$ light curves.}

ii) In different parts of the monitoring periods, mean {  observed} fluxes of the H$\alpha$,
H$\beta$ broad emission lines and blue/red continuum variations are
small (5-10\%, Table \ref{tab6}). But changes of the different mean
fluxes in the lines between 1987 and (1988--1994), and between  1998
and (1999--2010) are significantly larger than changes of continuum
fluxes. {   However,  after subtracting 
the host-galaxy contribution (in the the same periods) the amplitude of changes
of the mean continuum fluxes  (i.e. agn-continuum) is significantly
larger (1987/(1988--1994)$\sim$1.3 and 1998/(1999--2010)$\sim$1.6). After removing the narrow line 
contribution the line flux amplituda variation, the
 relative AGN continuum changes are comparable  (or even greater) with	
 the flux changes in lines.}
A rise in the  H$\alpha$ and H$\beta$ emission line fluxes
about 30--40\% in 1987 and 1998 is similar to a long flare duration
of about 80 days in 1998 (in 1987 observations are only 3 days).
However the mean fluxes during single years 1987, 1998 and during
periods 1988--1994 and 1999--2010 are constant inside this periods
within 5-10\% errorbars (Table \ref{tab6b}).

iii) The correlations between the observed fluxes of the  H$\alpha$
and H$\beta$ lines, as well as of their line segments, with
continuum fluxes (Figs. \ref{Hb_cnt}-\ref{Ha_cnt}) is very weak
(r$\sim$0.3--0.4). That points to additional sources of ionization
in the BLR apart from the central AGN  continuum source. But there
is a good linear relation between the observed fluxes in different
line segments (r$\sim$0.8), i.e. between blue and red wings, and
line core (Fig. \ref{wings}), and between the H$\alpha$ and H$\beta$
(Fig. \ref{hahb}) for total line and line segment fluxes, that
indicates the same geometry for both emission regions.  Also the
anticorrelation between the continuum flux and equivalent widths of
the H$\alpha$ and H$\beta$ lines (intrinsic Baldwin effect) is
observed (r$\sim$0.37 for H$\alpha$, r$\sim$0.49 for H$\beta$, see
Fig. \ref{BE}), as in some Seyfert galaxies \citep{gp03}.

iv) During the monitoring period, the broad H$\alpha$ and H$\beta$
lines of Arp 102B show double-peaked profiles.  The blue peak is
located around -5,000 km s$^{-1}$ and red around 6,000 km s$^{-1}$
from the line center. A big distance between the peaks ($\sim$
11,000 km s$^{-1}$) indicates a fast rotating disk, that is probably
close to the black hole. As it can be seen from Fig. \ref{mp} (mean
profiles and rms), the changes in the line profile have also
double-peaked rms, that indicates that changes in the blue wing is
significantly (some) bigger, than in the red wing. In the rms there
is one central peak, that may be caused by a central component in
the BLR \citep[][]{pop04,bo06,bo09}.

v) From the Lomb-Scargle periodogram method, applied on the measured 
red-to-blue peak flux ratio ($R$=$F$(red)/$F$(blue)) 
of H$\alpha$ and H$\beta$ lines, we found possible periodical variations 
(signals) in this ratio. We found a period of around 370 days, and two
additional peaks close to the period found by \cite{n97}.

vi) Several cross-correlation methods of the continuum and H$\alpha$,
H$\beta$ broad emission line fluxes indicates a lag of {  $\sim$15-20}
days with large errorbars, caused by the small flux variations in
period monitoring and not so good data sampling.

vii) From mean and rms profiles both emission lines we found
FWHM(rms)$\sim$15,000 km s$^{-1}$ and from the CCF analysis we
obtained a lag$\sim$20 days, thus we estimated the reverberation
central black hole mass of {  $M_{\rm rev}\sim 1.11\times10^8 M_{\odot}$,
that is smaller than previous estimates \citep{n97,s00}, but is
in an agreement with estimated black hole mass obtained from the $M-\sigma*$ relation.}

The main property of the double-peaked Seyfert galaxy Arp 102B is
the long-term variability on the timescale of months to some years,
which is consistent with the dynamical timescale of an accretion
disk. The observed long flare in H$\alpha$ and H$\beta$ with the
duration of about 80 days in 1998 cannot be only due to the
ionization by the AGN-continuum source. Also, due to the lack of
correlations with the AGN-continuum this variability can be
attributed to inhomogeneities in the line-emitting disk, i.e., hot
spots, spiral arms, eccentricity, and warps. We will consider in
more details the nature of the broad spectral shape variability and
consequently the BLR geometry in the forthcoming Paper II.

\section*{Acknowledgments}

This work was supported by INTAS (grant N96-0328), RFBR (grants
N97-02-17625 N00-02-16272, N03-02-17123, 06-02-16843, N09-02-01136,
12-02-00857a, 12-02-01237a), CONACYT research grants 39560, 54480,
and 151494, and PAPIIT-UNAM research grant IN111610 (M\'exico),
and the Ministry of Education and Science of Republic of Serbia through the project
Astrophysical Spectroscopy of Extragalactic Objects (176001). L. \v
C. P., W. K. and D. I. are grateful to the Alexander von Humboldt
foundation for support in the frame of program "Research Group
Linkage". W. K. is supported by the DFG Project Ko 857/32-1.
 We thank Moiseev A. for providing some spectra of Arp 102B,
obtained with the 6m telescope. {  We would like to thank to the anonymous referee
for very useful comments and suggestions.}



\begin{thebibliography}{}



\bibitem[\protect\citeauthoryear{Alexander}{2013}]{Al13} Alexander, T, 2013, http://adsabs.harvard.edu/abs/2013arXiv1302.1508A

\bibitem[\protect\citeauthoryear{Antonucci et al.}{1996}]{a96} Antonucci, R., Hurt, T.,  Agol, E. 1996, ApJ, 456, L20



\bibitem[\protect\citeauthoryear{Barth et al.}{2002}]{ba02} Barth, A. J., Ho, L. C., Sargent, W. L. 2002, AJ, 124, 2607




\bibitem[\protect\citeauthoryear{Bischoff \& Kollatschny}{1999}]{bk99}
Bischoff, K., \& Kollatschny, W. 1999, A\&A, 345, 49B



\bibitem[\protect\citeauthoryear{Bon et al.}{2009}]{bo09}
Bon, E., Popovi\'c, L. \v C., Gavrilovi\'c, N., La Mura, G., \&
Mediavilla, E. 2009, MNRAS, 400, 924

\bibitem[\protect\citeauthoryear{Bon et al.}{2006}]{bo06}
Bon, E., Popovi\'c, L., Ili\'c, D., Mediavilla, E. 2006, NewAR, 50, 716

\bibitem[\protect\citeauthoryear{Chen \& Halpern}{1989}]{ch89} Chen, K. \& Halpern, J. 1989, ApJ, 344, 115


\bibitem[\protect\citeauthoryear{Chen et al.}{1989}]{c89} Chen, K. Halpern, J. P., Filippenko, A. V. 1989, ApJ, 339, 742

\bibitem[\protect\citeauthoryear{Chen et al.}{1997}]{c97} Chen, K, Halpern, J. P., Titarchuk, L. G. 1997, ApJ, 483, 194

\bibitem[\protect\citeauthoryear{Dietrich et al.}{2012}]{di12} Dietrich, M., Peterson, B. M., Grier, C. J. et al. 2012, ApJ, 757, 53

\bibitem[\protect\citeauthoryear{Dimitrijevi\'c et al.}{2007}]{dim07} Dimitrijevi\'c, M. S., Popovi\'c,
 L. \v C., Kova\v cevi\'c, J., Da\v ci\'c, M., Ili\'c, D. 2007, MNRAS, 374,1181

\bibitem[\protect\citeauthoryear{Eracleous \& Halpern}{1994}]{eh94} Eracleous, M., \& Halpern, J. P. 1994, ApJS, 90, 1

\bibitem[\protect\citeauthoryear{Eracleous et al.}{1997}]{e97} Eracleous, M., Halpern, J., Gilbert, A., Newman, J. A.,
Filippenko, A.V. 1997, ApJ, 490, 216

\bibitem[\protect\citeauthoryear{Eracleous et al.}{2009}]{er09} Eracleous, M., Lewis,, K.T., Flohic, H. M.L.G.  2009, NewAR, 53, 133


\bibitem[\protect\citeauthoryear{Fathi et al.}{2011}] {fat11}  Fathi, K., Axon, D. J., Storchi-Bergmann, T., Kharb, P.,
Robinson, A., Marconi, A., Maciejewski, W., Capetti, A. 2011, ApJ,
736, 77

\bibitem[\protect\citeauthoryear{Flohic \& Eracleous}{2008}] {fl08}  Flohic, H. M. L. G.,  Eracleous, M. 2008, ApJ, 686, 138 


\bibitem[\protect\citeauthoryear{Gaskell}{2009}]{ga09} Gaskell, C. M. 2009, NewAR, 53, 140
\bibitem[\protect\citeauthoryear{Gilbert \& Peterson}{2003}]{gp03}
Gilbert, K. M., Peterson, B. M. 2003, ApJ, 587, 123


\bibitem[\protect\citeauthoryear{Gezari et al.}{2007}]{ge07} Gezari, S., Halpern, J.~P., \& Eracleous, M.\ 2007, \apjs, 169, 167

\bibitem[\protect\citeauthoryear{Gezari et al.}{2004}] {g04} Gezari, S., Halpern, J. P., Eracleous, M., Filippenko, A. V. 2004, IAUS, 222, 95

\bibitem[\protect\citeauthoryear{G\"ultekin et al.}{2009}]{gu09} G\"ultekin, K., Richstone, D. O., Gebhardt, K. et al. 2009, ApJ, 698, 198


\bibitem[\protect\citeauthoryear{Halpern et al.}{1996}]{h96} Halpern, J. P., Eracleous, M., Filippenko, A. V., \& Chen, K. 1996, ApJ, 464, 704



\bibitem[\protect\citeauthoryear{Jovanovi\'c et al.}{2010}]{jov10}
Jovanovi\'c, P., Popovi\'c, L. \v C., Stalevski, M., Shapovalova, A. I. 2010, ApJ, 718, 168


\bibitem[\protect\citeauthoryear{Kollatschny et al.}{2000}]{ko00} Kollatschny, W., Bischoff, K., Dietrich, M., 2000, \aa, 361, 901

\bibitem[\protect\citeauthoryear{Kollatschny \& Zetzl}{2010}]{ko10} Kollatschny, W., \& Zetzl, M.\ 2010, \aap, 522, A36

\bibitem[\protect\citeauthoryear{La Mura et al.}{2009}]{la09} 	La Mura, G., Di Mille, F., Ciroi, S., Popovi\'c, L. \v C., Rafanelli, P.
2009, ApJ, 693, 1437


\bibitem[\protect\citeauthoryear{Lewis \&  Eracleous}{2006}]{le06} Lewis, K. T., Eracleous, M. 2006, ApJ, 642, 711

\bibitem[\protect\citeauthoryear{Lewis et al.}{2010}]{le10} Lewis, K. T., Eracleous, M., Storchi-Bergmann, T,  2010, ApJS, 187, 416

\bibitem[\protect\citeauthoryear{Lomb}{1976}]{l76} Lomb, N.R., 1976, \apss, 39,  447

\bibitem[\protect\citeauthoryear{Miller \& Peterson}{1990}]{mp90} Miller, J. S., Peterson, B. M. 1990, \apj, 361, 98

\bibitem[\protect\citeauthoryear{Newman et al.}{1997}]{n97} Newman, J. A., Eracleous, M., Filippenko, A. V.,  Halpern, J. 1997, ApJ, 485, 570

\bibitem[\protect\citeauthoryear{O'Brien et al.}{1998}]{ob98} O'Brien P.T., Dietrich, M., Leighly, K et al. 1998, \apj, 509, 163

\bibitem[\protect\citeauthoryear{Onken et al.}{2004}]{on04}
Onken, C. A., Ferrarese, L., Merritt, D., Peterson, B. M., Pogge, R. W., Vestergaard, M., Wandel, A., 2004, ApJ, 615, 645

\bibitem[\protect\citeauthoryear{Peterson}{1993}]{pet93}  Peterson, B. M. 1993, \pasp, 105, 207

\bibitem[\protect\citeauthoryear{Peterson et al.}{1998}] {pet98} Peterson, B.~M.,
Wanders, I., Bertram, R., et al. 1998, \apj, 501, 82

\bibitem[\protect\citeauthoryear{Peterson et al.}{1999}]{pet99} Peterson, B.~M.,
Barth, A.~J., Berlind, P., et al.\ 1999, \apj, 510, 659

\bibitem[\protect\citeauthoryear{Peterson et al.}{1994}]{pet94} Peterson, B.~M.,
Berlind, P., Bertram, R., et al.\ 1994, \apj, 425, 622

\bibitem[\protect\citeauthoryear{Peterson et al.}{2002}]{pet02} Peterson, B.~M.,
Berlind, P., Bertram, R., et al.\ 2002, \apj, 581, 197

\bibitem[\protect\citeauthoryear{Peterson \& Collins}{1983}]{pc83} Peterson, B.M., \& Collins II, G.W. 1983, \apj,270,71

\bibitem[\protect\citeauthoryear{Peterson et al.}{1995}]{pet95} Peterson, B.~M., Pogge, R.~W., Wanders, I., Smith, S.~M.,
\& Romanishin, W.\ 1995, \pasp, 107, 579

\bibitem[\protect\citeauthoryear{Popovi\'c et al.}{2004}]{pop04}
Popovi\'c, L. \v C., Mediavilla, E., Bon, E., \& Ili\'c, D. 2004
A\&A 423, 909

\bibitem[\protect\citeauthoryear{Popovi\'c et al.}{2001}]{pop01} Popovi\'c, L. \v C.. Mediavilla, E. G.,  Mu\~noz, J. A.2001, A\&A, 378, 295


\bibitem[\protect\citeauthoryear{Popovi\'c et al.}{2011}]{pop11} Popovi\'c, L. \v C., Shapovalova, A. I., Ili\'c, D. et al.
2011, A\&A, 528A, 130




\bibitem[\protect\citeauthoryear{Scargle}{1982}]{sc82} Scargle, J.D., 1982, \apj,  263,  835

\bibitem[\protect\citeauthoryear{Shapovalova et al.}{2001}]{sh01} Shapovalova, A.~I., Burenkov,
A.~N., Carrasco, L., et al.\ 2001, \aap, 376, 775

\bibitem[\protect\citeauthoryear{Shapovalova et al.}{2004}]{sh04} Shapovalova, A.I., Doroshenko, V.T., Bochkarev, N.G, et al. 2004, \aap, 422, 925

\bibitem[\protect\citeauthoryear{Shapovalova et al.}{2010}]{sh10} Shapovalova, A. I., Popovi\'c, L.\v C., Bochkarev, N.G., et al. 2010, \aap, 517A, 42

\bibitem[\protect\citeauthoryear{Shapovalova et al.}{2009}]{sh09} Shapovalova, A. I., Popovi\'c, L.\v C., Bochkarev, N.G., et al. 2009, NewAR, 53, 191

\bibitem[\protect\citeauthoryear{Shapovalova et al.}{2012}]{sh12} Shapovalova, A.I., Popovi\'c, L.\v C., Burenkov, A. N., et al. \ 2012, \apjs, 202, 10

\bibitem[\protect\citeauthoryear{Shapovalova et al.}{2008}]{sh08} Shapovalova, A.I., Popovi\'c, L.\v C., Collin, S., et al. \ 2008, \aap, 486, 99

\bibitem[\protect\citeauthoryear{Sergeev et al.}{2000}]{s00} Sergeev, S. G., Pronik, V. I.,  Sergeeva, E. A. 2000, A\&A, 356, 41

\bibitem[\protect\citeauthoryear{Stauffer et al.}{1983}]{s83} Stauffer, J., Schild, R., Keel, W. 1983, ApJ, 270, 465

\bibitem[\protect\citeauthoryear{Sulentic et al.}{2000}]{sul00} Sulentic, J. W., Marziani, P., Dultzin-Hacyan, D.2000, ARA\&A, 38, 521

\bibitem[\protect\citeauthoryear{Sulentic et al.}{1990}]{su90} Sulentic, J. W., Zheng, W., Calvani, M., Marziani, P. 1990, ApJ, 355, 15


\bibitem[\protect\citeauthoryear{van Groningen \& Wanders}{1992}]{vg92} Van Groningen, E. \& Wanders, I.\ 1992, PASP, 104, 700

\bibitem[\protect\citeauthoryear{Vagnetti, et al.}{2013}]{vi13} Vagnetti, F.; Antonucci, M., Trevese, D.Veron, P., 
2013, A\&A, 550A, 71

\bibitem[\protect\citeauthoryear{Wandel et al.}{1999}] {Wa99} Wandel, A., Peterson,
B.~M., \& Malkan, M.~A.\ 1999, \apj, 526, 579

\bibitem[\protect\citeauthoryear{Wu \& Liu}{2004}]{wu04} Wu, X.-B., Liu, F. K. 2004, ApJ, 614, 91

\bibitem[\protect\citeauthoryear{Zu et al.}{2011}]{zu11} Zu, Y., Kochanek, C. S., Peterson, Bradley M. 2011, ApJ, 735, 80.

\end{thebibliography}
\end{document}